\documentclass{article}
\usepackage{epsfig}
\usepackage{hyperref}
\catcode`\@=11
\@addtoreset{equation}{section}
\def\theequation{\thesection.\arabic{equation}}

\def\@normalsize{\@setsize\normalsize{15pt}\xiipt\@xiipt
\abovedisplayskip 14pt plus3pt minus3pt%
\belowdisplayskip \abovedisplayskip
\abovedisplayshortskip \z@ plus3pt%
\belowdisplayshortskip 7pt plus3.5pt minus0pt}

\def\small{\@setsize\small{13.6pt}\xipt\@xipt
\abovedisplayskip 13pt plus3pt minus3pt%
\belowdisplayskip \abovedisplayskip
\abovedisplayshortskip \z@ plus3pt%
\belowdisplayshortskip 7pt plus3.5pt minus0pt
\def\@listi{\parsep 4.5pt plus 2pt minus 1pt
       \itemsep \parsep
       \topsep 9pt plus 3pt minus 3pt}}

\relax
\catcode`@=12
\catcode`\@=11
\def\section{\@startsection{section}{1}{\z@}{3.5ex plus 1ex minus
     .2ex}{2.3ex plus .2ex}{\large\bf}}
\def\thesection{\arabic{section}}

\def\appendix{\setcounter{section}{0}
   \def\thesection{Appendix \Alph{section}}
   \def\theequation{\Alph{section}.\arabic{equation}}}

\begin{document}

\begin{titlepage}
\begin{center}
{\Large   Dynamical  Symmetry Breaking   in      \\
  Supersymmetric
$SU(n_c)$  and $USp(2 n_c)$ Gauge
Theories   }
\end{center}

\vspace{1em}
\begin{center}
{\large    Giuseppe Carlino$^{(1)}$, Kenichi Konishi$^{(2,3,4)}$ \\
and \\ Hitoshi Murayama$^{(5,6) }$ }
\end{center}
\vspace{1em}
\begin{center}
{\it {
Department of Physics, University of Wales Swansea$^{(1)}$,
\\Singleton Park,
Swansea SA2 8PP, (UK)\\
Dipartimento di Fisica, Universit\`a di Pisa$^{(2)}$   and   Sezione di Pisa,
     \\    Istituto Nazionale di Fisica Nucleare$^{(3)}$, \,\,
Via Buonarroti, 2, Ed.B-- 56127 Pisa (Italy)\\
Department of Physics,  University of Washington$^{(4)}$, \,\,
Seattle,   WA 19185 (USA)   \\
Department of Physics$^{(5)}$ \\
University of California, Berkeley, CA 94720 (USA) \\
Lawrence Berkeley National Laboratory$^{(6)}$\\
1, Cyclotron Road, Berkeley, CA 94720 (USA)\\
E-mail: g.carlino@swansea.ac.uk;  konishi@phys.washington.edu;
murayama@lbl.gov
}   }
\end{center}
\vspace{3em}
\noindent
{\bf ABSTRACT:}
{ We find   the phase and  flavor symmetry breaking pattern   of each  $N=1$
    supersymmetric vacuum of  $SU(n_c)$ and $USp(2 n_c)$ gauge
theories,   constructed
   from the exactly solvable $N=2$
    theories  by perturbing them with  small adjoint and generic bare
    hypermultiplet (quark) masses.     In $SU(n_{c})$ theories with
$n_f \leq n_c$   the vacua are labelled by an integer  $r$, in
which the flavor
$U(n_{f})$
    symmetry  is dynamically broken to $U(r)
    \times U(n_{f}-r)$  in the limit of vanishing bare hyperquark
masses.  In the $r=1$ vacua
the dynamical symmetry breaking is caused  by the condensation of
magnetic  monopoles in the $\underline{n_{f}}$
    representation.  For general $r$, however, the monopoles in the
    $\underline{{}_{n_{f}}C_{r}}$ representation, whose condensation
    could  explain the flavor symmetry breaking but would  produce
    too-many Nambu--Goldstone multiplets, actually ``break up'' into
    ``magnetic quarks'':  the latter with nonabelian interactions
condense and induce confinement and dynamical
symmetry breaking.
   In $USp(2n_c)$ theories
    with $n_f \leq n_c + 1$, the flavor $SO(2n_f)$ symmetry is
    dynamically broken to $U(n_f)$, but with no description in terms of
    a weakly coupled local field theory.   In both $SU(n_c)$
    and $USp(2 n_c)$ theories,  with larger numbers  of quark flavors,
besides the vacua with these properties,
there exist
    also  vacua in free magnetic phase,  with unbroken global symmetry.  }

\vspace{1.5em}
\begin{flushleft}
IFUP-TH 2000-8; SWAT-00/255; UCB-PTH-00/07; LBNL-45323
\end{flushleft}

\begin{flushright}
May    2000
\end{flushright}
\end{titlepage}

\newcommand{\1}{{\Bbb I}}
\newcommand{\Z}{{\Bbb Z}}
\newcommand{\beq}{\begin{equation}}
\newcommand{\eeq}{\end{equation}}
\newcommand{\bea}{\begin{eqnarray}}
\newcommand{\eea}{\end{eqnarray}}
\newcommand{\beas}{\begin{eqnarray*}}
\newcommand{\eeas}{\end{eqnarray*}}
\newcommand{\defi}{\stackrel{\rm def}{=}}
\newcommand{\non}{\nonumber}
\def\dirac{{\cal D}}
\def\dplus{{\cal D_{+}}}
\def\dminus{{\cal D_{-}}}
\def\dbar{\bar{D}}
\def\L{{\mathcal L}}
\def\H{\cal{H}}
\def\de{\partial}
\def\si{\sigma}
\def\sb{{\bar \sigma}}
\def\rn{{\bf R}^n}
\def\r4{{\bf R}^4}
\def\s4{{\bf S}^4}
\def\ker{\hbox{\rm ker}}
\def\dim{\hbox{\rm dim}}
\def\sup{\hbox{\rm sup}}
\def\inf{\hbox{\rm inf}}
\def\infi{\infty}
\def\nrm{\parallel}
\def\nrmi{\parallel_\infty}
\def\om{\Omega}
\def\Tr{ \hbox{\rm Tr}}
\def\const{\hbox {\rm const.}}
\def\o{\over}
\def\th{\theta}
\def\im{\hbox{\rm Im}}
\def\re{\hbox{\rm Re}}
\def\bra{\langle}
\def\ket{\rangle}
\def\Arg{\hbox {\rm Arg}}
\def\Re{\hbox {\rm Re}}
\def\Im{\hbox {\rm Im}}
\def\diag{\hbox{\rm diag}}

\vspace{1.5em}
\section{Introduction  and    Summary  }

Many beautiful  exact results  on  supersymmetric   $4D$  gauge theories have
been obtained recently,   following   Seiberg and Witten's   breakthough  on
       $N=2$
supersymmetric   theories \cite{SW1}-\cite{SUN}.
One exciting related   development    is  Seiberg's $N=1$ non-Abelian
duality,  present in many cases, between a pair of
theories with different number of color, flavor, and  matter contents, which
describe the
same low-energy physics \cite{Sei}-\cite{Altri}.
Another   is the discovery  of  universal    classes of conformally
invariant  theories  (CFT)   for different values of  color, flavor,  and   in
some cases,   for appropriately tuned  values of the parameters of
the theory \cite{Sei}-\cite{EHIY}.

Still another concerns the microscopic   mechanism of confinement
(e.g.,  monopole condensation)  and dynamical
symmetry breaking, and study of other phases such as the oblique
confinement, upon addition of a
perturbation breaking the $N=2$  supersymmetry to $N=1$ and/or to
$N=0$ \cite{SW1,SW2},\cite{EHsu}-\cite{KT}.
   In fact,   an interesting phenomenon  has been observed   in  the
case of    $SU(2)$
gauge theories  with  various flavors  and    with adjoint mass
perturbation \cite{SW1,SW2}:  confinement is caused by
condensation of magnetic monopoles  carrying  nontrivial   flavor
quantum numbers.
More explicitly,   for $n_{f}=2$, monopoles in the
$(\underline{2},\underline{1})+(\underline{1},\underline{2})$
(spinor) representation of the
flavor $[SU(2)\times SU(2)]/Z_{2}=SO(4)$ group is found to  condense
upon     $N=1$
perturbation $\mu \,\Tr \, \Phi^2$: the flavor symmetry is
necessarily broken to $U(2)$.    For $n_{f}=3$, monopoles in the
$\underline{4}$ (spinor) representation of the flavor $SO(6)$ group
condense with $\mu\neq 0$ and the flavor symmetry is broken to $U(3)$.
In such  systems    spontaneous global
symmetry  breaking  is    caused by the same dyamical mechanism
responsible for confinement.

  In one
of the vacua with
$n_f=3$, though,  the
condensing entity carries magnetic number twice the basic unit   but
is  flavor neutral.  In this case
- which can be interpreted as  oblique confinement \`a la 't Hooft
\cite{TH} - confinement is not accompanied by flavor
symmetry breaking.   For $n_{f}=1,4$, there is no dynamical flavor symmetry
breaking.

   These  results naturally lead to a
conjecture that the condensation of magnetic monopoles with non-trivial flavor
transformation property might in a  general class of systems
explain   the confinement {\it \`a la}\/
`t Hooft, Nambu, Mandelstam \cite{TH,NM}    and the flavor symmetry
breaking,    simultaneously \footnote{Such a
possibility has been critically     discussed recently  in   QCD
\cite{TaKo}.  }.  However, a  simple thought reveals a
problem with this picture.  For instance,
   the monopoles in $USp(2n_{c})$ theories transform under the
spinor representation of $SO(2n_{f})$ flavor symmetry, and their
effective low-energy Lagrangian coupled to the magnetic $U(1)$ gauge
group would have a large  accidental $SU(2^{n_{f}-1})$ flavor symmetry:
their condensation would lead to far too many Nambu--Goldstone
multiplets.  The case of $SU(2)$ gauge theories was special because
the flavor symmetries of the monopole action precisely coincide with
the symmetry of the microscopic theories, somewhat accidentally,
due to the small number of
flavors.
It is not at all obvious   how such a paradox is avoided in
higher-rank theories.

Argyres, Plesser and Seiberg \cite{ArPlSei} studied higher-rank
$SU(n_{c})$ theories with $n_{f} \le 2n_{c}-1$ (asymptotically    free) in
detail.  They showed how the non-renormalization theorem of the
hyperK\"ahler metric on the Higgs branch could  be used to show the
persistence of unbroken non-abelian gauge group at the ``roots'' of
the Higgs branches (non-baryonic and baryonic branches) where they
intersect the Coulomb branch.  Some isolated points on the
non-baryonic roots with $SU(r)$ ($r \leq [n_{f}/2]$) gauge group as
well as the baryonic root (single point) with $SU(\tilde{n}_{c}) =
SU(n_{f}-n_{c})$ gauge group were  found to survive the $\mu \neq 0$
perturbation.
Their main focus, however, was an   attempt to ``derive'' Seiberg's
duality between $SU(n_{c})$ and $SU(\tilde{n}_{c})$ gauge theories
relying on the so-called baryonic root.  Their   ``derivation,'' however,
    was incomplete as it did not produce all components of the ``meson''
    superfield.  The effective
low-energy theory was perturbed
    by a relevant operator (the mass term for the mesons) and did not
    flow to the Seiberg's $N=1$   magnetic theory correctly. \footnote{
We thank P.
    Argyres  and  A. Buchel for discussions on this point.}   On the
other hand,    the issue of flavor
symmetry breaking was not studied at any depth  in    \cite{ArPlSei}.
Their   analysis also left a puzzle
why there were ``extra'' theories at the non-baryonic roots which
seemingly had nothing to do with  Seiberg's    dual theories.  Another
paper by Argyres, Plesser and Shapere addressed  similar questions,
and left puzzles,     in    $SO(n_{c})$ and $USp(2n_{c})$
theories \cite{APS2}.

We  investigate  here the microscopic mechanism of dynamical symmetry
breaking,
   taking    as  theoretical laboratory  the  same class of  theories
studied  by the above mentioned authors
\cite{ArPlSei,APS2},      namely,   theories     constructed from
exactly solvable $N=2$ $SU(n_c)$ and $USp(2n_c)$ gauge
theories with all possible numbers of flavor compatible with
asymptotic freedom, by
perturbing them with a small adjoint   mass (reducing supersymmetry to
$N=1$)   and quark masses.
The Lagrangian of the models has the structure
\beq
\L=     {1\over 8 \pi} \im \, \tau_{cl} \left[\int d^4 \th \,
\Phi^{\dagger} e^V \Phi +\int d^2 \th\,{1\o 2} W W\right]
+ \L^{(quarks)} + \Delta \L +  \Delta^{\prime} \L \, ,
\label{lagrangian}
\eeq
where
\beq
\Delta \L=   \int \, d^2 \theta \,\mu  \,\Tr \, \Phi^2
\label{N1pert}
\eeq
reduces the supersymmetry to $N=1$;
\beq \L^{(quarks)}= \sum_i \, [ \int d^4 \th \, \{ Q_i^{\dagger} e^V
Q_i + {\tilde Q_i}^{\dagger}  e^{ {\tilde V}}    {\tilde Q}_i \} +
\int d^2 \th
\, \{ \sqrt{2} {\tilde Q}_i \Phi Q^i    +      m_i   {\tilde Q}_i    Q^i   \}
\label{lagquark}
\eeq
describes the $n_{f}$ flavors of hypermultiplets (``quarks''),  and
\beq
\tau_{cl} \equiv  {\th_0 \o \pi} + {8 \pi i \o g_0^2}
\label{struc}
\eeq
is the bare $\theta$ parameter and coupling constant.
The $N=1$ chiral and gauge superfields $\Phi= \phi \, + \, \sqrt2 \,
\th \,\psi + \, \ldots \, $, and $W_{\alpha} = -i \lambda \, + \, {i
\o 2} \, (\si^{\mu} \, \sb^{\nu})_{\alpha}^{\beta} \, F_{\mu \nu} \,
\th_{\beta} + \, \ldots $ are both in the adjoint representation of
the gauge group, while the hypermultiplets are taken in the
fundamental representation of the gauge group.
We   shall consider,    besides  the adjoint mass,    small {\it
generic}      nonvanishing
   bare masses for the hypermultiplets
(``quarks'').  The advantage of doing so is   that the only
vacua  retained  are those in which the gauge coupling constant grows
in the infrared.    Another advantage   is that
    all flat directions are eliminated in this way   and one is left
with a finite number of isolated vacua;     keeping track
of this  number   allows us to do   highly nontrivial checks of our
analyses at various stages.

The most salient features of the  result of our analysis  will be
as follows.      The theories  studied  in different regimes,
semiclassical,  large and small adjoint and/or bare quark   masse,
give   a mutually   consistent picture as regards  the
number of the vacua and the dynamical properties  in each of them.
Various  dynamical possibilities, found to be
realized  in the
$SU(n_{c})$ and
$USp(2n_{c})$  theories with a small finite adjoint mass and in the
vanishing bare   quark mass limit,   are   summarized in   Tables
\ref{tabsun} and
\ref{tabuspn}.
\begin{table}[h]
\begin{center}
\vskip .3cm
\begin{tabular}{ccccc}
    label ($r$)    &   Deg.Freed.      &  Eff. Gauge  Group
&   Phase    &   Global Symmetry     \\
\hline
$0$  (NB)     &   monopoles   &   $U(1)^{n_c-1} $               &   Confinement
   &      $U(n_f) $            \\ \hline
$ 1$ (NB)            &  monopoles         & $U(1)^{n_c-1} $        &
Confinement       &     $U(n_f-1) \times U(1) $        \\ \hline
$ 2,.., [{n_f -1\o 2}] $ (NB)   &  dual quarks        &    $SU(r)
\times U(1)^{n_c-r}   $  &    Confinement
&          $U(n_f-r) \times U(r) $
\\ \hline
$ {n_f / 2}  $ (NB)   &   rel.  nonloc.     &    -    &    Almost SCFT
&          $U({n_f / 2} ) \times U({n_f/2}) $
\\ \hline
BR  &  dual quarks     &
$ SU({\tilde n}_c) \times  U(1)^{n_c -  {\tilde n}_c } $                &
Free Magnetic
&      $U(n_f) $         \\ \hline
\end{tabular}
\caption{  Phases of $SU(n_c)$ gauge theory with $n_f$ flavors.     ``rel.
nonloc."  means that
 relatively nonlocal monopoles and dyons   coexist  as low-energy effective
degrees of freedom.
``Confinement" and ``Free Magnetic" refer to phases with $\mu \neq 0$.
``Almost SCFT" means
that the theory is a non-trivial superconformal one when $\mu=0$ but
confines with $\mu \neq 0$.
  NB and BR stand for ``nonbaryonic roots" and
``baryonic roots"  (see Sec.~\ref{sec:EfLagdes})    respectively.
$ {\tilde n}_c
\equiv n_f-n_c$. }
\label{tabsun}
\end{center}
\end{table}

With small but generic bare   quark masses, the  order  parameter of
confining vacua    is  indeed the
condensation of magnetic monopoles for every $U(1)$ factor  \`a la 't
Hooft,  in all cases considered.
   The massless limit, however, is
non-trivial and   exhibits   a  much richer  range of  interesting
dynamical  possibilities.

In $   SU(n_{c})$ theories with  $n_f$ flavors,   the    following
diversity    of     dynamical scenarios  are  realized, according
to the number of flavors and to the particular  vacua considered.
In the first group of vacua  with finite meson or dual quark vacuum
expectation values   (VEVS), labelled by an integer $r$, $r
\le [n_f/2]$,   the system is in confinement phase.   The nature of
the   actual carrier of   the flavor quantum numbers
depends on
$r$.   In  vacua with $r=0$,     magnetic
monopoles  are singlets of  the global $U(n_f)$  group,  hence   no
global symmetry breaking accompanies confinement.
This is analogous to the oblique confinement 't Hooft  suggested for
QCD around     $\theta=\pi$.

In vacua with $r=1$,   the light particles are  magnetic  monopoles
in  the fundamental representation of $U(n_{f})$ flavor group,  and
are  charged
under one of the color  $U(1)$ factors.      Their condensation leads
to    the confinement and
flavor symmetry breaking,  simultaneouly.

In vacua   labelled by $r$,  $2\leq r < n_{f}/2$
but $r \neq n_{f} - n_{c}$,   the grouping of the associated
singularities  on the Coulomb
branch, with multiplicity, ${}_{n_f}\!C_r$,\footnote{In this  paper, we use the
traditional   notation
${}_{n}\!C_r$   for the binomial coefficient most of time: another
frequently  used symbol  is    ${n \choose r}$. }         at first
sight      suggests the condensation of monopoles in the rank-$r$
anti-symmetric tensor representation of the global
$SU(n_f)$ group.    Actually,  this does not occur.    The true
low-energy degrees of freedom of   these theories are
(magnetic) quarks
   plus a  number of singlet monopoles  of  an  effective
$SU(r)\times  U(1)^{n_c-r}$  gauge theory.
Monopoles in  higher representations of $SU(n_f)$ flavor group
probably  exist
semi-classically  as  seen in a Jackiw--Rebbi type   analysis
\cite{JR}.
Such monopoles can be interpreted    as  ``baryons''  made  of  the
magnetic quarks,  which,  interactions being   infrared-free,
however   break up into magnetic quarks    before they become
massless at  singularities on the Coulomb branch.     It is    the
condensation   of
these  magnetic  quarks  that    induces    confinement and  flavor
symmetry breaking,
$U(n_f) \rightarrow U(r) \times U(n_f-r)$,    in these vacua.   The system
thus   realizes  the exact global  symmetry
of the theory in a Nambu-Goldstone mode,
       without
having      unusually    many Nambu-Goldstone
bosons.    It  is a novel   mechanism   for confinement  and
dynamical symmetry breaking.

   In the special cases with
$r= n_f/2$,   still  another  dynamical scenario takes place.   In these
cases,   the interactions among the monopoles become  so strong
that  the low-energy theory describing them is a nontrivial superconformal
theory,  with  conformal invariance explicitly broken by the adjoint or quark
masses.     Although the symmetry breaking pattern is known ($U(n_{f})
\rightarrow U(n_{f}/2) \times U(n_{f}/2)$), the low-energy degrees of
freedom   involve relatively nonlocal   fields and   their
interactions cannot be described  in terms of    a local
action.

Finally,  in the group of vacua labelled by  $r=n_{f}-n_{c}$,    the
low-energy degrees of freedom are
again   magnetic   quarks
and   a  number of singlet monopoles
of       an effective
infrared-free $SU(n_f-n_c)\times U(1)^{2n_c- n_f}$   gauge theory.
   There are two physically distinct sub-groups    of vacua:
one in which   the magnetic quarks   condense   (i.e.   confinement
phase)   with the unbroken
symmetry $U(n_{f}-n_{c}) \times U(n_{c})$  is analogous to the ones
found for generic
$r$'s;    the other is vacua in which
magnetic-quark  do not  condense  and remain as  physically
observable particles at long distances  (free magnetic phase).
The   global  $U(n_{f})$ symmetry  remains    unbroken.

This   last  phase  is  related  to  the one    discovered
by Seiberg in  $N=1$ massless  SQCD  for  the  range of $n_f$,
$n_c +1 < n_f < 3 n_c/2$.         Nevertherless,    it  should be
emphasized that   the $m_i \to 0$   limit  here is a  smooth one and
the symmetry properties of the vacua  are     independent of
the way   the limit is taken, while  in  SQCD without the  adjoint
chiral superfield the vacuum properties depend critically on the order
in which  the
$m_i$'s approach zero,  showing typically  the phenomenon of the
run-away vacua.

All in all, we find  the number
\beq
{\cal N}_1= ( 2 \, n_c - n_f) \, 2^{n_f -1}
\label{vacfinitevev}\eeq
of $N=1$ vacua with finite flavor-carrying     VEVS and
\beq    {\cal N}_2= \sum_{r=0}^{ n_f-n_c-1}
\, ( n_f-n_c-r ) \cdot     {}_{n_f}\!C_ r,        \label{vaczerovev}   \eeq
of  them  with vanishing     VEVS.          The latter is present
only for  theories with  the large number of
flavors ($n_f \ge   n_c+1$).         Their sum,
${\cal N}=\sum_{r=0}^{{\hbox { min}} \, \{n_f, n_c-1\}} \, _{n_f}\! C_ r
\,  ( n_c -r   ) $,       correctly      generalizes \footnote { It
may be said that   the   fact that   the
generalization  is given by these formulas  and   analogous ones
Eq.(\ref{vacfinitevev2}), Eq.(\ref{vaczerovev2}),  and
not simply, e.g.,       by    $n_f + n_c$,       reveals   the
richness  of dynamical scenarios of these theories.}
    the well-known  number of the vacua in the $SU(2)$ gauge theory,
${\cal N}_{SU(2)} = n_f +2.$

\begin{table}[h]
\begin{center}
\vskip .3cm
\begin{tabular}{ccccc}
        &   Deg.Freed.      &  Eff. Gauge Group
&   Phase    &   Global Symmetry     \\
\hline
1st Group  &  rel.  nonloc.       &    -    &
Almost SCFT
&          $ U(n_f)  $
\\ \hline
2nd Group       &  dual quarks     &      $USp(2  {\tilde n}_c) \times
U(1)^{n_c -{\tilde n}_c} $               &  Free Magnetic
&      $SO(2n_f) $         \\ \hline
\end{tabular}
\caption{ Phases of $USp(2 n_c)$ gauge theory  with $n_f$ flavors  with
$m_i \to 0$.   $ {\tilde n}_c \equiv n_f-n_c-2$. }
\label{tabuspn}
\end{center}
\end{table}

\begin{table}[h]
\begin{center}
\vskip .3cm
\begin{tabular}{ccccc}
    Label ($r$)    &   Deg.Freed.      &  Eff. Gauge  Group
&   Phase    &   Global Symmetry     \\
\hline
$0$     &   monopoles   &   $U(1)^{ n_c} $               &   Confinement
   &      $U(n_f) $            \\ \hline
$ 1$           &  monopoles         & $U(1)^{n_c } $           &
Confinement       &     $U(n_f-1) \times U(1) $        \\ \hline
$ 2,\ldots ,    [{n_f -1\o 2}]  $  &  dual quarks        &    $SU(r)
\times U(1)^{n_c -r +1}      $  &    Confinement
&          $U(n_f-r) \times U(r) $
   \\ \hline
$ n_f/2  $  &  dual quarks        &    $SU(n_f/2)
\times U(1)^{n_c -n_f/2 +1}      $  &    Almost SCFT
&          $U(n_f/2) \times U(n_f/2) $
   \\ \hline
\end{tabular}
\caption{ The first group of vacua  of $USp(2  n_c)$    theory with $n_f$
flavors
 with $m_i=m \ne 0$.    }
\label{tabspnnzm}
\end{center}
\end{table}

In $USp(2n_c)$ theories,   again,    we find two groups of vacua,  whose
properties are shown in Table
\ref{tabuspn}.    The most interesting difference as compared to the
$SU(n_c)$ theory
is that here   the entire first group of vacua    correspond to
    SCFT.   As the superconformal theory is a nontrivial one,  one does
not have a local effective Lagrangian description   for those
theories.   Nonetheless,   the symmetry breaking pattern can be deduced,
from the analysis
 done at large $\mu$:
   $SO(2n_f) $  symmetry is always  spontaneously  to   $U(n_f)$.

It is most instructive    to consider the  equal but nonvanishing   quark
mass case, first.
 (See Table \ref{tabspnnzm}.)     The flavor symmetry
group of the underlying theory    is now broken explicitly   to $U(n_f)$.
The  first group of vacua
 split   into various
branches labelled by
$r$,   $r=0,1,2,\ldots, [{n_f -1\o 2}]$,    each of which   is described
by a {\it local}  effective gauge theory
of    Argyres-Plesser-Seiberg \cite{ArPlSei},     with     gauge group   $SU(r)
\times U(1)^{n_c -r-1}   $   and $n_f$ (dual)  quarks in the  fundamental
representation of   $SU(r)$.
Indeed, the gauge invariant composite VEVS  characterizing these  theories
differ   by  some powers of $m$, and
the validity of each effective theory is limited  by  small fluctuations
of order of $m$ around each vacuum.   In the
limit $m   \to 0$   these points  in the quantum moduli space (QMS)
collapse
into one single point.    Obviously,     a
smooth    $m_i\to 0$ limit is not possible.      The location of this
singularity
can be obtained exactly in terms of Chebyshev polynomials.     At the
singularity
there are   mutually non-local dyons and hence the theory is at a
non-trivial infrared
fixed point.
   In the example of $USp(4)$ theory with $n_f=4$, we have
explicitly verified this by determining the singularities and branch
points at finite equal mass $m$ and then    by studying the limit $m
\to 0$.

These  cases,  together with the special $r=n_f/2$  nonbaryonic root
for the $SU(n_c)$ theory,    constitute
another   new mechanism for dynamical symmetry breaking:   although
the global symmetry breaking
pattern  deduced  indirectly  looks familiar enough,  the low-energy
degrees of freedom are  relatively nonlocal dual
quarks and dyons.  It would be interesting to get a better
understanding   of    this phenomenon.

    For large
numbers of flavor,  there are  also     vacua,  just as in large
$n_f$  $SU(n_c)$  theories, with no confinement and no  dynamical
flavor symmetry breaking. The low-energy particles are  solitonlike
magnetic  quarks which weakly interact  with  dual (in
general) non-abelian gauge fields: the system  is in the free magnetic phase.

The vacuum counting gives,  for  $USp(2n_c)$ theories,
\beq    {\cal N}_1  =  ( 2 n_c + 2 - n_f ) \cdot  2^{ n_f -1}
\label{vacfinitevev2} \eeq
vacua  with finite VEVS (first group of vacua), and
\begin{equation}
    {\cal N}_2 = \sum_{r=0}^{n_f - n_c - 2}    \!  (n_f-n_c-1-r)  \cdot
{}_{n_f}\!C_r  \,
\label{vaczerovev2}    \end{equation}
of them  with vanishing VEVS  (second    group of vacua);  the latter  are
present only  for
$n_f  \ge   n_c +  2$.

\bigskip

The paper  will  be       organized as follows.
After discussing briefly the standard expectation on chiral symmety
breaking in Sec. \ref{sec:expect},
we  start in Sec. \ref{sec:classvac}  a preparatory analysis,
finding   all  isolated semi-classical vacua by minimizing
the scalar potential and determining the vevs of the adjoint scalar
   and of the squark  fields   in each of them.
This allows us to count the number of   all possible   vacua of the
theory, after taking
appropriate account of Witten's index corresponding to the unbroken
gauge group in each case.

The first major    step of our analysis is the  analytic,
first-principle   determination of  the global symmetry breaking
pattern in
each
   vacuum,           done
    by  studying
the  theories      at {\it large}  $\mu $   ($\gg \Lambda$),
and $m_i \to 0$  (Sec. \ref{sec:largem}).     Such a
determination is possible    because
   in this case  the effective superpotential    can be read off from the
bare Lagrangian by integrating out the heavy, adjoint fields and by
adding to  it   the known exact instanton--induced superpotentials of the
corresponding $N=1$ theories.
By minimizing the scalar  potential,   we reproduce in all cases  the
correct number of the     vacua
and  explicitly determine the pattern of global
symmetry breaking  in each them.  By $N=1$  supersymmetry and  holomorphic
dependence of physics on  $\mu$  the same   symmetry breaking pattern is
valid at any finite $\mu$.

In Section \ref{decouple} another related limit,  $\mu \to \infty$,
$m_i $  and    $  \Lambda_1\equiv  \mu^{n_c \o   3n_c-n_f}
\Lambda^{2n_c-
n_f  \o 3n_c-n_f   } $  fixed,   is studied
and   consistency with the known results in the  standard  $N=1$
theories without adjoint fields is  checked.

The properties of the  quantum vacua at {\it small}   $\mu$ and {\it
small}    $m_i$   are studied in detail,   in  Sections
\ref{sec:leeflag} - \ref{CKMmp}.      In  Section \ref{sec:leeflag}
we show  how
all of the  $N=1$
vacua     arise   from various classes of conformally  invariant
theories (CFT)  upon  perturbation in bare quark masses    on
Seiberg-Witten curves,      reproducing   the correct number of
$N=1$ vacua  found in the earlier analyses.      In Section
\ref{sec:flsymbr} we check and illustrate these  results in the cases
of rank-two gauge groups,
$SU(3)$ and  $USp(4)$ theories, by directly  finding the associated
singularities  numerically.
In  Section \ref{sec:EfLagdes}   the effective-Lagrangian description
of these  $N=1$  theories is analysed,
where   we verify that  the number of $N=1$ vacua  and the symmetry
breaking pattern in each of them
indeed agree with what we found earlier in Sections
\ref{sec:classvac}-\ref{sec:flsymbr}.         This leads  us to a
better,
more   microscopic understanding of the phenomena of confinement and
dynamical global symmetry breaking,  as summarized
above.   The actual perturbation theory in bare quark masses  on
Seiberg-Witten curves,  whose outcome is quite central to the
whole analysis but whose analysis {\it per se} is     independent of
the rest of the paper,     is  developed in  the last  section (Sec.
\ref{CKMmp}).

Several technical discussions are relegated to Appendices.
   \ref{grouptheory} gives a proof of $SO(2N)\cap  USp(2N) = U(N)$;
in \ref{sec:semicmon} we discuss the Jackiw-Rebbi
   construction of flavor multiplet structure of semiclassical
monopoles for $SU(n_c)$ and $USp(2n_c)$  gauge theories;
we list explicit  expressions for  $a_{Di}$,   $\,a_{i}$,  ${\de a_{Di} \o \de
u_j},$ and ${\de a_{i} \o \de u_j} $   in \ref{sec:formulas};
the proof of absence of the
``nonbaryonic branch root" with  $r={\tilde n}_c=n_f-n_c$          is
given in \ref{sec:absence};
     the
study of  the monodromy around the seventeen singularities  of
$SU(3)$,  $n_f=4$  theory  is discussed   in
\ref{sec:monodromy} .

A shorter version of  this work    has  appeared already \cite{CKM}.
The case of   $SO(n_c)$ theories, where some new  subtleties  are
present,   will be discussed in a
separate article.

\newpage

\section{Expected Pattern of Chiral Symmetry Breaking \label{sec:expect}}

In non-supersymmetric theories with fermions in the fundamental
representation of the gauge group, global symmetry can be broken
spontaneously if a fermion bilinear condensate
\beq
\bra  \psi \psi \ket  \sim \Lambda^{3}
\eeq
forms.
Apart from the requirement of gauge invariance there are no general
rules which fermion pairs condense, although in QCD at large $n_c$
limit one can argue \cite{ColWit} that $SU_L(n_f)\times SU_R(n_f)\times U_V(1)$
symmetry   is
broken to the diagonal $SU_V(n_f)\times U_V(1)$, which is the
observed pattern of
chiral symmetry breaking in Nature.

Nonetheless certain general considerations can be made.
For nonsupersymmetric $SU(2)$ theories, $2 n_f$ fermions transform as
the fundamental representation of the global $SU(2n_f)$ group:
bilinear condensate ($a,\,b = 1, 2$ are color indices; $i, \, j = 1,
2, \ldots 2n_f$ are flavor indices),
$\bra \epsilon^{ab} \psi_a^i \psi_b^j  \ket,$
is necessarily antisymmetric in $(i,j)$.
If these condensates can be put by an $SU(2n_f)$ transformation into
the ``standard'' form
\beq
\bra \epsilon^{ab} \psi_a^i \psi_b^j  \ket  = \const \, {J}^{ij},
\eeq
where $J = i \sigma_2 \otimes {{\bf I}}_{n_f \times n_f}$,
then such condensates  would leave $USp(2n_f)$ invariant.
Note that this is consistent with what was found in the $N=2$ $SU(2)$
QCD, broken to $N=1$ by a small adjoint mass term, in case of
$n_{f}=2$ and in one of the vacua of $n_{f}=3$, where the symmetry of
the vacuum was found to be $U(n_f)$ \cite{SW2,KT}.
The reason is that in the model of \cite{SW2,KT} the global symmetry
of the theory is reduced to $SO(2n_{f})$ due to the Yukawa
interaction, and the intersection of $USp(2n_f)$ and $SO(2n_{f})$ is
precisely $U(n_f)$ (see Appendix A).
For another vacuum of the $n_{f}=3$ theory (where oblique confinement
of 't Hooft takes place), the monopole that condenses is a flavor
singlet and chiral symmetry remains unbroken.

In the standard QCD  with $SU(n_c)$ gauge symmetry ($\, n_c \ge 3$),
$2 n_f$ fermions transform as  $({\underline n_f}, {\underline 1}) +
({\underline 1}, {\underline  n_f^*} )$
of $SU_L(n_f)\times SU_R(n_f)$;   condensates of the form
\beq
\bra {\bar \psi_R}^i \psi_{L j} \ket =  v \, \delta_j^i,
\eeq
is believed to form, at least for small $n_{f}$, leaving the unbroken
diagonal $SU(n_f)$ symmetry.
Unfortunately, in the corresponding $N=2$ theories (with a small
$N=1$ perturbation) the axial symmetry is explicitly broken at the
tree level by the characteristic Yukawa interactions so that the
global symmetry contains only the diagonal $SU(n_f)$, already at the
tree level.
Thus $SU(n_c)$ theories will be considered below mainly as a testing
ground of our approach, in correctly identifying the quantum vacua
which survive $N=1$ perturbation, matching the numbers of classical
and quantum vacua, and in verifying in each such vacua the 't
Hooft--Mandelstam mechanism for confinement.
In fact this study reveals   new,  unexpected ways confinement and dynamical
symmetry breaking are realized   in non-Abelian gauge theories.

The cases of $USp(2n_c)$ theories are more promising.
As noted above for $SU(2)$ gauge theories, in a nonsupersymmetric
theories with $2n_f$ fermions in the fundamental representation, the
global symmetry is $SU(2 n_f)$.
Bifermion condensates of the ``standard'' form
\beq
\bra  \psi_i^a \psi_j^b \, J_{ab} \ket= v \, J_{ij}
\label{fermcond}
\eeq
would break it to $USp \, (2 n_f)$.

On the other hand, the corresponding $N=2$ models (\ref{lagrangian})
have a smaller flavor group, $SO(2n_f) \subset  SU(2n_f)$ due to the
Yukawa interactions.
Nevertheless, there is a nontrivial overlap between its flavor group
$SO(2n_f)$ and the $USp(2 n_f)$ (expected invariance group for
nonsupersymmetric model), which is $U(n_f)$.
One then expects that the global chiral symmetry $SO(2n_f)$ is broken
spontaneously to $U(n_f) \subset SO(2n_f)$.
We shall see below that these expectations are indeed met  by  quantum
vacua of
the  $USp(2n_c)$   theories  (in the large flavor cases, these take
place in the first group of vacua,
while we find also a secon group of vacua in which the chiral
$SO(2n_f)$ symmetry remains unbroken).
The proof that  $SO(2N)\cap  USp(2N) = U(N)$ is given in Appendix A.

\newpage

\section{Semi-Classical Vacua \label{sec:classvac}}

In this section, we find all semi-classical  vacua in $N=2$ $SU(n_c)$ and
$USp(2n_c)$ theories with quark hypermultiplets   in their fundamental
representations,     perturbed by the quark masses as well as  that  of
the adjoint
fields    in the $N=2$ vector multiplet.  The analysis is quantum
mechanically   valid at large   $\mu$ and  $m_i$.
$N=1$ supersymmetry and holomorphy in $\mu$  and  $m_i$  forbid any
phase transition as one moves
to  smaller $|m_i|$    and  $|\mu|$
hence   the same number of  $N=1$ vacua  must be present   in the
different regimes to be considered in the
subsequent sections.   In particular,  this analysis allows to
determine the symmetry breaking pattern in the
equal (and nonvanishing)  mass case,   in which the classical  global
symmetry is  $U(n_f)$ in both  $SU(n_c)$ and
$USp(2n_c)$ theories.

\subsection {Semi-Classical Vacua in $SU(n_c)$}

In the  limit $m_i \to 0$ and $\mu \to 0$,          the global
symmetry of the model is
$U(n_f)\times  Z_{2 n_c - n_f}\times SU_R(2).$
    The superpotential of our theory,  with $N=2$ supersymmetry softly
broken to $N=1$  by  the adjoint mass,     is
given by:
\beq W= \mu \Tr \, \Phi^2  + \sqrt2 \,{\tilde Q}_i^a \Phi_a^b Q_b^i +
     m_i \, {\tilde Q}_i^a Q_a^i,\eeq
     where
\beq    \Phi  \equiv \lambda^A  \Phi^A, \quad (A=1,2,\ldots   N^2-1),
\eeq
\beq     \Tr \, ( \lambda^A \lambda^B ) = {1 \o 2} \delta^{AB}. \eeq
$i=1,2,\ldots n_f $ is the flavor index; $a,b=1,2,\ldots n_c$ are the
color
indices.
Note that with our normalization of the $SU(n_c)$ generators the color
Fierz   relation reads
\beq \sum_{A=1}^{n_c^2-1} (\lambda^A)_c^d (\lambda^A)_a^b = {1\o2} \,
\left[\delta_c^b
   \delta_a^d -
{ 1\o n_c}\delta_c^d \delta_a^b \right].    \eeq

The vacuum equations are
\beq   [ \Phi, \Phi^{\dagger}] = 0 \, ;
\label{D1}
\eeq
\beq
\nu \delta_a^b=  Q_a^i (Q^{\dagger})_i^b - ({\tilde Q}^{\dagger})_a^i
{\tilde Q}_i^b \, ;
\label{D2}
\eeq
\beq
Q_a^i {\tilde Q}_i^b - { 1\o n_c} \delta_a^b (Q_c^i {\tilde Q}_i^c) +
\sqrt2  \, \mu \Phi_a^b = 0 \, ;
\label{F1}
\eeq
\beq
Q_a^i m_i + \sqrt2 \,\Phi_a^b Q_b^i = 0  \qquad ( { \hbox {\rm no sum
over}} \, \,i) \, ;
\label{F2}
\eeq
\beq
m_i {\tilde Q}_i^a +  \sqrt2 \, {\tilde Q}_i^b \Phi_a^b = 0 \qquad (
{\hbox{\rm no sum over}} \,\,i) \, .
\label{F3}
\eeq
where the quark masses have been taken diagonal by flavor rotations.

Use first $SU(n_c)$ rotation to bring $\Phi$ into  diagonal form,
\beq
\Phi = \diag \, (\phi_1, \phi_2, \ldots  \phi_{n_c}) \, , \qquad \sum
\phi_a = 0 \, .
\label{phivev}
\eeq

Eq.~(\ref{F2}) and Eq.~(\ref{F3}) say that $Q_a^i $ and ${\tilde
Q}_i^b$
are either nontrivial eigenvectors of the matrix $\Phi $ with possible
eigenvalues $m_i$,  or null vectors.
With $\Phi$ put in the diagonal form, and with generic
(so unequal) masses,  the eigenvectors have  simple forms,
\beq
{Q}^i =  (0, \ldots, d_i, 0, \ldots ),
\eeq
each with only one nonvanishing component (similarly for ${\tilde
Q}_i^a $).
There can be at most $n_f $ nontrivial eigenvalues, chosen from $m_1,
m_2, \ldots , m_{n_f}$;
at the same time the form (\ref{phivev}) allows for at most $n_c-1$
nonzero
independent elements of $\Phi$.
The solutions can thus be classified by the number of nontrivial
eigenvectors, $r = 1, 2, \ldots, {\hbox {\rm min}} \{ n_f, n_c - 1
\}$.
There are
$\pmatrix{n_f \cr r}$
solutions for a given $r$, according to which $m_i$'s appear as
eigenvalues.

The solution with eigenvalues $m_1, m_2, \ldots, m_r$ is  \footnote{
These results  are slight generalization of the ones in
\cite{SW2,SUN,ArPlSei,Hira} to generic nonvanishing quark  and adjoint
masses.
Note that the flat directions are completely eliminated.}:
\beq  Q_a^i  = \pmatrix {d_1 \cr 0  \cr \vdots  \cr \vdots \cr  0},
\,\, \pmatrix {0 \cr d_2  \cr 0 \cr \vdots\cr 0},   \pmatrix {0 \cr \vdots
\cr  d_r \cr   0 \cr \vdots};  \qquad     Q_a^i   = 0, \,\,\,\,
i=r+1,\ldots, n_f \,  . \label{vevofq}\eeq
\beq  {\tilde Q}_i^a  = \pmatrix {\tilde d_1 \cr 0  \cr \vdots  \cr
\vdots \cr  0}, \,\, \pmatrix {0 \cr \tilde d_2  \cr 0 \cr \vdots\cr 0},
\pmatrix {0 \cr
\vdots
\cr  \tilde d_r \cr   0 \cr \vdots};  \qquad    {\tilde Q}_i^a  = 0,
\,\,\,\, i=r+1,\ldots, n_f \,  , \label{vevofqti}\eeq
where
\beq   r=0,1, \ldots, {\hbox {\rm min}} \, \{n_f, n_c-1\},
\eeq
\beq
\Phi = \frac{1}{\sqrt{2}}\diag \, (-m_1, -m_2, \ldots, -m_r, c,
\ldots , c) \, ; \qquad c = {1 \o n_c-r} \sum_{k=1}^r m_k \, .
\label{diagphi}
\eeq
$d_i$'s  can be chosen real (by residual $SU(n_c)$);   ${\tilde
d}_i$'s are in general  complex.
They are given by Eqs.~(\ref{solnd}) and (\ref{dedtild}) below.

Eq.~(\ref{D1}) is obviously satisfied.
The nondiagonal ($a \ne b)$ of Eq.~(\ref{D2}) is also obvious.
The first $r$ diagonal ($a=b)$ equations are:
\beq
\nu=  d_i^2- | {\tilde d}_i |^2 \,; \qquad (i=1,2, \ldots r)\, ;
\eeq
the others  give
\beq  \nu=0 \, , \eeq
hence
\beq
d_i^2 = | {\tilde d}_i |^2 \, .
\label{dedtild}
\eeq
Eq.~(\ref{F2}) and Eq.~(\ref{F3}) are satisfied by construction;
Eq.~(\ref{F1}) gives for $a=b =1,2, \ldots r$:
\beq
d_i {\tilde d}_i - { 1\o n_c} \sum_k d_k {\tilde d}_k =  \, \mu \,m_i
\, ;
\label{corres}
\eeq
from which one finds that
\beq
\sum_k d_k {\tilde d}_k= { n_c \o n_c -r}  \, \mu \,\sum_{k=1}^r m_k
\label{compa}
\eeq
and
\beq
d_i {\tilde d}_i =  \, \mu \,m_i  + { 1 \o n_c -r} \mu \,\sum_{k=1}^r
m_k \qquad   (d_{i} > 0)\, .
\label{solnd}
\eeq
On the other hand, Eq.~(\ref{F1}) for $a=b =r+1,r+2, \ldots n_c$
gives
\beq
\sum_k d_k {\tilde d}_k= n_c  \, \mu \, c \, .
\label{ddtilde}
\eeq
This is compatible with Eq.~(\ref{compa}) because of (\ref{diagphi}).

A solution with a given $r$ leaves a local  $SU(n_c-r)$ invariance.
Thus each of them counts as a set of $n_c-r$  solutions (Witten's
index).
In all, therefore, there are
\beq
{\cal N} = \sum_{r=0}^{{\hbox {\rm min}} \, \{n_f, n_c-1\}}\, (n_c-r)
\, \pmatrix{n_f \cr r}
\label{nofvac}
\eeq
classical  solutions.
(For $r=0$, $Q_a^i = {\tilde Q}_i^b =0,$ $\Phi=0$ is obviously a
solution with full $SU(n_c)$ invariance.)

For $n_c=2$ the formula (\ref{nofvac}) reproduces the known result
(${\cal N}= 2 + n_f$) as can be easily verified.

Note that when $n_f$ is equal to or less than $n_c$ the sum over $r$
is done readily, and Eq.~(\ref{nofvac}) is equivalent to
\beq
{\cal N}_1= ( 2 \, n_c - n_f) \, 2^{n_f -1}, \qquad (n_f \le n_c) \, .
\label{nofvacbis}
\eeq

\subsection{Semi-Classical Vacua in $USp(2n_c)$}

The superpotential reads in this case
\beq
W=  \mu \, \Tr \, \Phi^2  + { 1 \o \sqrt2} \, Q_a^i \Phi_b^a Q_c^i \, J^{bc} +
{m_{ij} \o 2} Q_a^i Q_b^j \, J^{ab} \, ,
\label{uspsuperpot}
\eeq
where $J = i\si_2 \otimes {{\bf 1}}_{n_c}$   and
\beq
m = - i \si_2 \otimes \diag \, (m_1, m_2, \ldots, m_{n_f}) \, .
\eeq
In the $m_i \to 0$ and $\mu \to 0$  limit, the global symmetry is
$SO(2n_f) \times Z_{2 n_c +2 -  n_f } \times  SU_R(2)$.

The vacuum equations are:
\beq
[ \Phi, \Phi^{\dagger}] = 0 \, ;
\label{D13}
\eeq
\beq
\sum_i( Q_a^{i \dagger} Q_b^i  - Q_{n_c+b}^{i \dagger} Q_{n_c+
a}^{i}) = 0 \, ;
\qquad   \sum_i Q_a^{i \dagger} Q_{n_c+b}^i = 0 \, ;
\label{D23}
\eeq
\beq
\sqrt2  \, \Phi_b^a Q_c^i \,J^{bc} + m_{ij} Q_b^j \,J^{ab} = 0 \, .
\label{F13}
\eeq
\beq
2 \, \mu \, \Phi_a^b+ {1 \o \sqrt2} \, Q_a^{i} \,J^{bc}Q_c^{i} = 0 \, .
\label{F23}
\eeq

To find the solutions of these equations diagonalize first $\Phi$ by
a unitary transformation:
\beq
\Phi = \diag \, (\phi_1, \phi_2, \ldots, \phi_{n_c}, -\phi_1,
-\phi_2, \ldots, -\phi_{n_c})  \, .
\eeq
Define
\beq
{\tilde Q}_a^i\equiv Q_{n_c+a}^i \, .
\eeq
The vacuum equations can be rewritten as:
\beq
\sum_i(Q_a^{i \dagger} Q_b^i  - {\tilde Q}_{b}^{i \dagger} {\tilde
Q}_{a}^{i}) = 0 \, ;
\qquad  \sum_i Q_a^{i \dagger} {\tilde Q}_{b}^i = 0 \, ;   \qquad
(i=1, \ldots 2n_{f}) \, ;
\eeq
and
\beq
\sqrt2 \, \phi_a {\tilde Q}_a^i
      -m_i   {\tilde Q}_a^{n_f+i} =0 \, ; \label{fromhere}
\eeq
\beq
\sqrt2  \,\phi_a {\tilde Q}_a^{n_f+i}
      + m_i   {\tilde Q}_a^{i} =0 \, ;
\eeq
\beq
\sqrt2  \, \phi_a {Q}_a^i
      + m_i   {Q}_a^{n_f+i} =0 \, ;
\eeq
\beq
\sqrt2  \, \phi_a {Q}_a^{n_f+i}
      -m_i   {Q}_a^{i} =0 \, ;
\eeq
\beq
{Q}_a^i  {\tilde Q}_b^{i} =0 \, , \qquad (a \ne b) \, ;
\eeq
and
\beq
2\sqrt2 \, \mu \, \phi_a +  Q_a^{i} {\tilde Q}_a^{i} +  Q_a^{n_f+i}
{\tilde Q}_a^{n_f+i} =0 \, .
\label{uptohere}
\eeq
In  Eq.~(\ref{fromhere}) -- Eq.~(\ref{uptohere})  the index $i$ runs
only over $i=1,2,\ldots, n_f$.

The solutions can again be classified by the number of the nonzero
$\phi$'s, $r=1, 2, \ldots, \min\, \{n_f, n_c\}$.
The one with $r$ masses $m_1, \ldots, m_r$ is
\beq
\phi = {1\o \sqrt2}  \, \diag \, (im_1, im_2, \ldots , im_r,0,\ldots,
-im_1, -im_2,
\ldots -im_r,0,\ldots ) \, ;
\eeq

\beq Q_a^i=
\pmatrix{  d_1 &&&&&& -i d_1 &&&&&  \cr
             &  d_2 &&&&&& -i d_2 &&&&   \cr
             && \ddots &&&&&& \ddots &&&  \cr
             &&&  d_r  &&&&&&  -i d_r &&  \cr
             &&&& 0    &&&&&&  0 &  \cr
             &&&&& \ddots  &&&&&& \ddots   \cr
             -i d_1    &&&&&&  d_1 &&&&&  \cr
             & - id_2  &&&&&&  d_2 &&&&   \cr
             && \ddots &&&&&& \ddots &&&  \cr
             &&&  -i d_r &&&&&& d_r &&  \cr
             &&&& 0 &&&&&& 0 &  \cr
             &&&&& \ddots  &&&&&& \ddots \cr}\eeq
where
\beq   d_i= \sqrt{m_i \mu }   \,. \eeq
Each solution with $r$ leaves unbroken  $USp(2(n_{c}-r))$   hence
counts as $n_{c}-r + 2 $  solutions.  The total number of the classical
vacua is then
\beq
{\cal N} = \sum_{r=0}^{min\{ n_c, \, n_f\}}  (n_c- r +1) \cdot
\pmatrix {n_f \cr  r} \, .
\label{Nvspnclass}
\eeq
Note that for smaller values of $n_f$, , the sum over $r$ is easily done
and an equivalent formula is
\beq   {\cal N} =(2 \, n_c+2-n_f) \, 2^{n_f-1},  \qquad   (n_f \le n_c).
\label{Nvspnclassbis} \eeq

It is amusing (and reassuring) that   different formulas,
Eq.(\ref{nofvac}) , Eq.(\ref{nofvacbis}),   Eq.(\ref{Nvspnclass})
and  Eq.(\ref{Nvspnclassbis})  found here    reproduce correctly  the formula
\beq    {\cal N}_{SU(2)} = n_f +2, \eeq
   for  the  $SU(2)$  gauge theory (which is the simplest case,
both of $SU(n_c)$ and $USp(2n_c)$),  for   $n_f=0\sim4$.

\newpage

\section{Determination of Symmetry Breaking Patterns  at Large $\mu$}
\label{sec:largem}

In this section,  we  determine the number  of $N=1$ vacua and the pattern of
    flavor symmetry breaking in each of them, in the limit $m_i \to 0$.
This can be done most easily  by studying
these $SU(n_c)$ and
$USp(2n_c)$ theories      at large adjoint mass $\mu$.  The
advantage of considering
the  theories     at large    $\mu$  is that   the adjoint field can
be  integrated
out from
the theory:  the resulting low-energy effective theory is an exactly
known $N=1$
supersymmetric gauge theory,    perturbed by certain  superpotential terms
suppressed  by $1/\mu$.  The dynamics of such a
theory is known,    either in terms of dynamical superpotential with
confined meson  or
baryon degrees of freedom \cite{Sei94},  or  in a   dual description
using magnetic degrees
of freedom \cite{Sei,IntSei}.           By minimizing the potential
of such low-energy effective
actions we find the  symmetry breaking pattern in each  $N=1$ vacua.
Supersymmetry and holomorphy
imply that there are no phase transition at finite $|\mu|$:   the
qualitative features  found here,
such as the unbroken symmetry and  the number of the Nambu-Goldstone
bosons, are valid also at small
nonvanishing $\mu$.

This method not only allows us to
cross-check the counting of the
number of vacua in the previous section,     but also enables us to
determine  the  pattern  of dynamical
symmetry breaking from the first principles,  as the limit of
massless   quarks can be
studied exactly.

To be precise,    we shall be  interested here in the limit
$ m_i \to 0$  with $ \mu $ and  $ \Lambda $   ($ \mu  \gg \Lambda
$)   fixed.    As long as  $\mu$ is kept fixed,  the limit  $ m_i \to 0$
is smooth and   it is possible to determine how the  exact flavor
symmetry  is realized
in each vacuum.

This is to be contrasted to  another    limit,
$\mu \to \infty \,\,$ with     $m_i $ fixed,   which will
be    discussed in the next section.
This latter limit,  which does not commute with the former
($m_i \to 0$ first),     is the relevant one      for studying the
decoupling  the adjoint field and verifying the consistency
with  the
   known results in the standard  $N=1$
theories.

The analysis  of this section   will be divided in two parts,
   for small  and for   large values of  $n_f$: this is  necessary
    due to the emergence of the dual  gauge group in the corresponding $N=1$
theories for relatively large values of the flavor
($n_f > n_c+1$ in $SU(n_c)$,  $n_f > n_c+2$ in $USp(2 n_c)$).

For the ease of reading, a  short summary  is given   at the end of
this section:  the reader who is more
interested  in the physics results
than the technical aspects of
the analysis,     might  well jump to it.

\subsection{$SU(n_c)$ theories: Small Numbers  ($n_f \le   n_c$)    of Flavor}
\label{sec:largemusmallf}

Let us first consider the cases,     $n_f <  n_c$.
At large $\mu$ one can integrate over $\Phi$:
\beq
\Phi^A=- {\sqrt2 \o \mu} {\tilde Q}_i^a (\lambda^A)_a^b Q_b^i \, ;
\eeq
resubstituting it into the superpotential one gets
\beq
W =  -{1 \o 2 \mu} \left[ \Tr M^2 - {1 \o n_c}(\Tr M)^2 \right]  + \Tr (M m
)  +
(n_c-n_f)  {\Lambda_1^{(3n_c - n_f)/(n_c-n_f)}  \o (\det M)^{1/(n_c-n_f)}} \, ,
\label{splarmusmnf}\eeq
where  $M_i^j \equiv {\tilde Q}_i^a Q_a^j$ and where  the known
instanton--induced effective superpotential of $N=1$ SQCD has been added.
\beq  \Lambda_1 \equiv  \mu^{n_c \o   3n_c-n_f}        \Lambda^{2n_c-
n_f  \o 3n_c-n_f   } \eeq
is the scale of the $N=1$  SQCD.

By making an ansatz   for $M$:
\beq
M = \diag \, (\lambda_1, \lambda_2,  \ldots, \lambda_{n_f}) \, ,
\eeq
   the equations for $\lambda_i$ are:
\beq
-{1 \o \mu} \left( \lambda_i - { 1\o n_c} \sum_j \lambda_j\right) + m_i -
{\Lambda_1^{(3n_c - n_f)/(n_c-n_f)}
\o (\prod_j \lambda_j)^{1/(n_c-n_f)}}  \lambda_i^{-1} =0 \,  .
\label{largemueq}
\eeq
We now study    the solutions of these equations    in the limit   $
m_i \to 0$,    with $\Lambda$ and
$\mu$ ($\mu \gg \Lambda $) fixed.
    By  making an ansatz,
\beq
M = \diag \, (\lambda_1, \lambda_2,  \ldots, \lambda_{n_f}) \, ,
\eeq and
upon multiplication  with $\lambda_i$   one finds (for each $i$)
\beq
-{1 \o \mu} ( \lambda_i^2  -  { \lambda_i  } Y  ) + m_i  \,\lambda_i
+  X  =0.  \,
\label{largemueqbis}
\eeq
where
\beq  X\equiv  - \Lambda_1^{(3n_c - n_f)/(n_c-n_f)}{ (\prod_j
\lambda_j)^{{1 \o   n_f-n_c}}}; \qquad Y
\equiv{1
\o n_c} \sum_{j}^{n_f} \lambda_j.
\eeq
One can take the limit  $m_i \to 0$   directly in
Eq.(\ref{largemueqbis}), which becomes
simply
\beq   \lambda_i^2 - Y \lambda_i -   \mu X=0.   \label{result421}\eeq
It can be solved by first assuming that $X$ and  $Y$ are  given:
\beq \lambda_i =  { 1\o 2 } (Y \pm \sqrt{Y^2 + 4 \mu  X}). \label{pmsigns} \eeq
In general,  $r$ of $\lambda_i$'s can take the upper sign and the
rest ($n_f-r$) of
   $\lambda_i$'s  the lower sign,  $r=0,1,2,\ldots, n_f$.
These solutions
\bea  && \lambda_1=\ldots =\lambda_r=
   { 1\o 2 } (Y + \sqrt{Y^2 + 4 \mu  X }); \non \\
&& \lambda_{r+1}=\ldots =\lambda_{n_f} =
   { 1\o 2 } (Y -  \sqrt{Y^2 + 4 \mu  X }),      \label{DSB}   \eea
must then be re-inserted into the definitions of $X$ and $Y$ to
determine the latter
quantities.   One finds two relations
\beq   ( 2 r- n_f ) \sqrt{Y^2 + 4 \mu X} = (2 n_c- n_f)  Y;
\label{usefuleq1} \eeq
and
\beq X= -  { \Lambda_1^{(3n_c - n_f)/(n_c-n_f)}\o 2^{n_f/(n_f-n_c)}}
   \left(Y + \sqrt{Y^2 + 4 \mu  X }\right)^{r\o n_f-n_c}
\left(Y -  \sqrt{Y^2 + 4 \mu  X }\right)^{n_f-r  \o
n_f-n_c}.\label{usefuleq2} \eeq
These two equations can be easily solved for $X$ and $Y$ and give
\beq  Y=  C \, {\Lambda_1^{(3n_c-n_f) /  (2n_c-n_f)}  \o
\mu^{(n_f-n_c)/(2 n_c- n_f)} }
  e^{  2\pi k i  /(2 n_c- n_f)} =  C \, \Lambda \mu  \,
  e^{  2\pi k i  /(2 n_c- n_f)} , \qquad  k=1,2, \ldots
     2 n_c- n_f;\eeq
\beq   X=  {C^{\prime} \o \mu  }   Y^2.     \label{sonforxs} \eeq
where $C, C^{\prime} $  are constants depending on $n_f$, $n_c$ and $r$.
   Note that  $X$ is   uniquely determined in terms of $Y$.   At given $r$,
then,
there are    $ 2 n_c- n_f$ solutions for $(X,Y) $  hence for $\{\lambda_i\}$.
By summing over $r$, taking into account  $_{n_f}C_{r}$  ways of distributing
$r$ solutions with positive sign among $n_f$ flavor,  one appears to end up
with
   $ \,( 2 n_c- n_f )  \sum_{r=0}^{n_f} {}_{n_f}\!C_{r}=   ( 2 n_c- n_f
) \cdot  2^{ n_f} \, $ vacua.

Actually,  one counts exactly twice each vacuum this way.
The solution for  $\{\lambda_i\}$'s  depends on the value of $r$ in a
non trivial manner, in general.   When $r$ is replaced by $n_f-r$,
   however,  $C$ remains unchanged:   the net change of
$(X,Y)$ and hence of $\{\lambda_i\}$'s, is  that the two types of
roots Eq.(\ref{pmsigns}) are  precisely interchanged, as can be  seen from
   Eq.(\ref{usefuleq1}), Eq.(\ref{usefuleq2}),  giving the same set of
$\{\lambda_i\}$'s.

We  find     therefore
\beq    {\cal N}_1  =  ( 2 n_c- n_f ) \cdot  2^{ n_f -1}
\label{slnsmz}    \eeq
solutions of this type (i.e., finite in the $m_i \to 0$ limit).
For $n_f < n_c$   these  exhaust all possible solutions (see
Eq.(\ref{nofvacbis}),
Eq.(\ref{numvacularmu})).   They are classified by
the value of an integer $r$:  in a vacuum characterized by $r$,
$U(n_f)$ symmetry of the theory is broken spontaneously to
$U(r)\times U(n_f-r)$   by the condensates, Eq.(\ref{DSB}).

The analysis   for the case of
$n_{f}=n_{c}$ is similar but can be made  by using the   superpotential
valid for  $n_{c}=n_{f}$ \cite{Sei94}
\beq
W=  -{1 \o 2 \mu} \left[ \Tr M^2 - {1 \o n_c}(\Tr M)^2 \right]  + \Tr (M m)  +
\Tr \, \kappa \, \{(\det M) - B {\tilde B} - \Lambda^{2 n_c} \} \, ,
\eeq
where $B$ and ${\tilde B}$ are baryonlike composite,
$B=\epsilon_{i_{1}i_{2}\ldots i_{n_{c}}} \epsilon^{a_{1}a_{2}\ldots
a_{n_{c}}}  Q^{i_{1}}_{a_{1}}Q^{i_{2}}_{a_{2}}\ldots
Q^{i_{n_{c}}}_{a_{n_{c}}}$  and analogously for ${\tilde B}$ in terms
of $  {\tilde Q}$'s, and $\kappa$ is a Lagrange multiplier.

\subsection{$SU(n_c)$:  $n_f =   n_c +1  $}

In the case  with  $n_{f}=n_{c} +1 $ the effective superpotential is
\bea
W&=&  -{1 \o 2 \mu} \left[ \Tr M^2 - {1 \o n_c}(\Tr M)^2 \right]  + \Tr (M m)
   \non \\ &+&
{1\o \Lambda_1^{2 n_f-3}} \{(\det M)   - B_i (M)^i_j {\tilde B}^j\},
\eea
where
\beq  B_i=\epsilon_{i  i_{1}i_{2}\ldots i_{n_{c}}} \epsilon^{a_{1}a_{2}\ldots
a_{n_{c}}}  Q^{i_{1}}_{a_{1}}Q^{i_{2}}_{a_{2}}\ldots
Q^{i_{n_{c}}}_{a_{n_{c}}}.       \eeq
Set
\beq M={\hbox{\rm diag}}   (\lambda_1, \ldots, \lambda_{n_f}), \eeq
\bea
W&=&  -{1 \o 2 \mu} \left[ \sum_i  \lambda_i^2  - {1 \o n_c}\left(\sum_i
\lambda_i \right)^2 \right]
   +\sum_i m_i \lambda_i    \non \\ &+&
{1\o \Lambda_1^{2 n_f-3}}   \{ (\prod_j \lambda_j)    - B_i {\tilde
B}^i  \lambda_i
\}.
\eea
Derivation with respect to   $B_j  $ and  ${\tilde B}^i$     yields
\beq
\lambda_i  \, {\tilde B}^i=0;\qquad  \lambda_i B_i  =0,
\eeq
while  derivation  with respect to   $\lambda_i$ leads to
\begin{equation}
   -{1 \o  \mu} \left[  \lambda_i  - {1 \o n_c}\sum_j\lambda_j   \right]
   +  m_i  +{1\o \Lambda_1^{2 n_f-3}}   \{ \prod_{j\ne i}  \lambda_j
- B_i {\tilde B}^i  \} =0.
\end{equation}
Multiplying with  $\lambda_i$ one gets (still no sum over $i$)
\begin{equation}
   -{1 \o  \mu} \left[  \lambda_i^2  - {1 \o n_c}\lambda_i  \sum_j\lambda_j
\right]
   +  m_i  \lambda_i  +{1\o \Lambda_1^{2 n_f-3}}   \prod_{j}  \lambda_j   =0.
\end{equation}
Set now $Y=  {1 \o n_c}(\sum_j\lambda_j) $, and $X=  \prod_{j}
\lambda_j,$  and set $\Lambda_1\equiv1$ from now on.
$\lambda_i$ satisfies
\beq  \lambda_i^2  - (Y + m_i \mu) \lambda_i - \mu X=0.\eeq
The solution is
\beq
\lambda_i= {1\o 2}  ( Y + m_i \mu  \pm  \sqrt {(Y + m_i \mu)^2 - 4 \mu X}).
\eeq
Note that when $m_i=0$,    $\lambda=B=0$, all $i$'s,  is a solution.
We now classify the solutions according to the number ($r$) of
$\lambda$' s which remain finite in the limit
$m_i=0$.

\noindent i)  First consider the solution   $\lambda_i\ne 0$, $\forall i$.
In this case $ B_i ={\tilde B}^i =0$ , $\forall i$.
$\lambda_i$ are given by
\beq
\lambda_i= {1\o 2}  ( Y \pm  \sqrt {  Y ^2 - 4 \mu X}).
\eeq
The rest of the argument is the same as the one given   Sec.
\ref{sec:largemusmallf}   so one has
$(2n_c-n_f) \cdot 2^{n_f-1}$ vacua of this kind.

\smallskip
\noindent ii)  Consider now    $\lambda_1 =0$,     $\,\,\lambda_{j }
\ne 0,   \,\, j \ne 1.$
One finds that \beq  B_j =  {\tilde B}^j=0, \quad j\ne 1;\eeq
\begin{equation}
    {1 \o n_c}\sum_{j\ne 1} \lambda_j
-  {\mu \o \Lambda^{2 n_f-3}}   B_1{\tilde B}^1=0;
\end{equation}
\begin{equation}
   \lambda_j  - {1 \o n_c}\sum_{k \ne 1}\lambda_k  - m_j \mu   =0.
\end{equation}
The last equation has the form,
   \begin{equation}
C \pmatrix {\lambda_2 \cr \lambda_3 \cr \vdots \cr \lambda_{n_f}}
-\mu  \pmatrix {m_2 \cr m_3 \cr \vdots
\cr m_{n_f}}=0,
\end{equation}
with
\beq \det  C=0.\eeq
For generic $m_i$'s therefore this equation has no solutions.  Here
it is important that we consider the  massless
limit of massive theory, not directly massless theory.

\smallskip
\noindent iii)   Consider now   $\lambda_i =0$,  $i=1,2,\ldots, r,$
and  $\,\,\lambda_{j } \ne 0,   \,\, j \ge {r+1}.$
One gets
    \beq  B_j =  {\tilde B}^j=0,  \quad  j=r+1, \ldots n_f\eeq
and
\begin{equation}
    {1 \o n_c}\sum_{j > r} \lambda_j
-  {\mu \o \Lambda^{2 n_f-3}}   B_i{\tilde B}^i=0,  \quad  i=1,2,\ldots, r;
\end{equation}
\begin{equation}
   \lambda_j  - {1 \o n_c}\sum_{k > r} \lambda_k - m_j \mu  =0,
\quad  j=r+1, \ldots n_f.
\end{equation}
The last equations now give  finite answers for $\lambda_{j }$: but
they are of $O(m_i)$
and approach $0$ as $ m_i \to 0$.   One gets then also,
\beq   B_i,  {\tilde B}^i  \to 0, \quad  m_i \to 0. \eeq
Therefore all these cases degenerate into one single solution,
\beq    \lambda=B={\tilde B}=0. \eeq
To conclude,    we find   in the massless limit
$(2n_c-n_f) \cdot 2^{n_f-1}$   solutions with finite $\lambda$'s
and   one solution  with
vanishing vevs.     The latter is consistent with  the general
formula Eq.(\ref{extrav})   for the second group of vacua
(with vanishing VEVS)    found for   $n_f > n_f+1$  below,     since
${\cal N}_2= \sum_{r=0}^{{\tilde n}_c-1} \,
_{n_f}\!  C_ r \, ( {\tilde n}_c-r ) =1,$  for  $n_f=n_c+1$.

\subsection{$SU(n_c)$:  large  numbers ($n_f >   n_c +1  $)    of flavor}
\label{sec:largemulargef}

In the cases $n_f > n_c +1  $,   the effective low energy degrees
of freedom are
dual quarks  and mesons  \cite{Sei}.
The effective superpotential   is given by
\beq    {\cal W} = {\tilde q} M q  +    \Tr ( m M )   -{1 \o 2 \mu} \left[ \Tr
M^2 - {1 \o n_c}(\Tr M)^2 \right],
\label{dualqeq}\eeq
where   $q$'s are $n_f$ sets of dual quarks in the fundamental
representation of the dual
gauge group $SU({\tilde n}_c)$,   with  ${\tilde n}_c = n_f-n_c$.
The vacuum equations following from Eq.~(\ref{dualqeq}) are:
\beq  M_{ij} q_j^{\alpha} =0 \, ; \qquad   {\tilde q}_{\alpha}^i
M_{ij}=0 \, ;  \label{first} \eeq
\beq      {\tilde q}_{\alpha}^i  q_j^{\alpha} + \delta_{ij} m_i  - { 1 \o
\mu} \left( M_{ij} - { 1\o n_c} (\Tr M) \delta_{ij} \right)=0 \, .
\label{second}\eeq
The first set of equations tell us that the meson matrix $M$ and the
dual squarks are orthogonal in the flavor space.  By using the dual
color and flavor rotations (and the use of Eq.~(\ref{second})) the dual
squarks can be taken to be nonvanishing in the first $r$ flavors and
of the form:
\beq   q_i^{\alpha} = \pmatrix {d_1 \cr 0  \cr \vdots  \cr \vdots \cr  0},
\,\, \pmatrix {0 \cr d_2  \cr 0 \cr \vdots\cr 0},   \pmatrix {0 \cr \vdots
\cr  d_r \cr   0 \cr \vdots};  \qquad     q_i^{\alpha}  = 0, \,\,\,\,
i=r+1,\ldots, n_f \,  . \label{vevofqLN}\eeq
\beq   {\tilde q}^i_{\alpha} = \pmatrix {\tilde d_1 \cr 0  \cr \vdots  \cr
\vdots \cr  0}, \,\, \pmatrix {0 \cr \tilde d_2  \cr 0 \cr \vdots\cr 0},
\pmatrix {0 \cr
\vdots
\cr  \tilde d_r \cr   0 \cr \vdots};  \qquad     \tilde q^i_{\alpha}  = 0,
\,\,\,\, i=r+1,\ldots, n_f \,  , \label{vevofqtiLN}\eeq
where\footnote{ Note that the value $r= {\tilde n}_c $ should be
excluded.  In this
case, $n_f- r= n_c$ and the nonvanishing meson submatrix is $n_c
\times n_c$.  Eq.~(\ref{second}) for $i,j=r+1, \ldots, n_f$ have no
solutions since the matrix $M_{ij} - { 1\o n_c} (\Tr M) \delta_{ij} $
is of rank $n_c-1$ while $\delta_{ij} m_i $ has rank $n_c$.  }
\beq   r=0,1,2,\ldots, {\tilde n}_c -1 \, ,  \eeq
and
\beq   d_i \tilde d_i + m_i  +   { 1\o  \mu \, n_c}  \cdot \Tr M =0 \, . \eeq
Eq.~(\ref{second})  implies   also    that the meson matrix is  diagonal,
\beq  M  = {\hbox {\rm diag}}  \,  (0,0,\ldots, 0, \lambda_{r+1},
\lambda_{r+2}, \ldots, \lambda_{n_f}) \, , \label{rankless} \eeq
\beq  \lambda_i -  { 1\o n_c}  \sum \lambda = m_i \mu \, .
\label{vevoflam} \eeq
Clearly the last equations determine uniquely all $\lambda$'s: on the
other hand there are $_{n_f} C_ r $ choices of masses which enter the
equations for nonzero squark VEVS.  Furthermore, because the vacua
with $r$ nonzero entries in the squark vevs leave an $SU({\tilde
n}_c-r)$ dual gauge group unbroken ($M$ being singlet), each such
vacuum must be counted as ${\tilde n}_c-r$ vacua (Witten's index).  In
all,  then, there are
\beq
{\cal N}_2= \sum_{r=0}^{{\tilde n}_c-1} \,
_{n_f}\!  C_ r \, ( {\tilde n}_c-r )
\label{extrav}
\eeq
vacua,  with ${\hbox {\rm Rank}} \, M < n_f$,
in which    the global $U(n_f)$  symmetry   remains  unbroken, in the
$m_i \to 0$ limit.

      We seem to face a difficulty, however.   The number of vacua
found here   is  less than the known
total  number of   vacua
${\cal N}$  (Eq.(\ref{nofvac})).   Where are  other vacua?

   This apparent puzzle can be solved once  the    nontrivial $SU({\tilde
n}_c)\,\,$
dynamics  are taken into account\footnote{ In fact,  a related puzzle
is   how Seiberg's dual Lagrangian \cite{Sei}
- the first two terms of  Eq.~(\ref{dualqeq})  -  can   give rise to   the
right number of vacua for  the   massive
$N=1$   SQCD  with  $n_f > n_c+1$.        By following the same method as below
but with   $\mu = \infty$,
   we do   find the correct
number ($n_c$)    of vacua.  }.    If the VEVS of the mesons have
Rank$\,M  = n_f,$  dual quarks are all massive.   The theory
becomes pure Yang-Mills type   in the infrared,  and  the strong
interaction effects of dual gauge dynaimcs   must be properly taken
into account.  By
integrating out  the dual quarks  out,   we find   the effective
superpotential,
\beq
W_{eff}  =  -{1 \o 2 \mu} \left[ \Tr M^2 - {1 \o n_c}(\Tr M)^2 \right]  +
\Tr (M m )  +
{ \Lambda}_1^{(3n_c - n_f)/(n_c-n_f)}   (\det M)^{1/(n_f-n_c)}.
\label{spotlarmu}    \eeq
The analysis of the vacua  of this   effective action   is similar
to   that of
the $n_f < n_c$ cases, Eqs.(\ref{largemueq})--(\ref{sonforxs}):
the superpotential
(\ref{spotlarmu})  is actually identical   to   (more precisely,
continuation of)   (\ref{splarmusmnf})!\footnote{
In the case of SQCD ($\mu = \infty$)   this observation
    is in agreement
with the well known fact that, in spite of  distinct physical features  at
$n_f \le  n_c $
and $n_f \ge n_c+1$, the results for the squark
and gaugino condensates,  $\bra m_i {\tilde Q}^i Q^i \ket = \bra \lambda
\lambda \ket  =  (\det m)^{1/n_c}  \,\,  \Lambda_1^{3- n_f/n_c}$
hold  true for all values of the
flavor and color  as long as      $n_f < 3 n_c$ \cite{AMKRV}.  }
One finds   therefore     ${\cal N}_1=    ( 2 n_c- n_f ) \cdot  2^{
n_f -1} $   solutions with finite VEVS in exactly the same way
as in   Eqs.(\ref{largemueq})--(\ref{sonforxs}).
In these vacua,   classified by
an integer $r$,
$U(n_f)$ symmetry of the theory is broken spontaneously to various
$U(r)\times U(n_f-r).$

It is now possible to make a highly nontrivial  consistency test,   by the
vacuum counting.
    By using   ${\tilde n}_c= n_f- n_c$ and changing the summation
index from   $r$ to $n_f -
r$,  it is easy to show that
\bea  {\cal N}_2  &= &    \sum_{r=0}^{n_f}  \, _{n_f}\! C_ r \,  ( r- n_c
)  -  \sum_{r=0}^{n_c}  \, _{n_f}\! C_ r \,  ( r- n_c  )
\non \\   &=& - \, ( 2 n_c -n_f ) \cdot 2^{n_f-1}    +   \sum_{r=0}^{n_c-1}  \,
_{n_f}\! C_ r \,  ( n_c -r   ).
\label{nontrivial1} \eea
   The total number of the $N=1$ vacua is  therefore    found to be
\beq    {\cal N}_1  + {\cal N}_2  =  \sum_{r=0}^{n_c-1}  \, _{n_f}\! C_ r
\,  ( n_c -r   ),  \label{nontrivial2}  \eeq
which is the correct answer  (see Eq.~(\ref{nofvac}))   for $n_f > n_c +1$!

\subsection{$USp(2n_c)$  Theories:   Small Numbers ($n_f \le n_c+1$)
of Flavor }

We start with the cases $n_f \le n_c+1$.
At large $\mu$ the equation of motion for  the $\Phi$ superfield is:
\beq
\Phi^{A} = - {1 \o \sqrt{2} \, \mu } (Q_{\alpha}^{i} S^{A}_{\alpha \beta}
Q^{i}_{\beta})\, ,
\eeq
where $\Phi = \Phi^{A} S^{A}$ and $S^{A}$ are the $USp(2n_c)$
   generators\footnote{
With respect to the $\hat{S}^{A}$ generators defined in App.~A, $
S^{A} = \hat{S}^{A} \, J$ with $S^{A}_{\alpha \beta} = S^{A}_{\beta
\alpha}$ and  $\Tr S^{A} \, J S^{B} \, J = {1 \o 2} \delta^{AB}$.}.
Resubstituting it into the superpotential (\ref{uspsuperpot}), and
accounting for the instanton--induced contribution for $n_f \leq
n_c$,    one gets:
\beq
W =  - {1 \o 8 \mu} \Tr MJMJ - {1 \o 2}\, \Tr \, mM
+ (n_c + 1 - n_f){ \Lambda_1^{3 +
2n_{f}/(n_{c}+1-n_{f})} \o ({\rm Pf}M)^{1/(n_{c}+1-n_{f})} } \, ,
\label{usplargemueflag} \eeq
where $M^{ij}=Q^{i}_{a} J^{ab} Q^{j}_{b}$'s are the meson--like
composite superfields,  $m = -i \sigma_{2} \otimes \diag \, (m_1,
\ldots,m_{n_f})$ is the mass matrix,  and
$\Lambda_1^{2(3n_c+3-n_f)} =  \mu^{2n_c+2}  \Lambda^{2(2n_c+2-n_f)}$.

With the ansatz $M = i\sigma_{2} \otimes {\rm diag}(\lambda_{1},
\lambda_{2}, \ldots, \lambda_{n_{f}})$, then the superpotential is
($\Lambda_1 = 1$ below):
\beq
W = {1 \o 4 \mu} \sum_{i=1}^{n_f} \lambda_{i}^{2} -  \sum_{i=1}^{n_f}
m_i \lambda_{i} + (n_c + 1 - n_f){1 \o (\prod_{i=1}^{n_f}
\lambda_{i})^{1/(n_c+1-n_f)}} \, ,
\eeq
and the vacuum equations become:
\beq
{1 \o 2 \mu} \lambda_{i} -  m_{i} - {1 \o \lambda_i}
{1 \o (\prod_{j}\lambda_{j})^{1/(n_c + 1 - n_f)} } = 0.
\label{uspvac}
\eeq
   Since the last term is common   to all
$i$,  we  find
\beq
\lambda_i = \mu ( m_i \pm \sqrt{m_i^2 + (2X/\mu)} ),  \qquad   X\equiv  {1
\o (\prod_{j}\lambda_{j})^{1/(n_c + 1 - n_f)} }
\label{noncommut}\eeq
We choose $r$ negative signs and $n_f - r$ positive signs in
the roots of $\lambda_i$.  In strictly massless limit
$m=0$, the definition of $X$ gives\footnote{ Eq.~(\ref{noncommut})  and
analogous relations in  $SU(n_c)$
case  clearly
show the non commutativity of the two limits,     $\mu \to \infty $  first
with
$\Lambda_1,    m_i $ fixed    to be studied in the next section,    and
$m_i \to 0$
first
with   $ \mu \gg \Lambda    $  fixed,     being   examined here. }
\beq
X \equiv {1 \o (\prod_{j}\lambda_{j})^{1/(n_c + 1 -
      n_f)} }
= {1 \o ((-1)^{r} (2X   \mu)^{n_f/2})^{1/(n_c + 1 -
      n_f)} }
\label{eqforX}\eeq
and hence
\beq
X_0 =\Lambda_1^{2(3n_c+3-n_f)\o 2n_c+2-n_f}   ( {2 \mu}
)^{-n_f \o 2n_c+2-n_f}  e^{2 \pi i k / (2n_c+2-n_f)}, \qquad k=1,2, \ldots,
2n_c+2-n_f,
\label{slnforx0}\eeq
where the subscript 0 indicates that this is for the massless quark
limit and we have reinstated  the dependence on  the scale $\Lambda_1$.
Note that
\beq
X_0 \propto   \mu \,  \Lambda^2\eeq
in terms of the $N=2$ scale factor $\Lambda $ and $\mu$.
 There are $(2n_c+2-n_f)$ roots for $X_0$.  Up to
$O(m)$, we find
\beq
X = X_0 \left[ 1 +
    \frac{2 ( \sum_{j=1}^r m_j   -  \sum_{j=r+1}^{n_f} m_j)}{2n_c+2-n_f}
    \left(\frac{\mu}{2X_0}\right)^{1/2} \right],
\eeq
and
\beq
\lambda_i = \pm \mu \left[ \left(\frac{2X_0}{\mu}\right)^{1/2}
    + \frac{ \sum_{j=1}^r m_j -  \sum_{j=r+1}^{n_f} m_j}{2n_c+2-n_f}
    \pm m_i \right]
\eeq
There appears to be $2^{n_f}$ choices for the signs among  $\lambda_i$'s;
actually,  the signs of $\lambda_i$ must be such that  Eq.~(\ref{eqforX}),
and not its $2(n_c+1-n_f)$
th power,  is satisfied.  This restricts the choices of the signs by half:
for a particular phase of $X_0$  with   $k$ even or odd,    the number of
minus signs
among $\lambda_i$ must be even or odd,  respectively.      In
total, there are
\beq   (2 \, n_c+2-n_f) \, 2^{n_f-1}\eeq
   vacua,    which is consistent
with the previous method as well as the semi-classical method in the
previous section (Eq.~(\ref{Nvspnclassbis})).

Since the massless limit gives $SO(2n_f)$ flavor symmetry, the choice
of different signs in $\lambda_i$ for each flavor strongly suggests
the vacua to form a spinor representation of $SO(2n_f)$.  The
constraint that the number of minus signs to be even  or odd   implies
that  it is a
spinor representation of a definite chirality.  In fact, all of these
vacua transform among each other under $SO(2n_f)$ group because it can
flip the signs of two eigenvalues at the same time, consistent with an
irreducible representation of $SO(2n_f)$.  The equal mass case $m_i =
m$ has $U(n_f)$ flavor symmetry and the vacua above form $_{n_f}C_r$
multiplets (with even or odd  $r$), consistent with the decomposition of the
$SO(2n_f)$ spinor to even or odd-rank $r$ anti-symmetric tensors under
$U(n_f)$. In each group of vacua, the global   symmetry is  broken to
$U(r) \times U(n_f-r)$.

One of the most important results of this section is  that the meson
condensates   in the massless
limit always break  $SO(2n_f)$ flavor symmetry as
\beq   SO(2n_f)   \to   U(n_f),\eeq     matching
nicely with the expectation discussed in Section 2.

When $n_f = n_c+1$, the large $\mu$ theory develops a quantum modified
constraint
\beq
W =  - {1 \o 8 \mu} \Tr M^{2} - {1 \o 2}\, \Tr \, mM + X ({\rm Pf} M -
\Lambda_1^{2n_f})\, .
\eeq
Following the similar analysis as above, we again find the total
number of vacua to be $(2\,n_c+2-n_f)  \,  2^{n_f-1}$, consistent with the
semi-classical method.

When $n_f = n_c+2$, the large $\mu$ theory develops a superpotential
\beq
W =  - {1 \o 8 \mu} \Tr M^{2} - {1 \o 2}\, \Tr \, mM
+ \frac{{\rm Pf} M}{\Lambda_1^{2n_f-3}}\, .
\eeq
Following the similar analysis as above, we again find the total
number of vacua to be $n_c  \, 2^{n_f-1}+1$, consistent with the
semi-classical method.  The last vacuum corresponds to the case
without symmetry breaking $\lambda_i = 2 m_i \mu \rightarrow 0$ in the
massless quark limit.

\subsection{$USp(2n_c)$  Theories:   Large   Numbers ($n_f > n_c+2$)
of Flavor \label{sec:Usplargef}}

Next  consider the cases  $n_f > n_c+2$.    The large $\mu$ theory has a
description
in terms of the dual magnetic gauge group $USp(2\tilde{n}_c) =
USp(2(n_f - n_c - 2))$ and magnetic quarks $q$,
\beq
W = - {1 \o 8
    \mu} \Tr M^{2} - {1 \o 2}\, \Tr \, mM + \frac{1}{\mu_m}   q  M q\, ,
\eeq
where the scale $\mu_m$ is the matching scale between the electric and
magnetic gauge  couplings.    As in the $SU(n_c)$   cases above,  it
is only consistent to use
this  effective action to get information on the vacuum properties   as long as
dual quarks   turn out to be    light. Otherwise,    the nontrivial dual
gauge dynamics must be taken into acoount.

   By minimizing the  potential of  the magnetic theory with dual
quarks directly,  we find
vacua  characterized by VEVS
\begin{equation}
    q_i = \sqrt{\mu m_i}, \qquad \lambda_i = 0,
\end{equation}
for $r$ flavors and
\begin{equation}
    q_i = 0, \qquad \lambda_i = m_i \mu,
\end{equation}
for the remaining $n_f - r$ flavors.  $r$ is restricted to be $r \leq
\tilde{n}_c = n_f - n_c - 2$.  For each value of $r$, there are
$_{n_f} C_r$ choices on which flavor to have non-vanishing $q_i$, and
there is unbroken $USp(2(n_f-n_c-2-r))$ gauge group.  It develops the
gaugino condensate, giving $n_f-n_c-1-r$ vacua each.  The number of
vacua of this type is
\begin{equation}
    {\cal N}_2 = \sum_{r=0}^{n_f - n_c - 2} {}_{n_f} \!C_r  \, (n_f-n_c-1-r).
\end{equation}
   Changing the variable $r$ to $n_f - r$, it is rewritten as
\begin{equation}
    {\cal N}_2 = \sum_{r=n_c+2}^{n_f}  {}_{n_f}\!C_r  \, (r-n_c-1)
    = \sum_{r=0}^{n_f} {}_{n_f}\!C_r \, (r-n_c-1)
    + \sum_{r=0}^{n_c+1} {}_{n_f}\!C_r  \, (n_c+1-r) .
\end{equation}
The first sum can be computed and gives
\begin{equation}
    {\cal N}_2 = - (2n_c+2-n_f) \, 2^{n_f-1}
    + \sum_{r=0}^{n_c} {}_{n_f}\!C_r  \,  (n_c+1-r) .
\end{equation}
Note that the sum in the second term can be stopped at $r=n_c$ because
the argument vanishes for $r=n_c+1$.  Therefore the total ${\cal N}=
{\cal N}_1 + {\cal N}_2$ agrees with the counting of classical vacua.
   On the other hand, this gives a nice interpretation of
the number that the ``extra'' contribution ${\cal N}_2$ signals the
emergence of the dual gauge group in the massless quark limit.
In this group   of     vacua  (present only for larger values of
$n_f$),  the chiral symmetry is
not spontaneously broken  in the limit, $m_i \to 0$.

In order to get the vacua  with   ${\hbox {\rm Rank}} M =  n_f$,
one must integrate out    the dual quarks first and consider the
resulting effective action:
\beq
W_{\it\!eff}= - {1 \o 8
    \mu} \Tr M^{2} - {1 \o 2}\, \Tr \, mM + (n_c +1 -n_f) \left[{\hbox {\rm
Pf}} \, {M \o \mu_m} \,
    {\tilde \Lambda}^{3(n_f-n_c -1) -n_f }\right]^{1/(n_f-n_c-1)}.
\label{efflaguspnf}\eeq
By making the ansatz, $M = i\sigma_{2} \otimes {\rm diag}(\lambda_{1},
\lambda_{2}, \ldots, \lambda_{n_{f}})$,
the vacuum equation becomes
\begin{equation}
    \left( \prod_i^{n_f} \frac{\lambda_i}{\mu_m} \right)^{1/(n_f-n_c-1)}
    \tilde{\Lambda}^{(3(n_f-n_c-1)-n_f)/(n_f-n_c-1)}
    =m_i\lambda_i - \frac{\lambda_i^2}{\mu}.
\end{equation}
Call the right hand side,  which  is flavor indepedent,  $X$.  We have two
solutions for $\lambda_i$ for given $X$,
\begin{equation}
    \lambda_i = \frac{\mu}{2} \left[ m_i \pm \sqrt{m_i^2 + 4
    X/\mu}  \right] .
\end{equation}
   In the massless limit, $\lambda_i = \sqrt{X \mu}$, which in turns gives
\begin{equation}
    X = (X \mu)^{n_f/2(n_f-n_c-1)}
    \tilde{\Lambda}^{(3(n_f-n_c-1)-n_f)/(n_f-n_c-1)}
    \mu_m^{-n_f/(n_f-n_c-1)}.
\end{equation}
The solution is given by
\begin{equation}
    X^{2n_c+2-n_f} =  \tilde{\Lambda}^{-2(3(n_f-n_c-1)-n_f)}
    \mu_m^{2n_f} \mu^{-n_f}.
\end{equation}
This obviously gives $2n_c+2-n_f$ solutions, for which there are
$2^{n_f-1}$ possibilities on the sign choices for each $\lambda_i$.
   Therefore we find
\begin{equation}
    {\cal N}_1 = (2n_c+2-n_f) \, 2^{n_f-1}
\end{equation}
   vacua.  In  this set of vacua     $X$ depends on $\tilde{\Lambda}$ and has a
finite  $m_i \to  0 $   limit ({\it i.e.}\/, $\lambda_i$ stay
non-vanishing, justifying the assumption of the maximal rank meson
matrix).   Since all  $\lambda_i$'s are equal in  the magnitude in this limit,
the chiral  $SO(2n_f)$  symmetry is spontaneously broken as
\beq    SO(2n_f) \to U(n_f) \eeq
in all vacua  belonging to this group.   Note that this number of
vacua  would  precisely corresponds to that
of monopole condensation in the $SO(2n_f)$ spinor representation at
the Chebyshev points of the curve (see below).

The total number of vacua  at large $\mu$  found,     $ {\cal N}_1 +
{\cal N}_2 =  \sum_{r=0}^{n_c} {}_{n_f} \!C_r \,  (n_c+1-r)$,
agrees with  that of the semiclassical theories.

\subsection{Summary of Section   \ref{sec:largem} }

The number of $N=1$ vacua and the pattern of symmetry breaking in each
of them has thus been determined in $SU(n_c)$ and $USp(2n_c)$ theories
at large $\mu$, from the first principles.  For small numbers of
flavor ($n_f \le n_c +1$ for $SU(n_c)$; $n_f \le n_c+2$ for
$USp(2n_c)$) the low--energy degrees of freedom are meson--like
(sometimes also baryon--like) composites: their condensation lead to a
definite pattern of symmetry breaking in each vacuum.

   $SU(n_c)$
theories  have  an
exact global $U(n_f)$ symmetry in the equal mass  (or massless)
limit, which is spontaneously broken
to $   U(r)\times
U(n_f-r)  $     in  $ \, (n_c -r)\,  {}_{n_f}\!C_ r\, $ vacua,
$r=0,1,\ldots, [n_f/2]$.
The number of the vacua with the particular pattern of
symmetry breaking will match exactly with those found from the
analysis of low-energy monopole/dual quark    effective action, to be
analyzed in
the next sections.

In $USp(2n_c)$ theories, for small numbers of flavor, {\it the chiral
symmetry ($SO(2n_f)$) in the massless limit is always spontaneouly
broken down to an unbroken $U(n_f)$.} This result nicely agrees with
what is expected generally from bifermion condensates of the standard
form in non supersymmetric theories.  This fact that   various vacua
have exactly the same
symmetry breaking pattern,  has an important consequence in the
physics at small $\mu$, to be studied
below.

The difference in the   symmetry breaking pattern in  $SU(n_c)$ and
$USp(2n_c)$   theories   reflects the
structures of the
low-energy effective actions of the respective theories,   which in
turn   is a direct consequence of  the different
structure of the two  types of gauge groups,   see
Eq.(\ref{splarmusmnf})  and Eq.(\ref{usplargemueflag}).

For larger numbers of flavor ($n_f > n_c +1$ for $SU(n_c)$; $n_f >
n_c+2$ for $USp(2n_c)$) the low-energy degrees of freedom are dual
quarks and gluons, as well as some mesons.  In these cases,  besides
the vacua with the properties
mentioned above, other vacua exist in which all VEVS  vanish and in
which the global symmetry ($SU(n_f)$ for $SU(n_c)$; $SO(2n_f)$ in
$USp(2n_c)$ theories) remains unbroken
   in the massless limit
($m_i \to 0,\, $ $\forall i$).

\newpage

\section{Decoupling of the Adjoint Fields       \label{decouple} }

In this section we discuss briefly  another  limit,  in which  $\mu$
is sent to $\infty$,     keeping    $m_i $ and $  \Lambda_1\equiv  \mu^{n_c
\o   3n_c-n_f}        \Lambda^{2n_c-
n_f  \o 3n_c-n_f   } $  fixed.    We first re-analyse the number of the
$N=1$ vacua in the
regime,  $\mu \gg \Lambda_1, m_i$,  and    reproduce the correct
multiplicity    of vacua.
 Since   the two limits
($\mu \to \infty $  first  and  $m_i \to 0$ first)  do not commute, this
provides for
 an independent check of the    vacuum counting.
Subsequently,     taking the decoupling limit, $\mu \to \infty$, we
identify the
standard  supersymmetric  vacua   of  the theories  without  the adjoint
field   $\Phi$ .

\subsection{$SU(n_c)$  }

First consider the cases  with  small number of flavors,    $n_f \le
n_c$, and go back to
the equations for $\lambda_i$
\beq
-{1 \o \mu} \left( \lambda_i - { 1\o n_c} \sum_j \lambda_j\right) + m_i - {1 \o
n_c-n_f} { \Lambda_1^{(3n_c - n_f)/(n_c-n_f)}  \o (\prod_j
\lambda_j)^{1/(n_c-n_f)}}  \lambda_i^{-1} =0 \,  .
\label{largemueqcop}
\eeq
following from
\beq
W =  -{1 \o 2 \mu} \left[ \Tr M^2 - {1 \o n_c}(\Tr M)^2 \right]  + \Tr (M m
)  +
{\Lambda_1^{(3n_c - n_f)/(n_c-n_f)}  \o (\det M)^{1/(n_c-n_f)}} \, ,
\label{splarmusmnfcop}\eeq
where  $M_i^j \equiv {\tilde Q}_i^a Q_a^j$;    $\bra M_i^j\ket =diag
(\lambda_1,\ldots, \lambda_{n_f})$.
The scale of the $N=1$  SQCD
\beq  \Lambda_1 \equiv  \mu^{n_c \o   3n_c-n_f}        \Lambda^{2n_c-
n_f  \o 3n_c-n_f   } \eeq
must be kept fixed in the $\mu \to \infty$ limit, to recover
 the standard $N=1$ SQCD.

In the large $\mu$   limit,    some of the $\lambda_i$'s  of
Eq.(\ref{largemueqcop})   are   of the order of
$\mu$, while others are much smaller.
The solutions can thus  be classified according to the number $r$
of the $ \lambda_i$'s  which are  of the order of $\mu$.
The large $\lambda_i$'s (say $\lambda_1, \lambda_2, \ldots,
\lambda_r$) satisfy (setting $\Lambda_1=1$  from now on)
\beq
\lambda_i- { 1\o n_c} \sum_{k=1}^r \lambda_k  -  {\mu \, m_i }
\simeq  0 \, ;
\label{largelameq}
\eeq
which nicely corresponds to Eq.~(\ref{corres}).
The justification for dropping the last term of Eq.~(\ref{largemueqcop})
will be shortly given.

The smaller eigenvalues $\lambda_p$ can be found as follows:
substituting the approximate solutions for the large $\lambda_i$'s
(see Eq.~(\ref{solnd}))
\beq
\lambda_i \simeq \, \mu \,m_i  + { 1 \o n_c -r} \mu \,\sum_{k=1}^r
m_k = (m_i + c) \mu
\eeq
into  Eq.~(\ref{largemueqcop}) with $p=r+1, r+2, \ldots, n_f \, $, one
finds
\beq
(c + m_p) \lambda_p ={1\o n_c-n_f} {1 \o (\prod_{i=1}^{r} \lambda_i
\cdot \prod_{q=r+1}^{n_f} \lambda_{q} )^{1/n_c-n_f} } \, .
\label{uniquely}
\eeq
This can be further put in the form,
\beq  (\lambda_p)^{n_c-n_f} \prod_{q}
\lambda_{q} = {A_p  \o \prod_{i=1}^{r} \lambda_i }=
{B_p \o \mu^{r}} \, ,
\eeq
where $A_p$ and $B_p$ are some finite constants depending on the
masses.
By taking the product over all $p=r+1, r+2, \ldots, n_f \, $, one gets
\beq
(\prod_{q} \lambda_{q})^{n_c-r}=   {\const \o  \mu^{r(n_f-r)}} \, ,
\eeq
therefore
\beq
\prod_{q} \lambda_{q}=   {\const \o  \mu^{r(n_f-r)/(n_c-r)}} \cdot
\exp {2\pi i k/ (n_c-r)} \, ,
\qquad   (k=1, 2, \ldots, n_c-r) \, .
\label{prodlam}
\eeq
Once $\prod_{q} \lambda_{q}$ is determined, each of the small
eigenvalues can be found uniquely from Eq.~(\ref{uniquely}), so that
each choice of $r$ large $\lambda_i$'s yields $n_c -r$ solutions.

It is  easy to see from Eq.~(\ref{prodlam})  that the last term of
Eq.~(\ref{largemueqcop}) behaves as
\beq
\mu^{-n_c/(n_c-r)}
\label{neglig} \eeq
and thus is indeed negligible as compared to the terms kept in
Eq.~(\ref{largelameq}), as   long as  $r     < n_c$.

The total number of the vacua at large  $\mu$  is thus
\beq
{\cal N} = \sum_{r=0}^{n_f}\, (n_c-r)
\, \pmatrix{n_f \cr r}
\label{numvacularmu}
\eeq
which coincides with the number of the classical vacua, Eq.~(\ref{nofvac}).

With
$n_f >n_c+1$,   a na\"{\i}ve use of Eq.(\ref{dualqeq})  in the limit $\mu
\to \infty$
would  lead to no supersymmetric vacua.   The correct vacua can be found by
taking into account    the   nontrivial   dual gauge dynamics and
consequently considering       the effective
action Eq.(\ref{spotlarmu}):  the analysis of the decoupling limit is then
similar to the   $n_f \le n_c$ cases
discussed above.    A subtle new point  however is  that now  the number of
``large'' eigenvalues
$r$,  can  {\it  a priori}  exceed $n_c$.   Actually, however,  for these
values of
$r$    the last term of   Eq.~(\ref{largemueqcop})    becomes dominant and
invalidates the solution (see Eq.(\ref{neglig})):  the sum over $r$
must be truncated at $r=n_c-1$.\footnote{From Eq.(\ref{largelameq})   it
can be   seen that  the case
$r=n_c $ is also excluded.  } One thus finds for the number of vacua
\beq
{\cal N} = \sum_{r=0}^{n_c-1 }\, (n_c-r)
\, \pmatrix{n_f \cr r} \qquad  {\hbox {\rm for}} \qquad n_f > n_c+1
\,  ,
\label{numvacularmularnf}
\eeq
which is indeed   the correct vacuum multiplicity   in this case
(Eq.~(\ref{nofvac})).

In the $\mu \to \infty$ limit,   all solutions except for  those  with
$r=0$  have  some VEVS  running  away
to infinity.   They do  not belong to  the space of vacua  of the
$N=1$ supersymmetric QCD.   Only  the  $r=0$  solutions  are
characterized by finite VEVS,
\beq    m_i \bra Q_i  {\tilde Q}_i \ket  = {\hbox {\rm indep. of }}
i  =   \Lambda_1^{3n_c-n_f \o  n_c}  (\prod_{j=1}^{n_f}    m_j^{1/n_c})
\cdot e^{ 2
\pi i k/n_c},     \quad  k=1,2,\ldots, n_c:
\eeq
they are indeed
the  well-known  $n_c$ vacua    of $N=1$ SQCD \cite{AMKRV}.

\subsection{$USp(2n_c)$  }

For small numbers of flavors  ($n_f \leq n_c$) the equations
for $\lambda_i$'s are
(Eq.~(\ref{uspvac})):
\beq
{1 \o 2 \mu} \lambda_{i} -  m_{i} - {1 \o \lambda_i}
{1 \o (\prod_{j}\lambda_{j})^{1/(n_c + 1 - n_f)} } = 0 .
\label{repeat}\eeq
     The large $\lambda_i$'s
(say $\lambda_1, \lambda_2, \ldots,\lambda_r$) satisfy:
\beq
\lambda_{i} \simeq  2 \,  m_i \mu \qquad \qquad i = 1, 2 , \ldots , r \, ,
\label{usplargel}
\eeq
where the last term of Eq.~(\ref{repeat})       is negligible, as  will
be  shown shortly.
The $n_f- r$ smaller $\lambda_p$'s are found  by  substituting
Eq.~(\ref{usplargel}) into Eq.~(\ref{repeat}):
\beq
   m_p \lambda_p  = {1 \o (\prod_{i} \lambda_{i}
\prod_{q} \lambda_q)^{1 / (n_c + 1 - n_f)}} \, ,
\eeq
where $i=1, 2, \ldots r$ runs over the large $\lambda$'s and $p,q =
r+1, r+2, \ldots n_f$ refer to  the smaller ones.  This can be further
rewritten  as:
\beq
(\prod_{p} \lambda_p)^{n_c+1-r} = {\const \o \mu^{r(n_f - r)}} \, ,
\label{rewritten}\eeq
and thus
\beq
\prod_p \lambda_p = {\const \o \mu^{r(n_f - r)/(n_c + 1 - r)} } \exp
2 \pi i k / (n_c + 1 - r) \qquad {\hbox{\rm with }}    \, \, \, k =
1,2,\ldots,n_c+1-r \, .
\eeq
One can see that each choice of $r$ large $\lambda_i$'s yields $n_c +1
-r$ solutions.
It is easy to show that the last term of
Eq.~(\ref{repeat})    is indeed  negligible, behaving as
\beq
\mu^{-(n_c + 1)/(n_c + 1 - r)} \,
\label{neglected}\eeq
as long as   $r \le n_c$ (see below).

Summing over  $r$  one finds for the total number of the vacua
\beq
{\cal N} = \sum_{r=0}^{n_f}\, (n_c +1 - r)
   \, \pmatrix{n_f \cr r}
\eeq
which   agrees (for $n_f \le n_c$) with the number of the classical
vacua,
Eq.~(\ref{Nvspnclass});     the above solutions  therefore exhaust  all
possible vacua of the theory.

The analysis for the cases with larger  number of flavor ($n_f >
n_c+1$)   is  quite    similar to
the one made above for smaller values of  $n_f$.   The only difference is that
for $r > n_c$ it is no longer correct to neglect the last term of
Eq.~(\ref{repeat}), as can be seen from
Eq.~(\ref{neglected}), hence the sum over $r$ must stop at $r=n_c$.   The
case $r=n_c+1$ might look subtle, but
it is clear from Eq.~(\ref{rewritten}) that no  solution  exists in this
case either.   One thus ends up with the
number of vacua,
\beq
{\cal N} = \sum_{r=0}^{n_c}  (n_c- r +1) \cdot
\pmatrix {n_f \cr  r},
\label{Nvspn2}
\eeq
which is the correct result    (see Eq.~(\ref{Nvspnclass})).

Again, in  the strict $\mu=\infty$  limit, only  the vacua  with
finite VEVS  must be retained.  They
are the solutions  correspondintg   to $r=0$ above:   we find
precisely  $n_c-1$  solutions
of the $N=1 $    $USp(2n_c-1)$ theory without the adjoint matter fields.

\newpage

\section{Microscopic Picture of  Dynamical Symmetry Breaking
\label{sec:leeflag}  }

       In this and in   three    subsequent sections  we   seek for a
microscopic  understanding of
   the  mechanism of    dynamical flavor symmetry breaking as well as
of confinement itself,    by   studying   these  theories at small
$\mu$ and small
$m_i$.   It is necessary to  analyze      the $N=2$ vacua on the
Coulomb branch which survive  the   $\mu \neq
0$ perturbation.
The auxiliary genus $g= n_c-1$ (or   $n_{c}$) curves for
$SU(n_c)$ ($USp(2 n_c)$) theories are given by
\begin{equation}
    y^{2} = \prod_{k=1}^{n_{c}}(x-\phi_{k})^{2} + 4 \Lambda^{2n_{c}-n_{f}}
    \prod_{j=1}^{n_{f}}(x+m_{j}), \qquad   SU(n_c), \, \,\,  n_f \le 2n_c-2,
\label{curve1} \end{equation}
and
\begin{equation}
    y^{2} = \prod_{k=1}^{n_{c}}(x-\phi_{k})^{2} + 4 \Lambda
    \prod_{j=1}^{n_{f}}\left(x+m_{j} + {\Lambda \o n_c} \right), \qquad
SU(n_c),
\, \,\,  n_f = 2n_c-1,
\label{curve2} \end{equation}
with $\phi_{k}$ subject to the constraint $\sum_{k=1}^{n_{c}}\phi_{k} =
0$, and
\begin{equation}
    x y^{2} = \left[ x \prod_{a=1}^{n_{c}} (x-\phi_{a}^{2})^{2}
      + 2 \Lambda^{2n_{c}+2-n_{f}} m_{1} \cdots m_{n_{f}} \right]^{2}
    - 4 \Lambda^{2(2n_{c}+2-n_{f})} \prod_{i=1}^{n_{f}}(x+m_{i}^{2}),
    \quad USp(2 n_c). \label{curve3}
\end{equation}
The connection between these genus $g$ hypertori and physics  is
made \cite{SW1}-\cite{SUN}      through the
identification of various   period integrals of   the { holomorphic
differentials}
on the curves with   $(d a_{D i} / du_j, d a_{ i} / du_j)$,   where
the gauge invariant parameters
$u_j$'s are defined by the standard relation,
\beq  \prod_{a=1}^{n_{c}}(x-\phi_{a})  =  \sum_{k=0}^{n_c}  u_k  \,
x^{n_c-k}, \qquad  u_0=1, \quad u_1=0,  \qquad
SU(n_c);
\eeq
\beq  \prod_{a=1}^{n_{c}}(x-\phi_{a}^2)  =  \sum_{k=0}^{n_c}  u_k  \,
x^{n_c-k}, \qquad  u_0=1,   \qquad
USp(2 n_c),
\eeq
and  $u_2\equiv \bra \Tr \Phi^2 \ket, $   etc.     The
VEVS of $a_{Di}, \,\, a_{i}$, which are    directly related to the
physical  masses of the
BPS  particles through the exact  Seiberg-Witten mass formula \cite{SW1,SW2}
\beq  M^{n_{mi}, n_{ei}, S_k} = \sqrt2   \,  \left|  \sum_{i=1}^g  (
n_{mi} \, a_{Di} + n_{ei}  \,  a_{i}  )     +   \sum_k   S_k   m_k
\right|, \eeq
   are
constructed as
   integrals   over the non-trivial cycles   of the  { meromorphic
differentials}
   on the  curves.  See \ref{sec:formulas}.

    We require that
the curve is maximally singular, i.e. $g= n_c-1$ (or $n_c$ for
$USp(2n_c)$) pairs of branch points
to coincide: this determines the possible values of $\{\phi_a\}$'s.
These points   correspond to the $N=1$ vacua, for   the particular $N=1$
perturbation, Eq.(\ref{N1pert}).     Note that as we work with
generic and nonvanishing quark masses  (and then  consider $m_i \to
0$  limit),  this is an unambiguous
procedure   to identify all the $N=1$ vacua of  our interest.
\footnote{There are other kinds of singularities of $N=2$ QMS at
    which, for instance, three of the branch points meet.  These
    correspond to $N=1$ vacua, selected out by different types of
    perturbations such as $\Tr \, \Phi^3$, which are not considered here. }

In fact,  near one of the singularities where  dyons  with quantum numbers
\beq  (n_{m1}, n_{m2},\ldots,n_{m g};
n_{e1},n_{e2},\ldots,n_{e g})=(1,0,\ldots,0; 0, \ldots ), \,\,
\ldots, \,\, (0,0,\ldots, 1; 0,\ldots)
\eeq
become  massless,  the effective superpotential reads
\beq
{\cal W} =  \sum_{i=1}^g  \left\{\sqrt2 \, a_{Di} \, {\tilde
M}_{i}  M_{i}  +    \sum_{k=1}^{n_f} S_{k}^{i}  \,
m_{k}  {\tilde M}_{i} M_{i} \right\} +  \mu  \, u_2(a_D, a)
\eeq
where  $S_{k}^{i}$ is the $k-$th   quark number charge of the
$i$ th
dyon \cite{SW2,KT}.
Treating $u_i$  as independent variables (or equivalently,
$a_{Di}$'s),     the equations of the minimum are
\bea
-{\mu \o  \sqrt 2} &=&    \sum_{i=1}^g   {\de a_{Di} \o \de u_2 }
{\tilde M}_{i} M_{i} \, ;  \qquad
0  =    \sum_{i=1}^g {\de a_{Di} \o \de u_j } {\tilde
M}_{i} M_{i},  \quad   j=3,4,\ldots,g+1;  \label{equmins} \eea
\bea
\left(\sqrt{2}\, a_{D1}+ \sum_{k=1}^{n_f}   S_{k}^{1}  \, m_{k} \right)
\,{\tilde M}_{1}  &=&  \left(\sqrt{2} \, a_{D1}+ \sum_{k=1}^{n_f}
S_{k}^{1}  \, m_{k} \right) \, {M}_{1}=0 \, ;   \non \\
       \ldots \qquad   \ldots  &=&  \ldots \qquad   \ldots  \non \\
   \left(\sqrt{2} \, a_{D g}+ \sum_{k=1}^{n_f}   S_{k}^{g}  \, m_{k} \right)
\,{\tilde M}_{g}  &=&  \left(\sqrt{2} \,a_{D   g}+ \sum_{k=1}^{n_f}
S_{k}^{g}  \, m_{k} \right) \, {M}_{g}=0 \, .
\label{vaceqef}
\eea
The $D$-term constraint gives    $|\tilde M_i| = |M_i|$.
For generic
   hyperquark masses,  Eqs.(\ref{equmins}) require  that
\beq   {\tilde M}_{i},  \,   {M}_{i}  \sim \sqrt {\mu \Lambda}   \ne
0, \quad     \forall i, \eeq
   since
${\de a_{Di} \o \de u_j } $ and $ {\de a_{Di} \o \de u_j }$  obey
no special relations.
This means that
\beq
\sqrt{2} \, a_{Di}+ \sum_{k=1}^{n_f}   S_{k}^{i}  \, m_{k} =
0 \,
\eeq
   for all $i$:
i.e.,   all  $g$ monopoles  are  massless simultaneously.
Condensation of each type of monopole,    ${\tilde M}_{i}\ne 0$;
${M}_{i}\ne
0$,   corresponding to  the  maximal Abelian subgroup
of $SU(n_c)$ or of $USp(2n_c)$, amounts   to    confinement \`a la 't
Hooft-Mandelstam-Nambu.

Actually, physical picture  in the  $m_i \to 0$ limit of these
theories  is more subtle and is far   more interesting,
as is  discussed especially    in Section \ref{sec:EfLagdes}.

   It is  in general  difficult to determine explicitly   the
configurations   $\{\phi_a\}$  which satisfy the   $N=1$ criterion
mentioned above, although in some special cases  (with $m_i=0$)  they
can be found  explicitly.
We approach  the problem  by first setting    $m_i=0,$  $\forall i$,
and   by perturbing the solutions
for    $\{\phi_a\}$    by  considering the effects of $m_i$ to first
nontrivial orders.

{\it  It turns out that
the $N=1$ vacua  of the ${ { SU(n_c)} }$  and $USp(2n_c)$      gauge
theories      can all be
generated from the various classes \cite{APSW,EHIY, ArPlSei,APS2}  of
superconformal theories with
$m_i=\mu=0$, by perturbing them  with   masses $m_i$        (as well
as with $\mu$). }

Some  qualifying  remarks are in order.      In the case of $SU(n_c)$
theories   we
find that  the  first  group of $N=1 $ vacua
surviving the  the adjoint mass  perturbation  ($\mu \Tr \Phi^2$)
and leading to finite VEVS,       arise  from  the class 1  ($r <
n_f/2$),
class  3  ($r=n_f/2$ with   $n_c-n_f/2$ odd)    and  class  4
($r=n_f/2$    with   $n_c-n_f/2$ even)    CFT,
according to the classification  of  Eguchi et. al \cite{EHIY}.
The second group of vacua,    present only for $n_f > n_c$,   on  the other
hand,  arise from the so-called baryonic branch root \cite{ArPlSei}
(see Eq.(\ref{bbranch})):      these   latter  CFT  can also  be
regarded
as special case of class 1 theories.  In these vacua the flavor symmetry is
unbroken  in the $m_i \to 0$ limit.

   It is important that the number of the vacua and  qualitative
results about symmetry breaking in each of them, can be obtained this
way   even when  the unperturbed solution for
$\{\phi_a\}$ is not  known explicitly.

\subsection{ Superconformal Points and   $N=1$ Vacua  for   ${ {
SU(n_c)} }$   \label{sec:masspertsun} }

It will be seen that  the first group    of
vacua (with multiplicity ${\cal N}_1$)    are  associated to   the   points,
\begin{equation}
    {\hbox {\rm diag}} \, \phi =(\,  {\underbrace  {0,0, \ldots, 0}_{r}},
    \phi^{(0)}_{r+1}, \ldots,  \phi^{(0)}_{n_c} ), \qquad
\sum_{a=r}^{n_c}   \phi^{(0)}_a=0,
\label{unpertphi}   \end{equation}
in Eq.(\ref{curve1}),  with $ \phi^{(0)}_a$'s chosen such that the
nonzero $2(n_{c}-r-1)$ branch points are
paired \footnote{Actually there is no vacuum of this
   type Eq.(\ref{rbranch})     with $r=\tilde{n}_{c}$.  This will be
shown in \ref{sec:absence}. }.      The    curves   are   of   the
form,
\begin{equation}
    y^2\sim   x^{2r} (x-\beta_{0 1})^{2} \cdots
        (x-\beta_{0, n_{c}-r-1})^{2} (x-\gamma_0) (x-\kappa_0),
        \qquad        r=0,1,2,\ldots, [n_f/2].
\label{rbranch}   \end{equation}
For  $r<n_f/2$   they   correspond   to the
so-called class 1 superconformal theories  \cite{EHIY}.

   The special cases   with $r=n_f/2$, with $n_c-n_f/2$ odd,
and   $r=n_f/2$,  $n_c-n_f/2$ even,     may be interpreted as
belonging to  class 3 and      class 4 \cite{EHIY},  respectively. In
fact,
in these special cases  the explicit configuration of $ \phi_a$'s
    can be found  by using the method of   \cite{DS}.
The curve for bare   quark masses  with   $r=n_f/2$   vanishing
$\phi_a$'s    is:
\begin{equation}
      y^{2} = x^{n_{f}} \left[
                  \prod_{k=1}^{n_{c}-n_{f}/2} (x-\phi_{k})^{2}
          - 4\Lambda^{2n_{c}-n_{f}} \right].
\end{equation}
We identify the first term in the square bracket with $(2
\Lambda^{n_{c}-n_{f}/2} T_{n_{c}-n_{f}/2} (x/2\Lambda))^{2}$,
where  $T_N(x)$  is  the   Chebyshev polynomial  of order $N$,
$T_N(x)=   \cos (N \cos^{-1} x), $      implying that
\beq   \phi_{k} = 2 \Lambda \cos \pi
(2k-1)/2(n_{c}-n_{f}/2), \qquad    k=1, \cdots, n_{c}-n_{f}/2.
\label{chebi1}\eeq
The two terms in the square bracket combine as
\begin{eqnarray}
      y^{2} &=& x^{n_{f}} \left[
          \left(2 \Lambda^{n_{c}-n_{f}/2} T_{n_{c}-n_{f}/2}
          \left(\frac{x}{2\Lambda}\right)\right)^{2}
          - 4\Lambda^{2n_{c}-n_{f}} \right]
                  \nonumber \\
          & = & - 4 x^{n_{f}} \Lambda^{2n_{c}-n_{f}}
                  \sin^{2} \left[\left(n_{c}-\frac{n_{f}}{2}\right)
                          \arccos \frac{x}{2\Lambda}
                  \right].
\label{chebi2}\end{eqnarray}
There are $n_{f}$-th order   zero at $x=0$, and double zeros at
$x=2\Lambda\cos \pi k/(n_{c}-n_{f}/2)$ for $k=1, \cdots,
n_{c}-n_{f}/2-1$, and single zeros at $x=\pm 2\Lambda$.
Note that    for $r=n_f/2$,  $n_c-n_f/2$ even,    the zero at
$x=0$ is actually of order  $n_f +1$.

   Since
these adjoint VEVS  leading to   Eq.(\ref{rbranch})  or
Eq(\ref{chebi2})    break the discrete symmetry spontaneously, they
appear in
$2n_c - n_f$ copies.\footnote{There is an exception to this.  In the case of
    $r=n_f/2$ with $n_f$ even, the explicit configuration of $ \phi_a$'s
    (Section~\ref{sec:pert1})   shows that  the vacuum
    respects $Z_2$ subgroup of the $Z_{2n_c-n_f}$ symmetry, showing that
    it appears in $n_c - n_f/2 $ copies rather than $2n_c - n_f.$ This
    fact is crucial in the vacuum counting below Eq.(\ref{vaccount1}).}
When (generic) quark masses are turned on, these vacua split into
${}_{n_f}\! C_r $-plet of single vacua.   The vacua
Eq.(\ref{rbranch}), Eq(\ref{chebi2}),  correspond  to what  is called
``nonbaryonic" branch  roots in \cite {ArPlSei}.

The second group   of vacua will  be   related   to    the (trivial)
superconformal theory
\begin{equation}
    y^2  =    x^{2{\tilde n}_c}
    ( x^{n_c - {\tilde n}_c } -   \Lambda^2   )^2, \qquad
    {\tilde n}_c = n_f - n_c,
   \label{bbranch}
\end{equation}
corresponding to the adjoint configuration
\begin{equation}
    {\hbox {\rm diag}} \, \phi =
    (\, {\underbrace  {0,0, \ldots, 0}_{{\tilde n}_c}  },
    \Lambda \,\omega^2,  \ldots,  \Lambda \, \omega^{2(n_c - {\tilde n}_c}))
   \label{bbranch2}  \end{equation}
with $\omega= e^{ \pi i / (n_c - {\tilde n}_c
    )}$.  This  is the ``baryonic" root of  \cite {ArPlSei}.

To justify     these statements   we must     solve the following
problem of purely mathematical nature.  Suppose that  a configuration
$\{\phi_a\}= \{\phi_a^{0}\}$  has been found     such that     the
curve of the $m_i=0$ theory  reduces to one of the
forms,  Eq.(\ref{rbranch}) or Eq.(\ref{bbranch}).   Now we  add small
generic bare hyperquark masses $m_i$;
we want to   determine the   shifts of  $\{\phi_a\}$,    $\{\phi_a\}=
\{\phi_a^{0} + \delta \phi_a \} $,  with constraints
$\sum_a  \delta \phi_a=0 $ and  $ \delta \phi_a \to 0 $ as  $ m_i \to
0$,     such that      the massive curve Eq.(\ref{curve1})
or    Eq.(\ref{curve2})  is maximally singular (with    $g=n_c-1$
double   branch points).
How   many  such solutions  are there?

It turns out  that this problem must be treated separately for
several   different cases:   the result of this mass perturbation
theory,     which will be   developed     in the last section of this
paper  (Section    \ref{CKMmp}),  can be summarized as follows.
For small number of  the flavors  ($n_{f}\leq n_{c}$) the total
number of the $N=1$ vacua generated  by the mass
perturbation  from  various   conformal invariant points
Eq.(\ref{rbranch}), Eq.(\ref{unpertphi}),
is:
\begin{equation}
   {\cal N}_1 =   (2 n_c - n_f)  \sum_{r=0}^{(n_f-1)/2}  {}_{n_f} C_r
    =   (2 n_c - n_f)  \, 2^{n_f-1}, \quad  (n_f = {\hbox {\rm odd}})
\label{vaccountnb}\end{equation}
\begin{equation}
   {\cal N}_1 =  (2 n_c - n_f) \sum_{r=0}^{{n_f}/2-1}  {}_{n_f} C_r
    \,   +  {2 n_c - n_f \o 2}\,  {}_{n_f} \!C_{n_f/2}
    =   (2 n_c - n_f) \, 2^{n_f-1},
    \quad  (n_f = {\hbox {\rm even}}),
    \label{vaccount1}
\end{equation}
which      exhausts ${\cal N}$,  Eq.(\ref{nofvacbis}).
In Eq.(\ref{vaccount1}) we have taken into
account the fact that for even $n_f$, the vacua with $r=n_f/2$ do not
transform under $Z_{2n_c-n_f}$ but only under $Z_{n_c-n_f/2}.  $
For larger $n_f$  ($n_f \ge n_c+1$),   there are also
   \begin{equation}
        {\cal N}_{2}
        = \sum_{r=0}^{\tilde{n}_{c}-1} (\tilde{n}_{c}-r) \,
{}_{n_{f}}\!C_{r}.
        \label{eq:N22}
\end{equation}
vacua coming from the ``baryonic root", Eq.(\ref{bbranch}),
Eq.(\ref{bbranch2}).   The total
$       {\cal N}_{1} +  {\cal N}_{2} $  correctly matches the known
total number of the vacua.  The arithmetics
is the same as in Eq.(\ref{nontrivial1}), Eq.(\ref{nontrivial2}), and
will not be repeated.

Actually,    there is an interesting    subtlety in this vacuum
counting.  In the  first group of vacua
Eq.(\ref{vaccountnb}),  Eq.(\ref{vaccount1}),      the term
$r={\tilde n}_c=n_f-n_c$   must    be dropped.     The vacuum
Eq.(\ref{rbranch})   turns out to be    nonexistent for $r={\tilde
n}_c=n_f-n_c$,   see \ref{sec:absence}.     At the same time,
however,     the mass perturbation  around the  baryonic branch root
Eq.(\ref{bbranch}),  Eq.(\ref{bbranch2}),   gives  (see Sections
\ref{sec:EfLagdes},
\ref{CKMmp})
\beq   {\cal N}_{2}   +
(n_{c}-\tilde{n}_{c}) \,  {}_{n_{f}}
\!  C_{\tilde{n}_{c}} \eeq
vacua,  the second term of which  compensates    precisely the
missing term   in the sum in ${\cal N}_1$.

\subsection  {Superconformal Points and   $N=1$ Vacua  of
$USp(2n_c)$  Theory   \label{sec:masspertusp}}

The general curve of $USp(2n_{c})$ theories is given by
\begin{equation}
      x y^{2} = \left[ x \prod_{a=1}^{n_{c}} (x-\phi_{a}^{2})
          + 2\Lambda^{2n_{c}+2-n_{f}} m_{1} \cdots m_{n_{f}} \right]^{2}
          - 4\Lambda^{2(2n_{c}+2-n_{f})} \prod_{i=1}^{n_{f}}(x+m_{i}^{2}).
      \label{eq:curve}
\end{equation}
When $2n_{c}+2-n_{f}=0$, the theory is superconformal and
$2\Lambda^{2n_{c}+2-n_{f}}$ in the above expression is replaced by
\begin{equation}
          g(\tau) =
          \frac{\vartheta_{2}^{4}}{\vartheta_{3}^{4}+\vartheta_{4}^{4}}.
\end{equation}
It turns out  that the first group of  $N=1$  vacua  can be generated
from   the  CFT  described by   Chebyshev solutions
for $m_i=0$   theory.

    When $n_{f}$ is odd, choose $(n_{f}-1)/2$ eigenvalues
$\phi_{1}=\cdots=\phi_{(n_{f}-1)/2}$ vanishing.  Then the curve
becomes
\begin{equation}
      y^{2} = x^{n_{f}-1} \left[
                  x \prod_{k=1}^{n_{c}-(n_{f}-1)/2} (x-\phi_{k}^{2})^{2}
          - 4\Lambda^{2(2n_{c}+2-n_{f})} \right].
\label{cheby0}   \end{equation}
We identify the first term      in the
square bracket with $(2
\Lambda^{2n_{c}+2-n_{f}} T_{2n_{c}+2-n_{f}} (\sqrt{x}/2\Lambda))^{2}$,
where $\phi_{k}^{2} = 4 \Lambda^{2} \cos^{2} \pi
(2k-1)/2(2n_{c}+2-n_{f})$ for $k=1, \cdots, n_{c}-(n_{f}-1)/2$.  Then
the two terms in the square bracket combine as
\begin{eqnarray}
      y^{2} &=& x^{n_{f}-1} \Lambda^{2(2n_{c}+2-n_{f})}\left[
                  4 T_{2n_{c}+2-n_{f}}^{2}
\left(\frac{\sqrt{x}}{2\Lambda}\right)
          - 4 \right]
                  \nonumber \\
          & = & - 4 x^{n_{f}-1} \Lambda^{2(2n_{c}+2-n_{f})}
                  \sin^{2} \left[(2n_{c}+2-n_{f}) \arccos
\frac{\sqrt{x}}{2\Lambda}
                  \right].
\label{chevy1} \end{eqnarray}
There are $n_{f}-1$ zeros at $x=0$, and double zeros at
$x=4\Lambda^{2}\cos \pi k/(2n_{c}+2-n_{f})$ for $k=1, \cdots,
n_{c}-(n_{f}-1)/2$, and a single zero at $x=4\Lambda^{2}$.

When $n_{f}$ is even, choose $n_{f}/2-1$ eigenvalues
$\phi_{1}=\cdots=\phi_{n_{f}/2-1}$ vanishing.  Then the curve
becomes
\begin{equation}
      y^{2} = x^{n_{f}-1} \left[
                  \prod_{k=1}^{n_{c}+1-n_{f}/2} (x-\phi_{k}^{2})^{2}
          - 4\Lambda^{2(2n_{c}+2-n_{f})} \right].
\label{chevy2} \end{equation}
We identify the first term in the square bracket with $(2
\Lambda^{2n_{c}+2-n_{f}} T_{2n_{c}+2-n_{f}} (\sqrt{x}/2\Lambda))^{2}$,
where $\phi_{a}^{2} = 4 \Lambda^{2} \cos^{2} \pi
(2k-1)/2(2n_{c}+2-n_{f})$ for $k=1, \cdots, n_{c}+1-n_{f}/2$.  Then
the two terms in the square bracket combine as
\begin{eqnarray}
      y^{2} &=& x^{n_{f}-1} \Lambda^{2(2n_{c}+2-n_{f})}\left[
                  4 T_{2n_{c}+2-n_{f}}^{2}
\left(\frac{\sqrt{x}}{2\Lambda}\right)
          - 4 \right]
                  \nonumber \\
          & &     = - 4 x^{n_{f}-1} \Lambda^{2(2n_{c}+2-n_{f})}
                  \sin^{2} \left[(2n_{c}+2-n_{f}) \arccos
\frac{\sqrt{x}}{2\Lambda}
                  \right].
\label{cheby3}      \end{eqnarray}
Since the $\sin^{2}$ factor gives a single zero at $x=0$,
there are $n_{f}$ zeros at $x=0$, and double zeros at
$x=4\Lambda^{2}\cos \pi k/(2n_{c}+2-n_{f})$ for $k=1, \cdots,
n_{c}-n_{f}/2$, and a single zero at $x=4\Lambda^{2}$.

In the absence of quark masses, the theory is invariant under
$Z_{2n_{c}+2-n_{f}}$ symmetry: $x \rightarrow e^{2\pi
i/(2n_{c}+2-n_{f})}x $, $\phi_{a}^{2} \rightarrow e^{2\pi
i/(2n_{c}+2-n_{f})} \phi_{a}^{2}$.  Therefore the Chebyshev solutions
discussed here appear $2n_{c}+2-n_{f}$ times.

$USp(2n_c)$ theories also have   special  Higgs branch roots   similar to the
baryonic roots of the $SU(n_c)$ theories.
This   baryonic-like root is obtained in the $m_i=0$    limit,  by setting
$\phi_{1}, \cdots, \phi_{n_{c}-\tilde{n}_{c}}\neq 0$,
$\phi_{n_{c}-\tilde{n}_{c}+1}= \cdots =\phi_{n_{c}}=0$.  Here and
below, $\tilde{n}_c = n_f - n_c - 2$.  The curve Eq.~(\ref{eq:curve})
becomes
\begin{equation}
          y^{2} = x^{2\tilde{n}_{c}+1}
          \prod_{k=1}^{n_{c}-\tilde{n}_{c}} (x-\Phi_{k})^{2}
                  - 4 \Lambda^{4n_c + 4 - 2 n_f} x^{n_f-1} .
\label{baryonliketext} \end{equation}
We take
\begin{equation}
          (\Phi_{1}^2, \cdots, \Phi_{k}^2, \cdots,
\Phi_{n_{c}-\tilde{n}_{c}}^2)
          = \Lambda^2 (\omega, \cdots, \omega^{2k-1}, \cdots,
          \omega^{2(n_{c}-\tilde{n}_{c})-1}),
\label{baryonliketext2} \end{equation}
where $\omega = e^{\pi i/(n_{c}-\tilde{n}_{c})}$.  Note that our
$\omega$ is the square root of $\omega$ in \cite{ArPlSei}  because of
later convenience.  Then the product
$\prod_{k=1}^{n_{c}-\tilde{n}_{c}} (x-\Phi_{k})$ can be rewritten as
$x^{n_{c}-\tilde{n}_{c}} + \Lambda^{2(n_{c}-\tilde{n}_{c})}$, and the
curve becomes
\begin{eqnarray}
    y^{2} &=& x^{2\tilde{n}_{c}+1}
    \left[ (x^{n_{c}-\tilde{n}_{c}} + \Lambda^{2(n_{c}-\tilde{n}_{c})})^{2}
    - 4 \Lambda^{4n_{c}+4-2n_{f}} x^{n_{c}-\tilde{n}_{c}} \right]
\nonumber \\
    &=& x^{2\tilde{n}_{c}+1}
    (x^{n_{c}-\tilde{n}_{c}} - \Lambda^{2(n_{c}-\tilde{n}_{c})})^{2}.
   \label{baryonliketext3}  \end{eqnarray}
The double zeros of the factor in the parenthesis are at
\begin{equation}
          x = \Lambda^2 \omega^2, \Lambda^2 \omega^{4}, \cdots,
                  \Lambda^2 \omega^{2k}, \cdots,
                  \Lambda^2 \omega^{2(n_{c}-\tilde{n}_{c})}=\Lambda^2  .
                  \label{eq:USpzeros}
\end{equation}

   When the quark masses are  turned on,   these   points split.    We
require again that  the shift of $\phi$'s  be such that the full
curve   remains  maximally sigular (with maximal possible number of
double roots).    The result of mass  perturbation analysis,
given in
Section \ref{CKMmp}, can be summarized   as follows.  There are
two groups of $N=1$ vacua  predicted by the   Seiberg-Witten
curve in
$USp(2n_c)$ theories.   The Chebyshev point  Eq.(\ref{chevy1}),
Eq.(\ref{chevy2}),  spawns
${\cal N}_1=(2n_{c}+2-n_{f}) \cdot   2^{n_f-1} $   vacua upon  mass
perturbation,  while   the special point
Eq.(\ref{baryonliketext}),  Eq.(\ref{baryonliketext2}),  splits
into   $ {\cal N}_2 =
\sum_{r=0}^{\tilde{n}_c} (\tilde{n}_c-r+1) \, {}_{n_f}\!C_r$  vacua.
Their sum coincides  with the total number of $N=1$ vacua
found from the semiclassical as well as from   large $\mu$ analyses.

\newpage

\section{Numerical  Study of $N=1$ Vacua  in
$SU(3)$ and $USp(4)$ Theories
\label{sec:flsymbr}}

As a way of checking  these results and  of illustrating   some of
their features,
we have performed a   study of the rank $2$  theories,  by
numerically  determing   the  points  in QMS   where  the curves
$y^2=G(x)$   of   Eq.(\ref{curve1})-Eq.(\ref{curve3})
become  maximally singular.  This has been done  by solving the equation
\beq   {\cal R} \left(G(x),  {dG(x) \o dx} \right)    =0, \label{resultant}\eeq
    where  $ {\cal R}$   stands for the resultant,      using Mathematica.

\subsection { $SU(3)$  theory  with  $n_{f}=1 \sim 5$ }

\begin{figure}[tb]
\begin{center}
      \epsfig{file=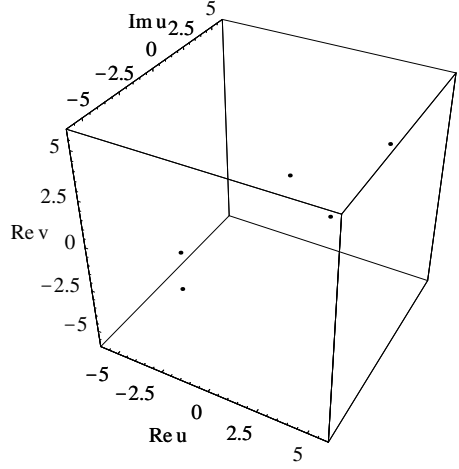,width=6cm}
\end{center}
\caption{Five vacua of the $SU(3)$, $N=1$ theory with $n_{f}=1$
flavors, plotted as the projection $(\Re \, u, \Im \, u, \Re \, v)$ of
the QMS.  ($\Lambda_{1} = 2$,   $m= 1/64$,
$u\equiv {1\o 2} \Tr \langle \Phi^{2} \rangle$, $v\equiv {1\o 3} \Tr
\langle \Phi^{3}
\rangle$).     }
\label{figsu3nf1}
\end{figure}

\begin{figure}[tb]
\begin{center}
      \makebox[\textwidth][s]{
      \epsfig{file=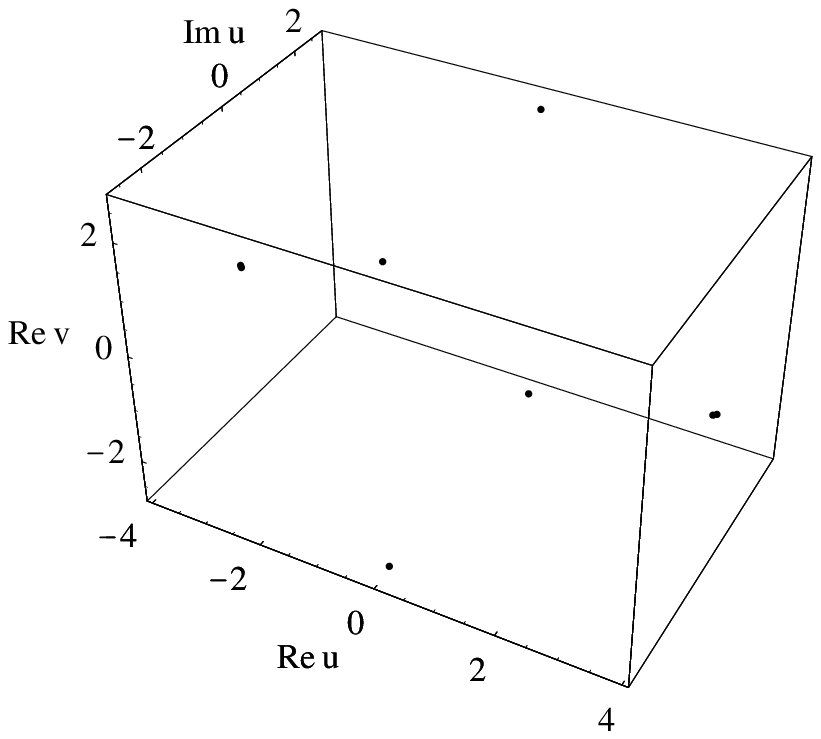,width=6cm}
      \epsfig{file=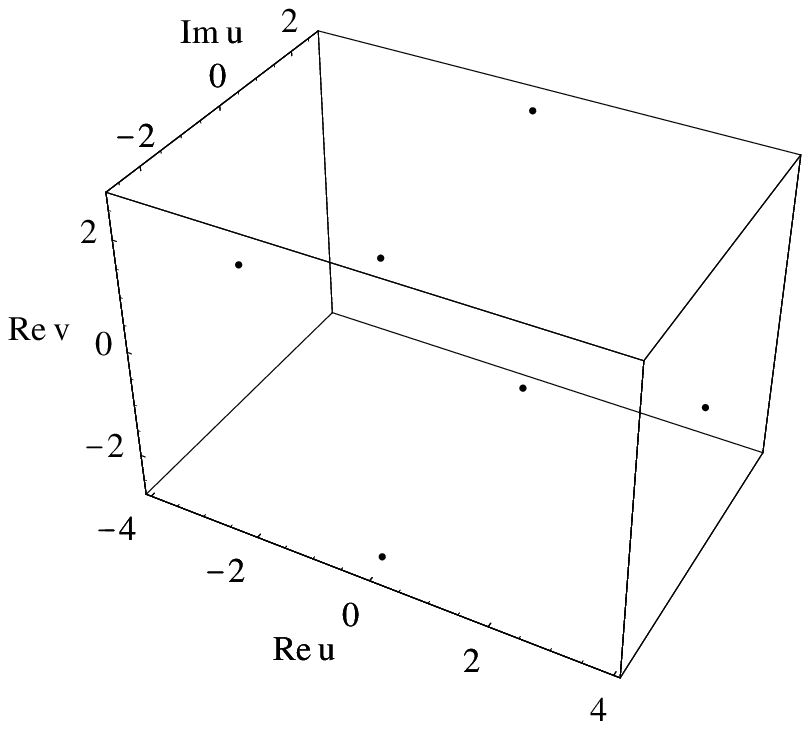,width=6cm}
      }
\end{center}
\caption{In the left figure are  the  eight vacua of the $SU(3)$ theory with
$n_{f}=2$, plotted as the projection $(\Re \, u, \Im \, u, \Re
\, v)$ of the QMS. ($\Lambda_{2} = 2, \, m_{1}=
1/64$,     $m_{2}= i/64$).  The  same in the right figure with    equal masses
$m_{1}=m_{2}=1/64$.  }
\label{figsu3nf2}
\end{figure}

\begin{figure}[tb]
\begin{center}
      \makebox[\textwidth][s]{
      \epsfig{file=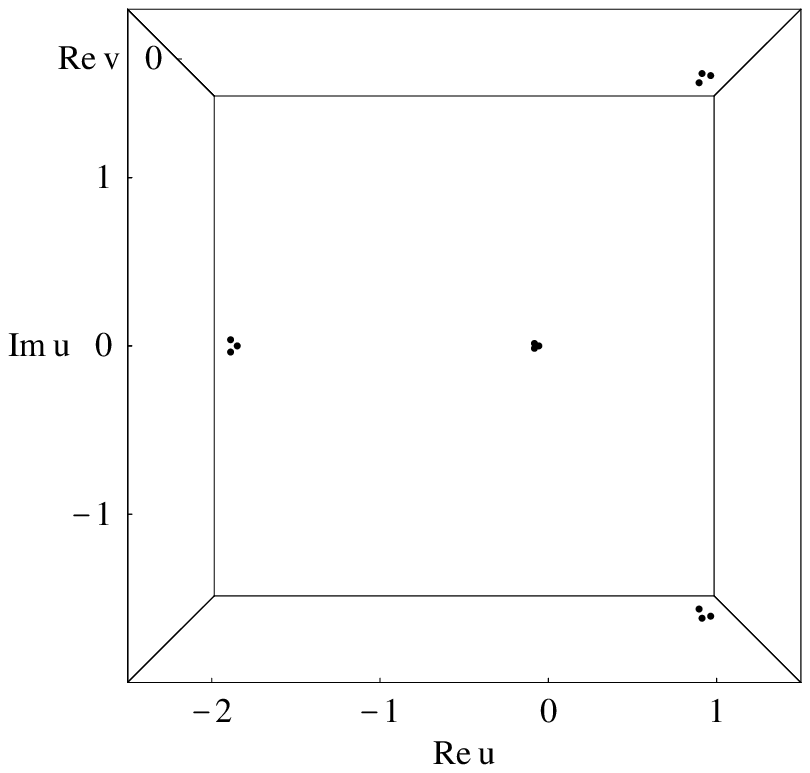,width=6cm}
      \epsfig{file=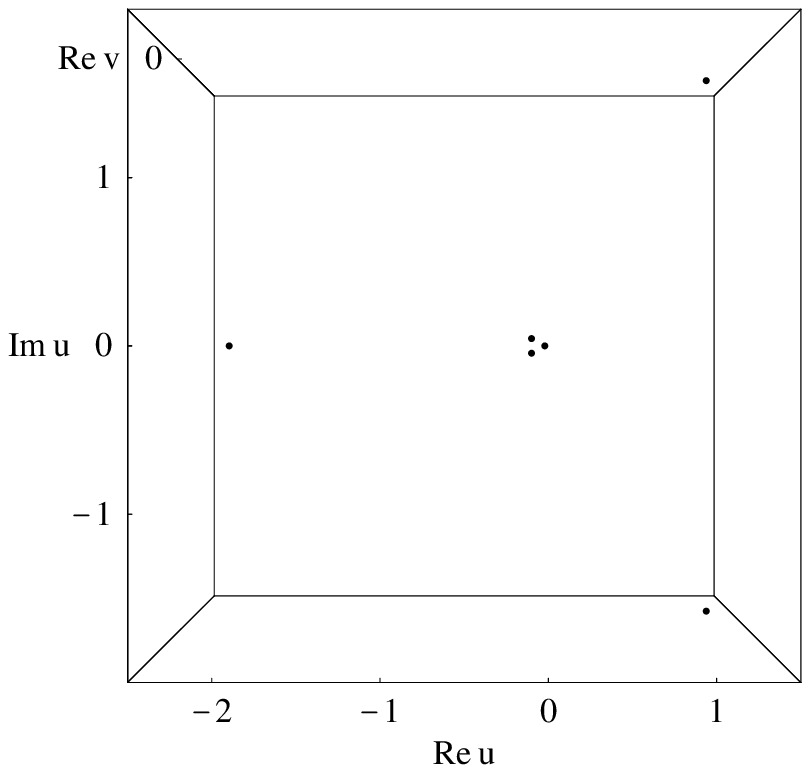,width=6cm}
      }
\end{center}
\caption{Twelve vacua of the $SU(3)$ theory with
$n_{f}=3$      in the projection $(\Re \, u, \Im \, u, \Re \,
v)$.    $\Lambda_{3} = 2, \, m_{1}= 1/64,
\, m_{2}= i/64$,     $m_{3} = - i/64$.
The same projection in the right with
equal masses: $m_i = 1/64$.  }
\label{figsu3nf3}
\end{figure}

\begin{figure}[tb]
\begin{center}
      \makebox[\textwidth][s]{
      \epsfig{file=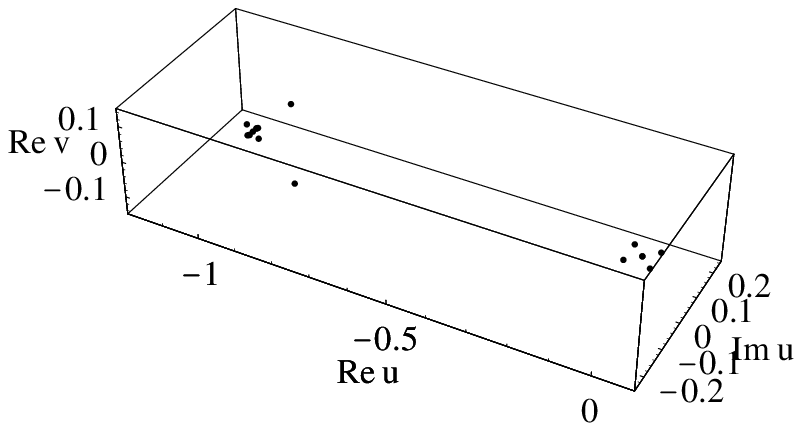,width=6cm}
      \epsfig{file=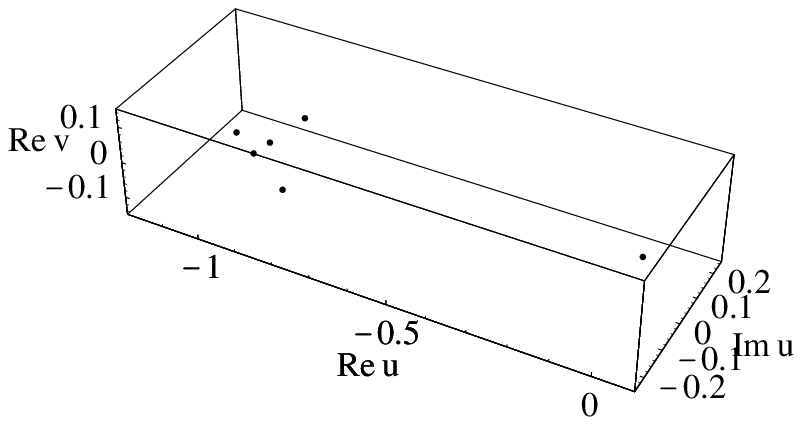,width=6cm} }
\end{center}
\caption{The seventeen vacua of the $SU(3)$ theory with
$n_{f}=4$       in the $(\Re \, u, \Im \, u, \Re \, v)$ projection.
$\Lambda_{4} = 2, \, m_{1}=
1/64, \, m_{2}= - 1/64, \, m_{3}= i/64$,  $m_{4}=-i/64$.  On the
right, the same plot    in the equal masses case with   $m_{i}= 1/64$.  }
\label{figsu3nf4}
\end{figure}

\begin{description}
\item{i)}    In the case with  $n_{f}=1$,   one expects from Eq.~(\ref{nofvac})
${\cal N} = 5$.   From the known  curve,
\beq
y^{2}=  (x^3 - u \, x - v)^2 - {\Lambda_{1}}^5 \, (x + m) \,  ,
\eeq
one finds indeed five vacua, related by an approximate $Z_{5}$
symmetry.  See Fig.~\ref{figsu3nf1} for $\Lambda_1=2$ and $m=1/64$.

\item{ii)}   In the case with $n_{f}=2$ one expects eight vacua, related
by an approximate $Z_{4}$ symmetry which transform $u$ and $v$ as $u
\rightarrow -u$ and $v \rightarrow {\rm exp}( \frac{3}{2} i \pi) v$.
One finds indeed eight singularities,
grouped into two approximate doublets, and four
singlets for generic small   masses.  (see Fig.~\ref{figsu3nf2}).

\item {iii) } For $n_f=3$,   one gets ${\cal N} = 12$, with $Z_{3}$ symmetry.
The numerical analysis
with
the curve:
\beq
y^{2}= (x^3 - u \, x - v)^2 -  {\Lambda_{3}}^{3} \, (x + m_1) (x +
m_2) (x + m_3).
\eeq
shows that there are indeed twelve vacua
satisfying the criterion of the two mutually local dyons becoming
massless, and they are found in roughly three groups of triplets and
three singlets of singularities.
In the equal mass limit each of the
three triplets coalesce to a point in the $(u,v)$ space, showing
that the massless monopoles there are in the representation
${\underline 3}$ of $SU(3)$, while  those at the other  vacua are
singlets (see Fig.~\ref{figsu3nf3}),    in complete  agreement   with
the analysis
of previous sections.

\item{iv)}  For $n_f=4$, Eq.~(\ref{nofvac})  gives ${\cal N} = 17$     vacua.
It is reassuring that one indeed finds from Eq.(\ref{resultant})
seventeen vacua
    for generic and unequal masses
\footnote{As a further check, we
verified the number of the vacua by  using another parametrization of
the curve given by Minahan et al. \cite{Neme}.}.
At small masses these vacua are
grouped into  an approximate  { sextet},    two   quartets   and
three singlets,
suggesting   the assignements   of rank $2$, $1$ and $0$
antisymmetric representation
of $SU(4)$ global flavor group.    The number of the vacua  (six) in
the limit of equal masses
is consistent with this assignment (see Fig.~\ref{figsu3nf4})

In a quartet vacuum, the condensation of the monopoles breaks the
$SU(4)$ symmetry to
$U(3)$,  while   in a singlet vacua  the flavor  symmetry remains   unbroken.

Something very interesting happens in the sextet vacua.   Namely, in
the equal mass
or massless limit,  we find that four branch points in the $x$  plane
coalesce, suggesting
conformal invariant vacuum.   In fact,  this result was to be expected, since
these ``sextet" vacua are examples of  the particular class (``class 3'' )
of  nontrivial   conformal invariant theories studied in \cite{EHIY},
where $2r$ branch points in
the $x$ plane coalesce, where  $ r= n_f/2.$  This is known to  occurs
for $SU(3)$ theories with
$n_{f}=4$ at special values of $u$ and $v$ when the quark masses are
all equal ($m$):       their precise position is \cite{EHIY}   $U=3
M^2, $ and $V=2
M^3$, where $u= (7 + 24 M + 36 M^2) /12$ and $v= (11 + 45 M + 54 M^2 +
54 M^3)/27$ and $M=m-1/3, $ and the curve becomes
\beq   y^2= (x+m)^4
(x+1-2m) (x-1-2m).\eeq
The values of $u, \, v$   found by us numerically   from the
criterion of $N=1$ vacua
   precisely match  these  values,    showing that   this particular
conformal vacuum  survives the
$N=1$ perturbation.

In order to determine  which particles  are actually present, it is
necessary to study
    the monodromy transformation properties
of $(a_{D1}, a_{D2},a_1,a_2)$
\cite{SW1,SW2,SUN}  around this particular singularity.    In
\ref{sec:monodromy}
   we present   such an analysis    (the analysis is
actually  done  for all seventeen vacua  of   $SU(3)$,
$n_f=4$ model).     Our result shows  that   the massless particles
(in the $\mu =0$ limit) present  at this singularity
have quantum numbers
\beq   (n_{m1}, n_{m2}, n_{e1}, n_{e2})= (0,1,0,1),
\,\,(1,2,2,0),\,\, (1,1,0,  1).     \eeq
As these particles are relatively nonlocal,  such a vacuum is
conformal invariant \cite{APSW}.

This is an example of a very general phenomenon, already discussed in
the previous section.

\item{v)}     Finally, for $n_{f}=5$ we verified (for large masses)
the presence of
twenty--three quantum vacua, in accordance with Eq.~(\ref{nofvac}).
  From the curve:
\beq
y^{2}=\prod_{a=1}^{n_c} (x - \phi_a)^2  - 4 \Lambda_{5} \,
\prod_{i=1}^{5} \, \left(x + m_i + {\Lambda_{5} \o n_{c}}\right)  =  (x^3 -
u \,
x - v)^2 - 4 \Lambda_{5} \,
\prod_{i=1}^{5} \, \left(x + m_i + {\Lambda_{5} \o n_{c}}\right) \, ,
\label{auxcurvenf5}
\eeq
in the small mass  limit, the grouping of the singularities found   is
compatible with the assignment  into  a decuplet, two quintets and three
singlets of (approximate) global $SO(5)$ symmetry.  The number of
vacua (six) in the equal mass case, is in
agreement with this structure.

In particular,   the decuplet vacua corresponds to the curve,
\beq     y^2 =  \left(x + m + {\Lambda_{5} \o n_{c}}\right)^4  \, \left[
\left( x -  {  2
\Lambda_{5} \o  n_c}  -  2 m \right)^2   -4 \Lambda_{5}  \left(x + m +
{\Lambda_{5}
\o n_{c}}\right)  \right], \eeq and corresponds to   the class 1 (trivial)
conformal  field
theory \cite{EHIY}.

Furthermore, since $n_{f} > n_{c}+1$ in this case, the theory belongs
to the ``large $n_{f}$''
class of Sec.~\ref{sec:largemulargef}
we expect  ${\cal N}_2=  7$  (see Eq.~(\ref{extrav}))  of  vacua to
show  particular properties.
      We find that
indeed seven of the vacua  (in the equal mass limit,  a quintet and
two singlets)
   approache   the form
\beq
y^2 = x^4 \,  (x - \Lambda)^2  \,  \label{auxcurvebis}
\eeq
in the $m_i \to 0$  limit, after a shift in  $x$.
This is exactly the structure of the singularities at the root of the
baryonic branch\cite{ArPlSei}.

\end{description}

\subsection {$USp(2 n_c)$ with  $2n_f$ Flavors}

\begin{figure}[tb]
\begin{center}
      \epsfig{file=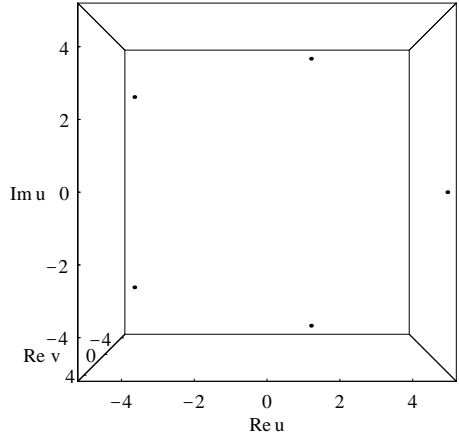,width=6cm}
\end{center}
\caption{Five vacua of the $USp(4)$ with $n_{f}=1$ flavors,
plotted   in  $(\Re u, \Im u, \Re v)$ projection.      $\Lambda_{1} =
2^{1/5}$,   $m = 1/64$,
while $u=\phi_1^2 + \phi_2^2$ and $v=\phi_1^2 \phi_2^2$.  }
\label{figusp4nf1}
\end{figure}

\begin{figure}[tb]
\begin{center}
      \makebox[\textwidth][s]{
      \epsfig{file=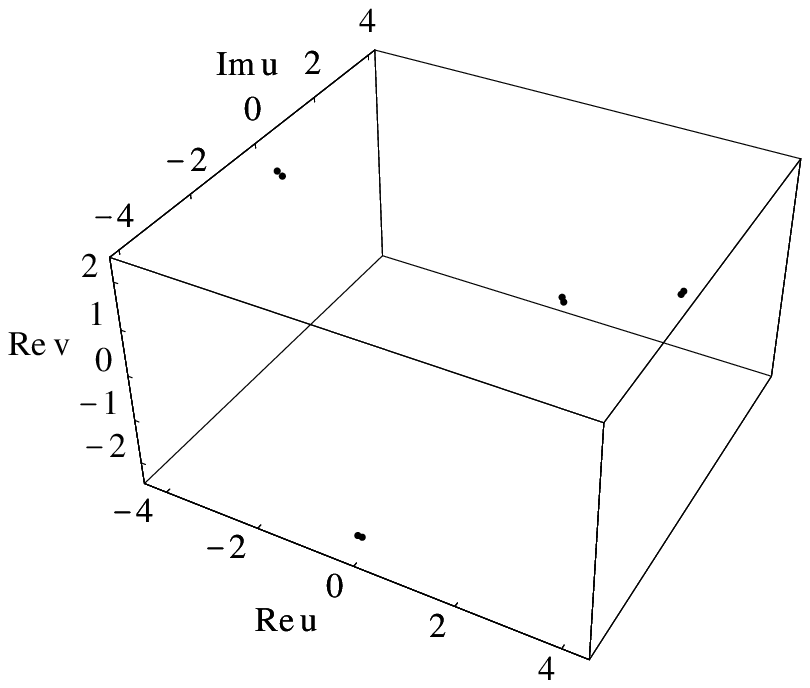,width=6cm}
      \epsfig{file=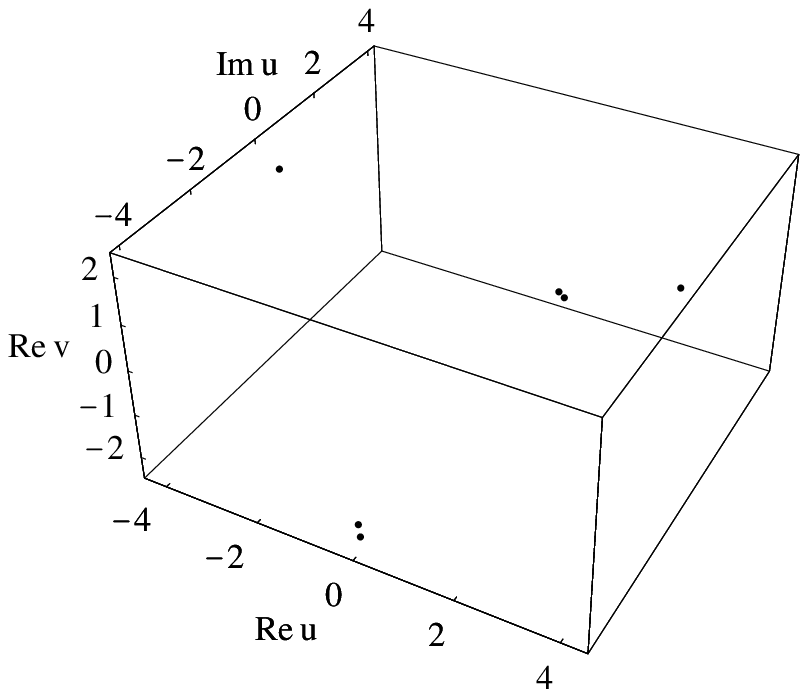,width=6cm}
      }
\end{center}
\caption{In the left figure are  the eight singularities of the $USp(4)$
theory with $n_{f}=2$ are plotted in the in  $(\Re u, \Im u, \Re v)$
projection.
   $\Lambda_{2} = 2^{1/4}, \, m_{1} =
1/64$,  $m_{2}= i/64$.  In the right, the same plot   with equal
masses  ($m= 1/64$).  }
\label{figusp4nf2}
\end{figure}

\begin{figure}[tb]
\begin{center}
      \makebox[\textwidth][s]{
      \epsfig{file=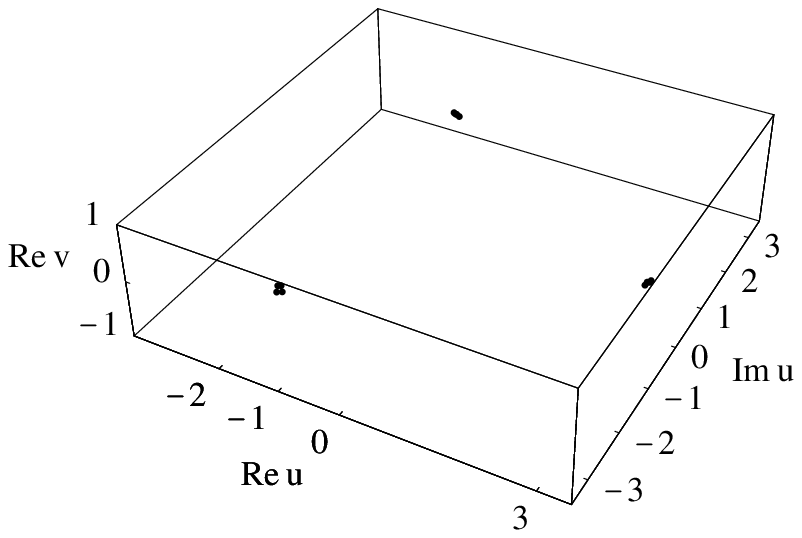,width=6cm}
      \epsfig{file=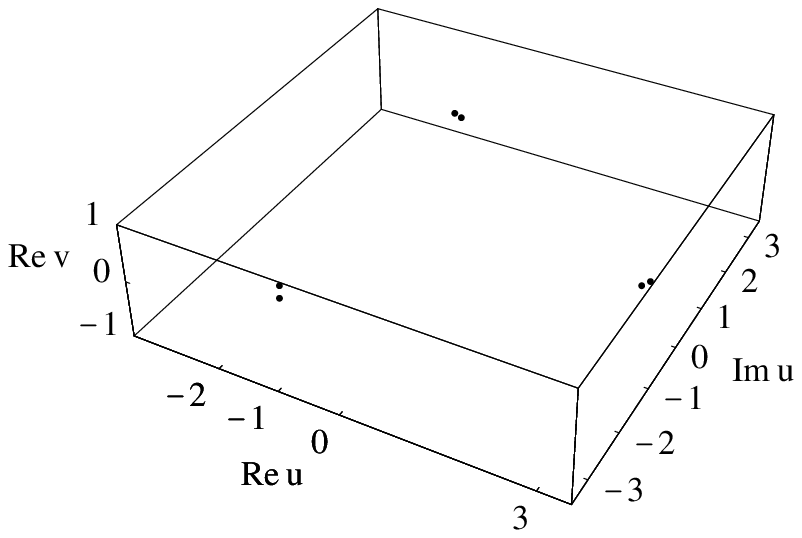,width=6cm}
      }
\end{center}
\caption{The twelve singularities of the $USp(4)$
theory with $n_{f}=3$, plotted in the $(\Re \, u, \Im \, u , \, \Re v)$
projection of the QMS.      $\Lambda_{3} =
2^{1/3}, \, m_{1}= 1/64, \, m_{2}= i/64$,  $m_{3}= i/256$.  On the
right,   the same plot    with equal  masses ($m=1/64$).  }
\label{figusp4nf3}
\end{figure}

\begin{figure}[tb]
\begin{center}
      \makebox[\textwidth][s]{
      \epsfig{file=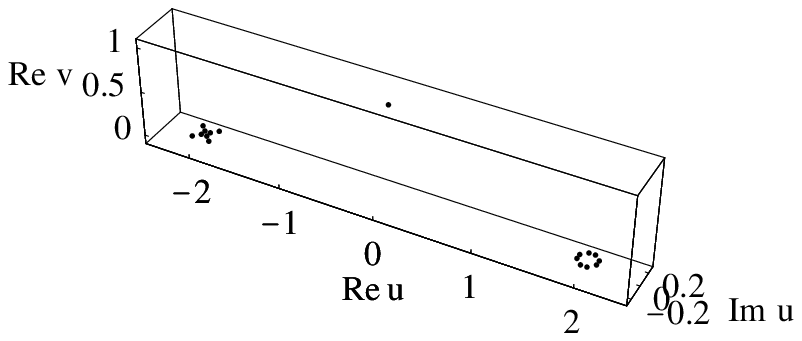,width=6cm}
      \epsfig{file=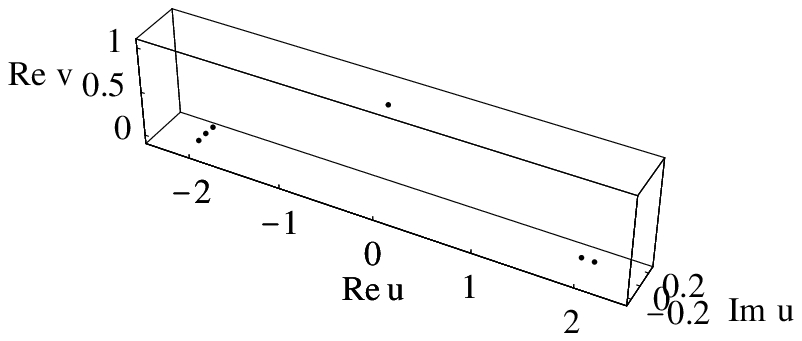,width=6cm}
      }
\end{center}
\caption{The seventeen vacua of the $USp(4)$ with
$n_{f}=4$,     plotted in the $(\Re , u, \Im \, u , \Re \, v)$
projection.     $\Lambda_{4} =
\sqrt{2}, \, m_{1}= 1/64, \, m_{2}= i/64, \, m_{3}= 1/32$,   $m_{4}=
i/32$.  On the right, the same plot   in the case of equal
masses, $m_{i}= 1/64,   \, \, \, \forall i$.  }
\label{figusp4nf4}
\end{figure}

\begin{description}

\item{i)}
For $USp(4)$ with $n_f=1$, we do find five vacua, as shown in
Fig.~\ref{figusp4nf1}, consistently with the approximate $Z_{5}$
symmetry.

\item{ii)}
For $n_f=2$ with the same gauge group, eight quantum vacua are found
to group into four doublets, showing that monopoles appear in the
fundamental representation $(2,1) $ and $(1,2)$, of the flavor group
$SO(2 n_f) \sim SU(2)\times SU(2)$.  Their condensation leaves a $U(2)$
subgroup invariant, in accordance with the na{\"{\i}}ve expectation.
In the equal mass limit one of the $SU(2)$ symmetries is exact.  It is
seen that in this case the $SU(2)$ symmety is broken spontaneously in
four of the vacua, while in four others it is not, in perfect
agreement with what was found in Sec.~\ref{sec:largem} (see
Fig.~\ref{figusp4nf2}).

\item{iii)}
For the $USp(4)$ theory with $n_{f}=3$, the formula   (\ref{Nvspnclassbis})
and the discrete symmetry suggests that monopoles form a quartet ($12$
vacua grouped into three quartets of nearby singularities).  This is
indeed the case as shown in Fig.~\ref{figusp4nf3}.  The condensation
of ${\underline 4}$ of $SO(6) \sim SU(4)$ breaks the chiral symmetry
to $U(3)$, as expected.

\item{iv)}
The theory with four flavors (and the same gauge group) has seventeen
vacua.    In the massless
limit, they group in two spinors (octets) and one singlet of the
gloabal symmetry $SO(8)$.  In the case of degenerate but nonvanishing
masses, the
spinors $\mathbf{8}$ of $SO(8)$ can split in two possible ways:
$\mathbf{1} + \mathbf{6} + \mathbf{1}$ of $U(4)$ in one case, and with
$\mathbf{4} + \mathbf{4}^{*}$ of $U(4)$ in the other.  This is indeed
the situation shown in Fiq.~\ref{figusp4nf4}.   This shows the
correctness of our assignment and
   that in the massless limit the
condensation of the monopole in the spinor representation of $SO(8)$
breaks the chiral symmetry to $U(4)$.

\item{v)}    Finally,    we verified    that the theory $USp(4)$ with
five flavors has
indeed   twenty-three quantum vacua.

\end{description}

\newpage

\section{Effective Lagrangian  Description of   $N=1$
Vacua at Small $\mu$  \label{sec:EfLagdes}   }

A deeper insight   into   physics   at  small $m_i$  and $\mu$       can be
obtained  by re-examining the
works of    Argyres, Plesser and Seiberg \cite{ArPlSei} and  Argyres,
Plesser and Shapere \cite{APS2},
who  showed how the non-renormalization theorem of the
hyperK\"ahler metric on the Higgs branch could  be used to show the
persistence of unbroken non-abelian gauge group at the ``roots'' of
the Higgs branches  where they
intersect the Coulomb branch.  In fact,    they found    two kinds of such
submanifolds, called ``non-baryonic branch" (or mixed-branch) roots,
and ``baryonic branch"
roots  (these terminologies  refer    specifically to the $SU(n_c)$
theory,  but   the situation is similar
also in $USp(2n_c)$  theory).   The latter is
present only for larger values of the flavor  ($n_f > n_c$) while the
former exists always.

Below, we show  how   the   low-energy  effective action description
of  \cite{ArPlSei,APS2}    match      our findings    of Sections
\ref{sec:classvac} -
\ref{sec:flsymbr},   after
correcting  a few errors  and clarifying some   issues   left unclear
there.      In doing so,  a very clear
and interesting    picture  of the infrared dyamics of our theories
emerges,    which was summarized in the Introduction.

   Let us discuss the  $SU(n_c)$ theories first.

\subsection{$SU(n_c)$ \label{shown} }

The non--baryonic roots are further classified into sub--branches
characterized by an unbroken $SU(r)\times U(1)^{n_c-r}$ gauge
symmetry for $r \le [n_f/2]$, with $n_f$ flavor of massless
hypermultiplets in the fundamental representation of $SU(r)$, as well
as $n_c-r-1$ singlet ``monopole" hypermultiplets having charges only in
the $U(1)^{n_c-r}$ gauge sector.  Their quantum numbers are shown in
Table \ref{tabnonb}  taken   from \cite{ArPlSei}.
\begin{table}[b]
\begin{center}
\vskip .3cm
\begin{tabular}{ccccccc}

&   $SU(r)  $     &     $U(1)_0$    &      $ U(1)_1$
&     $\ldots $      &   $U(1)_{n_c-r-1}$    &  $ U(1)_B  $  \\
\hline
$n_f \times  q$     &    ${\underline {\bf r}} $    &     $1$
&     $0$
&      $\ldots$      &     $0$             &    $0$      \\ \hline
$e_1$                 & ${\underline {\bf 1} } $       &    0
&
1      & \ldots             &  $0$                   &  $0$  \\ \hline
$\vdots $  &    $\vdots   $         &   $\vdots   $        &    $\vdots   $
&             $\ddots $     &     $\vdots   $        &     $\vdots   $
\\ \hline
$e_{n_c-r-1} $    &  ${\underline {\bf 1}} $    & 0                     & 0
&      $ \ldots  $            & 1                 &  0 \\ \hline
\end{tabular}
\caption{The effective degrees of freedom and their quantum numbers at the
``nonbaryonic root". }
\label{tabnonb}
\end{center}
\end{table}

Upon turning on the $\mu \Phi^2$ perturbation, the effective superpotential of
the theory is, according to Argyres, Plesser and Seiberg
\cite{ArPlSei},
\beq
W_{non ~bar} = \sqrt2 \Tr (q \phi {\tilde q}) + \sqrt2 \psi_0 \Tr (q
{\tilde q}) + \sqrt 2 \sum_{k=1}^{n_c-r-1} \psi_k e_k {\tilde e}_k +
\mu \left(\Lambda \sum_{i=0}^{n_c-r-1} x_i \psi_i + {1 \o 2} \Tr \phi^2\right),
\label{nonbaryonic}
\eeq
where the last term arises from $\mu \Phi^2$ perturbation, $\phi$
referring to the $SU(r) $ part of the adjoint field and $\psi_i$ being
the $N=2$ partner of the dual $ U(1)_i $ gauge field; $x_i$ are some
constants.  By minimizing the potential, one finds that supersymmetric
$N=1$ vacua exist for any $r$.
(Actually, one assumed here that the maximum number of massless
monopole--like particles $e_k$ exist; such vacua are called in
\cite{ArPlSei} ``special points" of the non baryonic roots.  Only these
vacua survive the $N=1$ perturbation.  Some flat directions remain, as
expected.)

We now perturb these theories at the nonbaryonic  branch roots further with
small hypermultiplet (quark) masses, $m_i$.
Add
the mass terms
\beq
\Delta W_{non ~bar} = m_i q_i \tilde q_i + \sum_{k=1}^{n_c-r-1} S_k^j
m_j e_k {\tilde e}_k ,
\label{masses}\eeq
where $S_k^j$ represents the $j-th$ quark number of the ``monopole"
$e_k$ \cite{SW2,KT}.
   The part involving  $e_k$, ${\tilde e}_k$   and $\psi_k$,
$k=1,2,\ldots n_c-r-1$ is trivial and gives
\beq   \psi_k \sim \{m_i\}; \quad    e_k= {\tilde e}_k \sim \sqrt{\mu
\Lambda}. \eeq
   The vacuum equations for other components are
\beq  0=  [ \phi, \phi^{\dagger}]; \label{effD1}\eeq
\beq \nu \delta_a^b=  q_a^i (q^{\dagger})_i^b - ({\tilde
q}^{\dagger})_a^i
   {\tilde q}_i^b; \label{effD2}\eeq
\beq 0=  q_a^i (q^{\dagger})_i^a - ({\tilde
q}^{\dagger})_a^i
   {\tilde q}_i^a; \label{effD2bis}\eeq
\beq q_a^i {\tilde q}_i^b - { 1\o r} \delta_a^b (q_c^i {\tilde q}_i^c)
+ \sqrt2  \, \mu \phi_a^b =0; \label{effF1} \eeq
\beq   0= \sqrt2 \,\phi_a^b q_b^i + q_a^i  (m_i  + \sqrt2  \, \psi_0)
; \quad ( {\hbox{\rm no
sum
over}} \,\,i,\,\, a) \,
   \label{effF2} \eeq
\beq  0=  \sqrt2 \, {\tilde q}_i^b
\phi_b^a +   (m_i  + \sqrt2  \, \psi_0) \,  {\tilde q}_i^b
\quad ( {\hbox{\rm no sum over}}\,\,i, \,\, a).  \, \label{effF3} \eeq
\beq  \sqrt2 \, \Tr ( q {\tilde q})  + \mu \Lambda =0.  \label{effF4} \eeq
First  diagonalize the Higgs scalar by color rotations.
\beq   \diag \, \phi = ( \phi_1, \phi_2, \ldots  \phi_{r}), \quad
\sum \phi_a=0. \label{effphivev} \eeq
The equations  Eq.(\ref{effD1})-Eq.(\ref{effF3}) are formally
identical to  Eq.(\ref{D1})-Eq.(\ref{F3}),
   with the replacements,
\beq   Q \to q; \quad, m_i \to m_i +  \sqrt2  \,\psi_0; \quad  n_c \to r, \eeq
therefore we immediately  find   solutions     classified by an
integer $\ell$  ($\ell=0,1,\ldots, r-1$)   where
\beq   \phi ={1 \o \sqrt2} \diag \,  ( -m_1\!- \!\sqrt2  \psi_0 ,
-m_2- \sqrt2 \psi_0 , \ldots  -m_{\ell}- \sqrt2  \, \psi_0 ,
c,c,\ldots c),   \quad
\sum \phi_a=0;    \label{phiveveff} \eeq
\beq  q_a^i  = \pmatrix {d_1 \cr 0  \cr \vdots  \cr \vdots \cr  0},
\,\, \pmatrix {0 \cr d_2  \cr 0 \cr \vdots\cr 0},   \pmatrix {0 \cr \vdots
\cr  d_{\ell}   \cr   0 \cr \vdots};  \qquad     q_a^i   = 0, \,\,\,\,
i=\ell +1,\ldots, n_f \,  . \label{vevofqEL}\eeq
\beq  {\tilde q}_i^a  = \pmatrix {\tilde d_1 \cr 0  \cr \vdots  \cr
\vdots \cr  0}, \,\, \pmatrix {0 \cr \tilde d_2  \cr 0 \cr \vdots\cr 0},
\pmatrix {0 \cr
\vdots
\cr  \tilde d_{\ell}  \cr   0 \cr \vdots};  \qquad    {\tilde q}_i^a  = 0,
\,\,\,\, i=\ell +1,\ldots, n_f \,  , \label{vevofqtiEL}\eeq
where
\beq
c = {1 \o r-\ell } \sum_{k=1}^{\ell} (  m_k  + \sqrt{2}  \psi_0 )
\eeq
and  $d_i$, ${\tilde d}_i$'s   are  of order of    $O(\sqrt{\mu
(\psi_0 + m_i)} )$.      Note however that   Eq.(\ref{effF4})   is
new:  it  fixes
$\psi_0$    such   that
\beq   { \sqrt2 r \mu     \o r - \ell} \sum_{j =1}^{\ell}  (\psi_0 +
m_j) = -\mu \Lambda. \eeq
These are the first group of $N=1$ solutions found in \cite{ArPlSei}.
    The fact that in the limit $m_i \to 0$,
\beq  \psi_0 \sim \Lambda,  \label{faraway}  \eeq
however,   shows that    {\it        these  solutions, involving
fluctuations much larger than both $m_i$
    and $\mu$,     lie   beyond the validity of the low-energy
    effective Lagrangian}.    They are   therefore    an  artifact
    of the approximation, and must be   discarded.

The correct  $N=1$  vacua  are found instead      by choosing
$\ell=r$  in  Eq.(\ref{phiveveff})  and
by selecting VEVS,
\beq  \psi_0= -{  1 \o \sqrt{2} \,   r}    \sum_i m_i , \label{corrvacua}  \eeq
\beq    d_i  {\tilde d}_i  = \mu \left( m_i - \frac{1}{r}\sum_j^r
m_j\right) - \frac{\mu
\Lambda}{\sqrt{2} r};\eeq
\beq  e_k {\tilde e}_k  =  -\mu \Lambda,  \eeq
with no $c$'s in Eq.(\ref{phiveveff}) \footnote{ Note that an
analogous solution  was   not possible for
Eq.(\ref{D1})-Eq.(\ref{F3}),
since the quark masses are generic and   the adjoint  field must be traceless.
}.
These  satisfy clearly all of the vacuum equations.
These vacua are smoothly related to the unperturbed ones and
therefore are reliable.
In the massless limit this gives
\beq  d_i \sim  {\tilde d_i}  \sim   \sqrt{ \mu \Lambda}.
\label{corrvacua2} \eeq
We find   thus   ${}_{n_f}\! C_r \cdot (2 n_c- n_f)  $    vacua with
the correct   symmetry breaking pattern,
\beq   U(n_f) \to  U(r) \times U(n_f-r), \eeq
which is exactly what is expected from the   analysis  made at large $\mu$.
The multiplicity ${}_{n_f} C_r $ arises from the choice of $r$ (out of
$n_f$) quark masses used to construct the solution.

For $r=n_{f}/2$, the theory at the singularity
becomes a non-trivial superconformal theory.
There is no description of this singularity
in terms of weakly coupled local field theory.  The monodromy around
the singularity shows that
the theory is indeed   superconformal (we checked this explicitly for $n_c=3$
and $n_f=4$).   Careful
perturbation of the curve by the quark masses made in Section
\ref{CKMmp} shows    that there are
$(n_{c}-n_{f}/2) \, {}_{n_{f}}\!C_{n_{f}/2}$ vacua.

The total number of $N=1$ vacua  is then given by  ${\cal N}_1$  of
Eq.(\ref{vaccountnb}), Eq.(\ref{vaccount1}),
after summing over   $r$.   Actually,  one must subtract the   term
$r={\tilde n}_c$,  since the nonbaryonic root
with $r={\tilde n}_c$    does not exist (see \ref{sec:absence}).

The baryonic root is an $SU({\tilde n}_c) \times U(1)^{n_c - {\tilde
n}_c}$ theory (${\tilde n}_c \equiv n_f- n_c$), with $n_f$ massless
quarks $q$ and $n_c- {\tilde n}_c = 2n_c-n_f$ massless singlets.
Their charges are summarized in Tab.~\ref{tabbar}.
\begin{table}[h]
\vskip .3cm
\begin{center}
\begin{tabular}{cccccc}
&   $SU({\tilde n}_c)  $     &     $U(1)_1$    &       $\ldots $
&   $U(1)_{n_c-{\tilde n}_c}$    &  $ U(1)_B $  \\
\hline
$n_f \times  q$     &    ${\underline {\tilde n}_c}$                &
$1/{\tilde n}_c$
&      $\ldots$      &     $1/{\tilde
n}_c$             &
$-n_c/{\tilde n}_c$      \\ \hline
$e_1$                 & ${\underline {\bf 1} } $
&      $ -1$
& \ldots             &  $0$                   &  $0$  \\ \hline
$\vdots $  &    $\vdots   $             &    $\vdots   $         &
$\ddots $     &     $\vdots   $        &     $\vdots   $    \\ \hline
$e_{n_c-{\tilde n}_c} $    &  ${\underline {\bf 1}}   $                     & 0
&      $ \ldots  $            & $-1$                 &  0 \\ \hline
   \end{tabular}
\caption{The effective degrees of freedom and their quantum numbers at the
``baryonic root", taken from \cite{ArPlSei}. }
\label{tabbar}
\end{center}
\end{table}
The effective action in this case, upon the $N=1$ perturbation, is
\beq
W_{bar} = \sqrt2 \Tr (q \phi {\tilde q}) + {\sqrt2 \o {\tilde n}_c}
\Tr (q {\tilde q}) \left( \sum_{k=1}^{n_c - {\tilde n}_c} \psi_k
\right) - \sqrt 2 \sum_{k=1}^{n_c-{\tilde n}_c} \psi_k e_k {\tilde
e}_k + \mu \left(\Lambda \sum_{i=1}^{n_c-{\tilde n}_c} x_i \psi_i + {1 \o
2} \Tr
\phi^2\right) \, .
\label{baryonic}
\eeq
Again, we add the quark mass terms
\beq
\Tr ( m q) - \sum_{k,i} S_k^i m_i e_k {\tilde e}_k
\label{baryomas}
\eeq
and minimize the potential.  The equations are:
\begin{equation}
q_a^i q_{i}^{\dagger b} - {\tilde q}_a^{\dagger i} {\tilde q}_i^b =
\nu \delta_a^b \, ;
\label{DD1}
\end{equation}
\beq
q_a^i  {\tilde q}_i^b   -  {1 \o {\tilde n}_c} \delta_a^b  (q_c^i  {\tilde
q}_i^c  ) + \mu \phi_a^b=0; \label{FF1}
\eeq
\beq
q_a^i m_i + \sqrt2 \phi_a^b q_b^i   + { \sqrt2 \o   {\tilde n}_c } q_a^i
\sum_k \psi_k =0; \label{FF2}
\eeq
\beq
{\tilde  q}_i^a m_i + \sqrt2 \phi_b^a {\tilde  q}_i^b   + { \sqrt2 \o
{\tilde n}_c } {\tilde  q}_i^a  \sum_k \psi_k =0; \label{FF3}
\eeq
\beq
{ \sqrt2 \o {\tilde n}_c } \Tr ( {\tilde q} q) - \sqrt2 e_k {\tilde
e}_k + \mu \Lambda x_k=0; \label{FF4}
\eeq
\beq
(\psi_k - S_k^i m_i) e_k =0; \quad (\psi_k - S_k^i m_i) {\tilde e}_k
=0 \, ,
\label{FF5}
\eeq
and
\beq
\phi = \frac{1}{\sqrt{2}}\diag \, (-m_1, \ldots, -m_r, c, \ldots , c)
\, ; \qquad c = {1 \o {\tilde n}_c-r} \sum_{k=1}^r m_k \, .
\label{diagphi2}
\eeq
We find two types of vacua.
The first type has $e_{k}=\tilde{e}_{k} = (\mu \Lambda
x_{k}/\sqrt{2})^{1/2}$ for all $k=1, \cdots n_c-\tilde{n}_c$.
Minimizing the potential in this case, we find
\begin{equation}
    {\cal N}_2   = \sum_{r=0}^{\tilde n_c-1} ( \tilde n_c-r) \,\,
    {}_{n_f}\!C_r
    \label{nmvcbar}
\end{equation}
$N=1$ vacua, characterized by the VEVS  Eq.(\ref{diagphi2})  and
\begin{equation}
    d_i ,   {\tilde d_i}   \sim  \sqrt {\mu    m}
    \stackrel{ m_i \to 0}{\longrightarrow}   0,
    \qquad \qquad e_k,   {\tilde e}_k  \sim  \sqrt{\mu\Lambda}.
\end{equation}
The unbroken $SU(\tilde{n}_{c}-r)$ gauge group gives
$\tilde{n}_{c}-r$ vacua each.
These vacua describe the vacua with unbroken $U(n_f)$ symmetry,
which  are known to exist from the
large $\mu$ analysis.   The total number of vacua of   this group found
here agrees with     Eq.(\ref{eq:N22})
at the large $\mu $  regime.

The second type of vacua in Eqs.~(\ref{baryonic},\ref{baryomas}) has
one of the $e_{k}=\tilde{e}_{k}=0$ (hence $n_c-{\tilde
n}_c=2n_{c}-n_{f}$ choices) while $\partial W/\partial\psi_{k}=0$
requires quarks to condense with $q = \tilde{q} \sim \sqrt{\mu
\Lambda}$.  Dropping $e_{k}=\tilde{e}_{k}=0$ from the Lagrangian, it
becomes the same as that of the non-baryonic root
Eqs.~(\ref{nonbaryonic},\ref{masses})     and gives
$(2n_{c}-n_{f}) \, {}_{n_{f}}\!C_{\tilde{n}_{c}}$ vacua.  This precisely
compensates the exclusion of $r=\tilde{n}_{c}$ in the sum for the
non-baryonic roots and the correct total number of vacua ${\cal
N}_{1}+{\cal N}_{2}$ is obtained.

\bigskip

We thus find that both the number and the symmetry properties of the
$N=1$ theories at small adjoint mass $\mu$   match   exactly    those found
at large $\mu.$
For vacua with $r=1$, the ``quarks'' in the effective
Lagrangian (\ref{nonbaryonic})
are nothing but the $U(1)^{n_c-1}$      monopoles in the fundamental
representation of $U(n_f)$; this is checked by studying the monodromy
around the singularity which showed that the ``quarks'' are indeed
magnetically charged.  Therefore the standard   picture of confinement and
flavor symmetry breaking by condensation of flavor-non-singlet
monopoles is valid   for these vacua.

   For general vacua $r>1$  associated with various    nonbaryonic
roots,  the effective Lagrangian (\ref{nonbaryonic})
describes   correctly the physics of
$N=1$  vacua   at small $\mu$,   in terms of   {\it  magnetic quarks}
of a non Abelian  $SU(r)\times  U(1)^{n_c-r}$ theory.  In
contrast to  the $r=1$ case,  these quarks   cannot be identified with  the
semiclassical monopoles of  the maximally Abelian $U(1)^{n_c-1}$ group.
Note  that  the condensation of  such  monopoles in   the
rank-$r$ anti-symmetric tensor representation,  which might  be
suggested from the number of the singularities
which group into a nearby cluster in the small $m_i$ limit and at the
same time   from the semiclassical analysis
   (see \ref{sec:semicmon}),
would have yielded   the correct pattern of symmetry breaking;  at the
same time,  however,     it   would have     led to an uncomfortably
large number of Nambu-Goldstone bosons associated to the  accidental
$SU({}_{n_f}\! C_r)$ symmetry.            The system avoids  this
paradox elegantly,     by having   magnetic quarks as  low-energy
degrees of freedom   and   having these   condensed.
These facts, and the comparison of their quantum numbers,
lead us      to conclude that, as we approach the
non-baryonic roots from semi-classical (large VEV) region on the
Coulomb branch,      the semi-classical monopoles in the rank-$r$
anti-symmetric tensor representation are smoothly matched to
``baryons'' of the   $SU(r)$ theory,
\beq       \epsilon^{a_1   \ldots  a_r} {q_{a_1}^{i_1}  q_{a_2}^{i_2}
\ldots q_{a_r}^{i_r} },        \eeq
      and     break up (the system    being infrared-free)   into weakly
coupled magnetic quarks, before   becoming massless.

   The case $r=n_f/2$ is exceptional and    highly
non-trivial.  Although the analysis leading to
Eq.(\ref{corrvacua}),  Eq.(\ref{corrvacua2}), is formally valid in
this case
also,   physics is  really  different.  The low energy degrees of
freedom  involve  relatively  nonlocal states,
arising from the nontrivial, class 3   CFT \cite{EHIY}.  In this case,
the theory is in the same universality class as the finite $SU(n_f/2)$
theories.
For an  explicit check with  monodromy,    see
\ref{sec:monodromy},     for the simplest example of this type,
$r=2$ vacua  of the $SU(3)$ theory with $n_f=4$.

As for the second group of vacua with vanishing VEVS   found at the
``baryonic root",    they are  in
the  free-magnetic phase, with
observable   (magnetic)   quarks,    weakly interacting    with
$SU(n_f - {n}_c)  $  gauge fields.

\subsection{ ${   {USp(2 n_c)} }$  }

    In the ${   {USp(2 n_c)} }$
   gauge theories, the first type of vacua can be
identified more easily by first considering the equal but novanishing
quark masses.  The adjoint VEVS in the curve Eq.(\ref{eq:curve}) can
be chosen so as to factor out the behavior
\begin{equation}
    y^2 = (x+m^2)^{2r}\, [\ldots],    \quad       r=1,2,\ldots
\end{equation}
which describes an $SU(r)\times U(1)^{n_c-r+1}$ gauge theory with $n_f$ quarks.
See  Section~\ref{sec:masspert2}.  These
(trivial) superconformal theories belong in fact to the same universality
classes
as those   found   in the $SU(n_c)$ gauge theory,    as pointed out
by Eguchi and others
\cite{EHIY}.  They are therefore described by exactly the same
Lagrangian Eq.(\ref{nonbaryonic}).  At each vacuum with $r$, the
symmetry (of equal mass theory, $U(n_f)$) is broken spontaneously as
\begin{equation}
    U(n_f) \to U(r) \times U(n_f-r).
    \label{funny}
\end{equation}
    When a small mass splitting is added
among $m_i$'s, each of the $r$ vacuum splits into ${}_{n_f}\!C_r$
vacua, leading to the total of
\begin{equation}
    (2 n_c +2 - n_f)  \sum_{r=0}^{(n_f-1)/2}  {}_{n_f} C_r
    =   (2 n_c +2 - n_f)  \, 2^{n_f-1}, \quad  (n_f = {\hbox {\rm odd}})
\end{equation}
\begin{equation}
    (2 n_c +2 - n_f) \sum_{r=0}^{{n_f}/2-1}  {}_{n_f} C_r
    \,   +  {2 n_c +2 - n_f \o 2}\,  {}_{n_f} \!C_{n_f/2}
    =   (2 n_c +2 - n_f) \, 2^{n_f-1},
    \quad  (n_f = {\hbox {\rm even}}),
\end{equation}
vacua of this type,      consistently    with Eq.~(\ref{Nvspnclassbis}).
These $N=1$ vacua seem to have been
    overlooked in \cite{APS2} altogether\footnote{With a hindsight,  we
see that  this was inevitable:
as  we show below,   there is no local effective Lagrangian
description  of low-energy  physics  in the   $m_i  = 0$ limit
   and
Argyres, Plesser and Shapere
\cite{APS2}  worked  precisely in such    a  regime.    }.

In the massless limit the underlying theories possess a larger, flavor
$SO(2n_f)$ symmetry.  We know also from the large $\mu$ analysis that
in the first group    of vacua (with finite vevs), this symmetry is
broken spontaneously to $U(n_f)$ symmetry always.  How can such a
result be consistent with Eq.(\ref{funny}) of equal (but nonvanishing)
mass theory?

What happens is that in the massless limit various $N=1$ vacua with
different symmetry properties Eq.(\ref{funny}) (plus eventually other
singularities) coalesce.  The location of this singularity
   can be obtained exactly in terms of Chebyshev polynomials,  and  is
given by   Eqs. (\ref{cheby0})-(\ref{cheby3}).
         At the singularity
there are   mutually non-local dyons and hence the theory is at a
non-trivial infrared
fixed point.     In the example of $USp(4)$ theory with $n_f=4$, we have
explicitly verified this by determining the singularities and branch
points at finite equal mass $m$ and by studying the limit $m \to 0$.
There is no description
in terms of a weakly coupled local field theory, just as in     the case
$r=n_f/2$ for $SU(n_c)$ theories.
Since the global flavor symmetry is $SO(2n_f)$,      {\it these superconformal
theories belong to a different universality class as compared to those
at finite mass or    to those of $SU(n_c)$  theories}.  When $n_f$ is even,
the theory is in the same universality class as the finite $USp(n_f-2)$
theories.  When $n_f$ is odd, the theory is in a different universality
class of ``strong-coupled conformal theories."

We find this behavior resonable because the
semi-classical monopoles are in the spinor representation of the
$SO(2n_f)$ flavor group and, in contrast to the situation in
$SU(n_c)$  theories,
cannot ``break up'' into quarks in the
vector representation.     They  are  therefore  likely to  persist at the
singularity and in general make the theory superconformal.
We indeed checked that there are  mutually non-local degrees of
freedom using monodromy of the curve for $USp(4)$ theory with
$n_f=5$.\footnote{The case $n_f=3$, however, is special.
In this case, the analysis of the curve in Section 9.2 tells us that
the theory at the Chebyshev vacuum is the same as the $SU(2)$
theory with $n_f=3$.  Seiberg and Witten studied this case \cite{SW2}
and found that the singularity can be described in terms of massless
monopoles in the spinor representation of the flavor $SO(6)$ group
interacting locally with the magnetic photon.  Note that there is
no problem with unwanted Nambu--Goldstone multiplets in this case
(see Section 8 in \cite{SW2}).}   We believe that
the situation with higher $n_f$ is also non-local; this is because a local
effective
low-energy Lagrangian of massless monopoles in the spinor representation
of the flavor group would have an accidental $U(2^{n_f-1})$ symmetry
and would lead to an unacceptably large number of Nambu--Goldstone
multiplets.  Once the quark
masses are turned on, however, the flavor group reduces to (at least)
$U(n_f)$ and it becomes possible for monopoles to break up into
quarks; this explains the behavior in the equal mass case.

As for the second group of vacua, the situation is more analogous to
the case of $SU(n_c)$ theories.  The superpotential reads in this case
(by adding mass terms to Eq.(5.10) of \cite{APS2}):
\begin{eqnarray}
    W &=& \mu \,
    \left(\Tr \, \phi^2 + \Lambda \sum_{a = 1}^{2 n_c+2-n_f} x_a
    \psi_a\right)
    + { 1  \o \sqrt2} \, q_a^i \phi_b^a q_c^i \,
    J^{bc} + {m_{ij} \o 2} q_a^i q_b^j \, J^{ab}    \nonumber \\
    &+& \sum_{a = 1}^{2 n_c+2-n_f} \left(\psi_a \, e_a \, {\tilde e}_a +
    S_a^i m_i \, e_a \, {\tilde e}_a \right)\, ,
\end{eqnarray}
where $J = i\si_2 \otimes {{\bf 1}}_{n_c}$   and
\begin{equation}
    m = - i \si_2 \otimes \diag \, (m_1, m_2, \ldots, m_{n_f}) \, .
\end{equation}
By minimizing the potential, we find
\begin{equation}
    {\cal N}_2 =   \sum_{r=0}^{{\tilde n}_c}  ({\tilde n}_c- r +1)
    {}_{n_f} C_{r}\,
\end{equation}
vacua, which precisely matches the number of the vacua of the second
group, with squark VEVS behaving as
\begin{equation}
    q_i, {\tilde q}_i \sim \sqrt {\mu m_i}
    \stackrel{ m_i \to 0}{\longrightarrow}   0.
\end{equation}
These are the desired   $SO(2n_f)$  symmetric vacua.

\newpage

\section{Quark-Mass Perturbation    on  the  Curve \label{CKMmp}}

In this section    we develop   a   perturbation  theory in the quark
masses  for the   singularities
of the Seiberg-Witten curves  of $SU(n_c)$ and $USp(2n_c)$  theories,
around certain conformal points.    The results of this
section    allows us,   on the one hand,   to establish the
connection  between the classes of CFT  singularities of $N=2$
space of vacua  and  the   $N=1$ vacua surviving the perturbation
$\mu \Phi^2$    (as
discussed in Section
\ref{sec:leeflag}),  and on the other,    to identify the   $N=1$
vacua  at small $\mu$  (whose  physics
was discussed  in  the previous  section)  with  those found  at large $\mu$.

   \subsection{Perturbation  around  CFT  Points:    $SU(n_c)$
\label{sec:pert1}}

\begin{description}
\item{{\bf i)} }  Generic $r$     ($r < {n_f \o 2})$ and
formulation of the problem.

Suppose         the conformal point (Eq.(\ref{unpertphi})),   $
{\hbox {\rm diag}} \, \phi =(\,  {\underbrace  {0,0, \ldots,
0}_{r}},
    \phi^{(0)}_{r+1}, \ldots,  \phi^{(0)}_{n_c} ), \,\,
\sum_{a=r}^{n_c}   \phi^{(0)}_a=0,  $
where  the full curve   with $m_i=0$
\begin{equation}
    y^{2} = \prod_{k=1}^{n_{c}}(x-\phi_{k})^{2} + 4 \Lambda^{2n_{c}-n_{f}}
    \prod_{j=1}^{n_{f}}(x+m_{j}) |_{m_i=0}
\label{curve111} \end{equation}
takes  the form
\begin{equation}
    y^2 =   x^{2r} (x-\beta_{0 1})^{2} \cdots
        (x-\beta_{0, n_{c}-r-1})^{2} (x-\gamma_0) (x-\kappa_0),
        \qquad        r=0,1,2,\ldots, [n_f/2]
\label{rbranch111}   \end{equation}
is  given.   For nonzero  and small      bare quark masses $m_i$  the
multiple zero at the origin will split,
and other  zeros will be shifted.     We require    that  the
perturbed  singularity,
\beq
{\hbox {\rm diag}} \, \phi =(\,  \lambda_1,\lambda_2, \ldots, \lambda_r,
    \phi^{(0)}_{r+1} +\delta \phi_{r+1} , \ldots,  \phi^{(0)}_{n_c}
+\delta \phi_{n_c} ), \,\,
\eeq
\beq \sum_{a=1}^r  \lambda_a + \sum_{a=r+1}^{n_c} \delta \phi_a =0,
\eeq
be such that    the
curve
\beq  y^2 = F(x)  =\prod_{a=1}^r (x-\lambda_a)^2 \prod_{a=r+1}^{n_c}
(x-\phi_{0  a}  -
\delta \phi_a)^2  - \prod^{n_f}_{i=1}  (x+ m_i),
\label{difficult1} \eeq
   is maximally singular, i.e.,
\beq  y^2 =  \prod_{a=1}^r   (x-\alpha_a)^2 \{\prod_{b=1}^{n_c-r-1}
(x-\beta_b)^2 \}   (x-\gamma) (x-\kappa); \label{difficult2}\eeq
\beq  \
beta_i = \beta_{0 i}+\delta \beta_i ; \quad \gamma = \gamma_0 +\delta
\gamma,  \quad \kappa= \kappa_0 +\delta
\kappa.
\eeq
The problem is to determine how many such sets $\{\lambda_i,
\alpha_i,  \delta \beta_i,  \delta \phi_i, \, \delta\gamma,
\delta\kappa
\}$ exist.
The condition that the curve is maximally singular (maximal number of
double branch points)  can be expressed as
($ F^{'} \equiv  d F(x) / dx $):
\bea   F(\alpha_a) &= &  F^{'}(\alpha_a)=  0, \quad a=1,2,\ldots r;  \quad
\quad  F(\gamma) =  F(\kappa) =0  \non\\  F(\beta_{i}) &=&
F^{'}(\beta_{i}) = 0, \quad i=1,2,\ldots  n_c-r-1,\,
\label{condgeneral}  \eea
There are  $2 n_c$ relations for $2 n_c$ unknowns.

Let us consider the case   $r=1$ first.  In this case there is only
one set of  $\{\alpha, \lambda\}$.    Consider first the
   $2  n_c-2$   relations
\beq   F(\beta_{i}) =  F^{'}(\beta_{i}) = 0, \quad i=1,2,\ldots  n_c-2,\,
\quad  F(\gamma) =  F(\kappa) =0, \label{condreqone}
   \eeq
and expand each equation in the small quantities $\eta_{A}\equiv
\{\lambda,  \alpha,  \delta \beta_i,  \delta \phi_i, \, \delta\gamma,
\delta\kappa
\}$.   By assumption   these quantities are zero if $m_i=0$.  To
zeroth order,    (\ref{condreqone})   is satisfied by assumption.
    To first order we find a  {\it linear}  system of   $2  n_c-2$   equations,
\bea  \sum_A{d F(\beta_{i}) \o d\eta_A}  \eta_A  &= &   \sum_A{d
F^{'}(\beta_{i})\o d\eta_A}  \eta_A = 0, \quad i=1,2,\ldots  n_c-2,\,
\non \\ \sum_A{d F(\gamma) \o d\eta_A}  \eta_A &=&  \sum_A{d
F(\kappa) \o d\eta_A}  \eta_A=0,  \label{condreqonebis}
   \eea
which   determine   $\delta \beta_i,  \delta \phi_i, \, \delta\gamma,
\delta\kappa$    uniquely,   in terms of $m_i$'s and of   $\lambda,  \alpha$.

\smallskip

The
   two   equations
\beq    F(\alpha)=   F^{'}(\alpha)  =  0  \label{small}   \eeq
(which  have no zeroth order  counterpart) cannot be linearized.
In fact,  they   give
\beq     C (\alpha -\lambda)^2  - \prod_{i=1}^{n_f} (\alpha + m_i)=0,
\quad  C=O(1);   \label{funny1} \eeq
\beq      2 \, C \,  (\alpha -\lambda) +  C^{'}  (\alpha -\lambda)^2
-\sum_{i=1}^{n_f} \prod_{j\ne i}^{n_f}  (\alpha + m_i)=0,  \quad
C^{'}=O(1).
\label{funny2}
   \eeq
      Eq.(\ref{funny1})  and   Eq.(\ref{funny2})   must be solved for
$\alpha, \, \lambda
$.

       We see immediately that  $\alpha $  must be very special.    Set
    \beq m_i  =O(\epsilon)   \ll   1,   \eeq
   and  suppose     $\alpha \sim
O(\epsilon^{'}).  $     In either cases,  $\epsilon \gg
\epsilon^{'}$   or  $\epsilon \ll  \epsilon^{'}$,   we get
from   Eq.(\ref{funny1})  and   Eq.(\ref{funny2})
\beq    |(\alpha -\lambda)^2|   \sim   \epsilon^{n_f}; \quad
|\alpha -\lambda|  \sim   \epsilon^{n_f-1};\eeq
or an analogous relation with $\epsilon \to   \epsilon^{'}$.   These
cannot be satisfied (in other words,  there are no solutions
of this  type).
   The only way out (to get solutions)    is to assume that   $\epsilon
\sim  \epsilon^{'}$  and take
\beq  \alpha = - m_i + \Delta,\quad      \Delta \le  O(\epsilon^2),
\quad i=1,2,\ldots, n_f.\eeq
There are obviously $n_f$  such possibilities.
We find now from  Eq.(\ref{funny1})  and   Eq.(\ref{funny2}),
\beq    |(\alpha -\lambda)^2|   \sim   \Delta \cdot
\epsilon^{n_f-1}; \quad   |\alpha -\lambda|  \sim
\epsilon^{n_f-1}.\eeq
    Note that in   Eq.(\ref{funny2})  the terms containing $\Delta$ are
indeed   smaller than the term kept.   Now these  can be solved
and give
\beq   \Delta \sim  \epsilon^{n_f-1}; \quad      |\alpha -\lambda|
\sim   \epsilon^{n_f-1}, \eeq
in other words,
\beq     \alpha = -m_i +  O(\epsilon^{n_f-1});   \quad  \lambda=-m_i
+  O(\epsilon^{n_f-1}). \eeq
We found obviously   $n_f$ solutions   with $r=1$,  according to
which $m_i$ is used to make the solution.

A straightforward generalization  to generic $r$, $r< {n_f \o 2}$
leads to the result   that
$\lambda_a \simeq \alpha_a,$    $a=1,2,\ldots, r$  must be chosen to
be equal to  $r$ out of $n_f$  masses, $m_i$.
There are thus   ${}_{n_f}\!C_r$   solutions of this type, according
to which masses are chosen to construct the solution.
In order to see  the order of magnitude of  $\lambda_a - \alpha_a,$
let us write  the low-energy curve   (at $x \ll \Lambda$)    as
\begin{equation}
        y^{2}  \equiv   F(x) =    \prod_{a=1}^{r} (x-\lambda_{a})^{2} +
        \frac{4}{\Lambda^{n_{f}-2r}} \prod_{i=1}^{n_{f}} (x+m_{i}),
        \label{eq:IRfreecurve}
\end{equation}
   and require   that this curve behaves as
\begin{equation}
        F(x) = \prod_{a=1}^{r} (x-\alpha_{a})^{2},
\end{equation}
if we neglect zeros at $x \sim O(\Lambda)$.  Namely, we require
\begin{eqnarray}
        F(\alpha_{1}) & = & \prod_{a=1}^{r} (\alpha_{1}-\lambda_{a})^{2} +
        \frac{4}{\Lambda^{n_{f}-2r}} \prod_{i=1}^{n_{f}} (\alpha_{1}+m_{i})
        = 0,
        \label{eq:F}
         \\
        F'(\alpha_{1}) & = & 2 \sum_{b=1}^{r} (\alpha_{1}-\lambda_{b})
                \prod_{a\neq b}(\alpha_{1}-\lambda_{a})^{2}
                + \frac{4}{\Lambda^{n_{f}-2r}} \sum_{j=1}^{n_{f}}
                \prod_{i\neq j} (\alpha_{1}+m_{i}) = 0, \nonumber \\
        \label{eq:F'}
\end{eqnarray}
and similarly for $\alpha_{2}, \cdots, \alpha_{r}$.  To satisfy the
first equation (\ref{eq:F}) down to $O(m^{n_{f}+1})$, all we need is
to set $\alpha_{1} = \lambda_{1}$, etc.  The second equation
(\ref{eq:F'}) is then approximated by keeping only the first power in
$(\alpha_{1}-\lambda_{1})$ (in other words, only keeping $b=1$ in the
sum),
\begin{equation}
                F'(\alpha_{1}) = 2 (\alpha_{1}-\lambda_{1})
                \prod_{a\neq 1}(\alpha_{1}-\lambda_{a})^{2}
                + \frac{4}{\Lambda^{n_{f}-2r}} \sum_{j=1}^{n_{f}}
                \prod_{i\neq j} (\alpha_{1}+m_{i}) = 0.
                \label{eq:F'2}
\end{equation}
Barring an  accidental cancellations
between $\alpha_{1}$ and other $\lambda_{a}$ ($a \neq 1$), we find
\beq  \alpha_{1}-\lambda_{1} \sim
O(m^{n_{f}-1}/\Lambda^{n_{f}-2n_{c}}/m^{2(r-1)}) =
O(m^{n_{f}-2r+1}/\Lambda^{n_{f}-2r}).  \label{nvforrnf2}\eeq
    Using this fact, the first
term in (\ref{eq:F}) is $O(m^{n_{f}-2r+1}/\Lambda^{n_{f}-2r})^{2}
\times O(m^{2(r-1)}),    $
   which is much smaller than the second term of
$O(m^{n_{f}})$.  Therefore, the second term must vanish by itself,
which requires $\alpha_{1} = -m_{1}$ etc.  Repeating the same
analysis for every $\alpha_{a}$, we need to choose $r$ masses out of
$n_{f}$ and assign $\alpha_{a} = -m_{a}$ etc.  Then we should retain only
the term $j=1$ in the second term in Eq.~(\ref{eq:F'2}), and we find
\begin{equation}
        \alpha_{1}-\lambda_{1} = \frac{4}{\Lambda^{n_{f}-2r}}
                \prod_{i\neq 1}^{n_{f}} (-m_{1}+m_{i})
                \frac{1}{2 \prod_{a\neq 1}^{r}(m_{1}-m_{a})^{2}}.
\end{equation}
Obviously the case of equal masses is singular and beyond the validity
of this analysis.

\item {\bf ii)}      $r={n_f \o 2}$   with   odd $n_{c}-n_{f}/2$

For special cases with   $r=n_f/2$   some of the considerations above
are not valid (e.g.,  Eq.(\ref{nvforrnf2})),  and  the
analysis must be done  ad hoc.  Fortunately,  in these cases  it is
possible to find the  unperturbed configuration $\{\phi_0\}$
explicitly  as in  Eq.(\ref{chebi1}), Eq.(\ref{chebi2}).  When
$n_{f}$ is even, choose $n_{f}/2$ eigenvalues
$\phi_{1}=\cdots=\phi_{n_{f}/2}$ vanishing.  Then the curve
becomes
\begin{equation}
      y^{2} = x^{n_{f}} \left[
                  \prod_{k=1}^{n_{c}-n_{f}/2} (x-\phi_{k})^{2}
          - 4\Lambda^{2n_{c}-n_{f}} \right].
\end{equation}
By identifying  the first term in the square bracket with $(2
\Lambda^{n_{c}-n_{f}/2} T_{n_{c}-n_{f}/2} (x/2\Lambda))^{2}$,
where
\beq   \phi_{k} = 2 \Lambda \cos \pi
(2k-1)/2(n_{c}-n_{f}/2), \qquad    k=1, \cdots, n_{c}-n_{f}/2,
\label{chebi111}\eeq
one easily finds   that
\begin{eqnarray}
      y^{2} &=& x^{n_{f}} \left[
          \left(2 \Lambda^{n_{c}-n_{f}/2} T_{n_{c}-n_{f}/2}
          \left(\frac{x}{2\Lambda}\right)\right)^{2}
          - 4\Lambda^{2n_{c}-n_{f}} \right]
                  \nonumber \\
          & = & - 4 x^{n_{f}} \Lambda^{2n_{c}-n_{f}}
                  \sin^{2} \left[\left(n_{c}-\frac{n_{f}}{2}\right)
                          \arccos \frac{x}{2\Lambda}
                  \right].
\label{chebi222}\end{eqnarray}
There are $n_{f}$ zeros at $x=0$, and double zeros at
$x=2\Lambda\cos \pi k/(n_{c}-n_{f}/2)$ for $k=1, \cdots,
n_{c}-n_{f}/2-1$, and single zeros at $x=\pm 2\Lambda$.

In the absence of quark masses, the theory is invariant under
$Z_{2n_{c}-n_{f}}$ symmetry: $x \rightarrow e^{2\pi i/(2n_{c}-n_{f})}x
$, $\phi_{a} \rightarrow e^{2\pi i/(2n_{c}-n_{f})} \phi_{a}$.
However, the gauge invariant symmetric polynomials of $\phi_{k}$
vanish for odd powers because of the equal number of positive and
negative ones with the same absolute values.  Therefore the Chebyshev
solutions discussed here appear only $n_{c}-n_{f}/2$ times.
This point is crucial  for the vacuum counting (see Eq.(\ref{vaccount1}))
to work out correctly.

We   first study the specific case of $2n_{c} = n_{f}+2$.
More general cases will be seen to reduce to   this  case.
The Chebyshev solution is obtained in the massless limit by setting
all of $\phi_{a}=0$:
\begin{equation}
      y^{2} = x^{2n_{c}}-4\Lambda^{2} x^{n_{f}}
          = x^{n_{f}} (x+2\Lambda) (x-2\Lambda).
\end{equation}
The zero at $x=0$ is of degree $n_{f}$, and there are other
isolated zeros at $x=\pm 2\Lambda$.
Under the perturbation by generic quark masses, we go back to the
original curve.  The only way that the curve can be arranged to have
$n_{c}-1$ double zeros as
\begin{equation}
      y^{2} = \prod_{a=1}^{n_{c}-1} (x-\alpha_{a})^{2}
          (x+2\Lambda-\beta) (x-2\Lambda-\gamma)
\end{equation}
is by assuming
\begin{equation}
      \alpha_{a} \sim m, \qquad
      \phi_{1}\sim \cdots \sim \phi_{n_{c}-2} \sim m, \qquad
      \phi_{n_{c}-1} = - \phi_{n_{c}} \sim (m \Lambda)^{1/2}.
\end{equation}
(These behaviors have been suggested by the   numerical solution  of
several explicit examples.)
Studying the region $x\sim m$, and neglecting $\beta,\gamma \ll
\Lambda$, $x \ll \phi_{n_{c}}$, we need to solve the
equation
\begin{equation}
          \phi_{n_{c}}^{4}
          \prod_{a=1}^{n_{c}-2} (x-\phi_{a})^{2}
          - 4\Lambda^{2}\prod_{i=1}^{2n_{c}-2}(x+m_{i})
          = - 4\Lambda^{2}\prod_{a=1}^{n_{c}-1} (x-\alpha_{a})^{2}.
\end{equation}
Note that the traceless condition for $\phi_{a}$ is not a stringent
constraint because $\phi_{n_{c}-1} = - \phi_{n_{c}} =
O(m\Lambda)^{1/2}$ can shift by small amount to absorb the trace of
$\phi_{a}$.  By moving terms around, we find
\begin{eqnarray}
      \lefteqn{
      \prod_{i=1}^{2n_{c}-2}(x+m_{i})
      } \nonumber \\
      &=& \left[ \prod_{a=1}^{n_{c}-1} (x-\alpha_{a})
          + i \frac{\phi_{n_{c}}^{2}}{2\Lambda}
          \prod_{a=1}^{n_{c}-2} (x-\phi_{a}) \right]
          \left[ \prod_{a=1}^{n_{c}-1} (x-\alpha_{a})
          - i \frac{\phi_{n_{c}}^{2}}{2\Lambda}
          \prod_{a=1}^{n_{c}-2} (x-\phi_{a}) \right] .
          \nonumber \\
\end{eqnarray}
For this identity to hold for any $x$, $2n_{c}-2$ zeros of l.h.s.
$(-m_{i})$ must coincide with $n_{c}-2$ zeros in each of square
brackets in the r.h.s. of the equation.  Therefore, we divide up
$2n_{c}-2$ masses into to sets $\{p_{i}, i=1,\cdots,n_{c}-1\}$ and
$\{q_{i}, i=1,\cdots,n_{c}-1\}$.  This gives us $_{n_{f}}C_{n_{f}/2}$
choices.  By identifying $p_{i}$ to the first square bracket and
$q_{i}$ to the second square bracket, we find
\begin{eqnarray}
      \prod_{i=1}^{n_{c}-1}(x+p_{i})
      &=& \prod_{a=1}^{n_{c}-1} (x-\alpha_{a})
          +i \frac{\phi_{n_{c}}^{2}}{2\Lambda}
          \prod_{a=1}^{n_{c}-2} (x-\phi_{a}),  \non     \\
      \prod_{i=1}^{n_{c}-1}(x+q_{i})
      &=& \prod_{a=1}^{n_{c}-1} (x-\alpha_{a})
          -i \frac{\phi_{n_{c}}^{2}}{2\Lambda}
          \prod_{a=1}^{n_{c}-2} (x-\phi_{a}).
\end{eqnarray}
Expanding both sides in terms of symmetric polynomials,
\begin{eqnarray}
      \sum_{k=0}^{n_{c}-1} s_{k}(p) x^{n_{c}-1-k}
      &=& \sum_{k=0}^{n_{c}-1}(-1)^{k}
          s_{k}(\alpha) x^{n_{c}-k-1}
          + i \frac{\phi_{n_{c}}^{2}}{2\Lambda} \sum_{k=0}^{n_{c}-2}
          (-1)^{k} s_{k}(\phi) x^{n_{c}-k-2}, \nonumber \\
      \sum_{k=0}^{n_{c}-1} s_{k}(q) x^{n_{c}-1-k}
      &=& \sum_{k=0}^{n_{c}-1}(-1)^{k}
          s_{k}(\alpha) x^{n_{c}-k-1}
          - i \frac{\phi_{n_{c}}^{2}}{2\Lambda} \sum_{k=0}^{n_{c}-2}
          (-1)^{k} s_{k}(\phi) x^{n_{c}-k-2}.\nonumber \\
\end{eqnarray}
Finally, we find the solutions
\begin{eqnarray}
      (-1)^k s_{k}(p)
      &=& s_{k}(\alpha) - i \frac{\phi_{n_{c}}^{2}}{2\Lambda}
          s_{k-1}(\phi),  \non    \\
      (-1)^k s_{k}(q)
      &=& s_{k}(\alpha) + i \frac{\phi_{n_{c}}^{2}}{2\Lambda}
          s_{k-1}(\phi).
\end{eqnarray}
Here and below, we use the notation that $s_{-1}=0$.  By
setting $k=1$ and subtracting both sides, we find $\phi_{n_{c}}^{2} =
-i\Lambda (s_{1}(p) - s_{1}(q)) = -i\Lambda \sum_{k=1}^{n_{f}/2}
(p_{i}-q_{i})$.  All $s_{k}(\alpha)$ and $s_{k-1}(\phi)$ are then given
in terms of $s_{k}(p)$ and $s_{k}(q)$, and hence we can write order
$n_{c}-1$ ($n_{c}-2$) polynomial equation for $\alpha$ ($\phi$) which
can always be solved to find all $\alpha$'s ($\phi$'s).

For smaller even $n_f$ with $n_c-n_f/2$ odd, the problem   reduces the one
with $2n_c-n_f=2$.  To see this, we write the curve around the
Chebyshev point as
\begin{equation}
    y^2 = \prod_{a=1}^{n_f/2} (x-\phi_a)^2
    \prod_{k=1}^{n_c-n_f/2} (x-\phi_k)^2
    - 4 \Lambda^{2n_c-n_f} \prod_{i=1} (x+m_i),
\end{equation}
where $\phi_k = 2 \Lambda \cos \pi (k-1/2)/(n_c-n_f/2)$.  When
$k=(n_c-n_f/2+1)/2$, $\phi_k=0$.  Therefore, neglecting the
fluctuations of $\phi_k$ ($k\neq(n_c-n_f/2+1)/2$), the curve reduces
to
\begin{equation}
     y^2 = \prod_{a=1}^{n_f/2+1} (x-\phi_a)^2
    \prod_{k=1, k\neq (n_c-n_f/2+1)/2}^{n_c-n_f/2} \phi_k^2
    - 4 \Lambda^{2n_c-n_f} \prod_{i=1} (x+m_i),
\end{equation}
where $\phi_a$ with $a=n_f/2+1$ is $\phi_k$ with $k=(n_c-n_f/2+1)/2$.
The product of non-vanishing $\phi_k$'s can be obtained as follows.
Recalling $T_N(x) = 2^{N-1} \prod_{k=1}^{N} (x-w_k)$ with $w_k = \cos
\pi (k-1/2)/N$, we obtain for odd $N$, $T'_N(0) = 2^{N-1} \prod_{k\neq
    (n_c-n_f/2+1)/2} (-w_k)$, while $T'_N(x) = N\sin(N \arccos
x)/\sqrt{1-x^2}$ and hence $T'_N(0) = N \sin N\pi/2 = N
(-1)^{(N-1)/2}$.  Therefore, $\prod_{k=1, k\neq
    (n_c-n_f/2+1)/2}^{n_c-n_f/2} \phi_k^2 = (T'_N(0))^2 \Lambda^{2N-2} =
N^2 \Lambda^{2N-2}$.  The low-energy curve is then
\begin{equation}
     \frac{y^2}{(n_c-n_f/2)^2 \Lambda^{2n_c-n_f-2}}
     = \prod_{a=1}^{n_f/2+1} (x-\phi_a)^2
     - 4 \frac{\Lambda^{2}}{(n_c-n_f/2)^2} \prod_{i=1} (x+m_i).
\end{equation}
This is nothing but the curve of $SU(n_f/2+1)$ theory with the
dynamical scale $\Lambda/(n_c-n_f/2)$.

\item{\bf iii)}  $r={n_f\o 2} $  with      even     $n_{c}-n_{f}/2$

The  Chebyshev solution  Eq.(\ref{chebi111}), Eq.(\ref{chebi222})  is
valid in this case also.
Again,    we  first study the special  case with      $2n_{c} =
n_{f}+4$.     The Chebyshev solution is obtained in the massless
limit by
setting all but two of $\phi_{a}=0$:
\begin{equation}
      y^{2}
      = x^{n_{f}} \left[ (x-\phi_{n_{c-1}})^{2} (x-\phi_{n_{c}})^{2}
          - 4 \Lambda^{4} \right]
          = x^{n_{f}} (x^{2}-\phi_{n_{c}}^{2}-2\Lambda^{2})
          (x^{2}-\phi_{n_{c}}^{2}+2\Lambda^{2}) .
\label{specialcase}\end{equation}
By choosing $-\phi_{n_{c}-1} = \phi_{n_{c}} = \sqrt{2}\Lambda$, the curve
becomes
\begin{equation}
      y^{2} = x^{2n_{c}}-4\Lambda^{2} x^{n_{f}+2}
      = x^{n_{f}+2} (x-2\Lambda) (x+2\Lambda).
\end{equation}
The zero at $x=0$ is of degree $n_{f}+2$, and there are other
isolated zeros at $x=\pm 2\Lambda$.  There is another solution with
$-\phi_{n_{c}-1} = \phi_{n_{c}} = i \sqrt{2}\Lambda$ as required by
the discrete $Z_{4}$ symmetry under which the Chebyshev solution
transforms as a doublet.

Under the perturbation by generic quark masses, we go back to the
original curve.  The only way that the curve can be arranged to have
$n_{c}-1$ double zeros as
\begin{equation}
      y^{2} = \prod_{a=1}^{n_{c}-1} (x-\alpha_{a})^{2}
          (x+2\Lambda-\beta) (x-2\Lambda-\gamma)
\end{equation}
is by assuming
\begin{equation}
      \phi_{a} \sim m, \qquad
      \alpha_{1}\sim \cdots \sim \alpha_{n_{c}-2} \sim m, \qquad
      \alpha_{n_{c}-2} = - \alpha_{n_{c}-1} \sim (m \Lambda)^{1/2}.
\end{equation}
(The choice  $\alpha_{n_{c}-1} = -
\alpha_{n_{c}}$  has been suggested  from several explicit
examples studied numerically.)  Studying the region $x\sim m$, and neglecting
$\beta,\gamma \ll \Lambda$, $x \ll \alpha_{n_{c}-1}$, we must   solve the
equation
\begin{equation}
      4\Lambda^{4} \prod_{a=1}^{n_{c}-2} (x-\phi_{a})^{2}
          - 4\Lambda^{4}\prod_{i=1}^{2n_{c}-4}(x+m_{i})
          = - 4\Lambda^{2} \alpha_{n_{c}-1}^{4}
          \prod_{a=1}^{n_{c}-3} (x-\alpha_{a})^{2}.
          \label{eq:tosolve0}
\end{equation}
Note that the traceless condition for $\phi_{a}$ is not a stringent
constraint because $\phi_{n_{c}-1} = - \phi_{n_{c}} =
O(\Lambda)$ can shift by small amount to absorb the trace of
$\phi_{a}$.  By moving terms around, we find
\begin{eqnarray}
      \lefteqn{
      \prod_{i=1}^{2n_{c}-4}(x+m_{i})
      } \nonumber \\
      &=& \left[ \prod_{a=1}^{n_{c}-2} (x-\phi_{a})
          + i \frac{\alpha_{n_{c}-1}^{2}}{\Lambda}
          \prod_{a=1}^{n_{c}-3} (x-\alpha_{a}) \right]
          \left[ \prod_{a=1}^{n_{c}-2} (x-\phi_{a})
          - i \frac{\alpha_{n_{c}-1}^{2}}{\Lambda}
          \prod_{a=1}^{n_{c}-3} (x-\alpha_{a}) \right] .
          \nonumber \\
\end{eqnarray}
For this identity to hold for any $x$, $2n_{c}-4$ zeros of l.h.s.
$(-m_{i})$ must coincide with $n_{c}-1$ zeros in each of square
brackets in the r.h.s. of the equation.  Therefore, we divide up
$2n_{c}-4=n_f $ masses into to sets $\{p_{i}, i=1,\cdots,n_{c}-2\}$ and
$\{q_{i}, i=1,\cdots,n_{c}-2\}$.  This gives us $_{n_{f}}C_{n_{f}/2}$
choices.  By identifying $p_{i}$ to the first square bracket and
$q_{i}$ to the second square bracket, we find
\begin{eqnarray}
      \prod_{i=1}^{n_{c}-2}(x+p_{i})
      &=& \prod_{a=1}^{n_{c}-2} (x-\phi_{a})
          + i \frac{\alpha_{n_{c}-1}^{2}}{\Lambda}
          \prod_{a=1}^{n_{c}-3} (x-\alpha_{a}),  \non   \\
      \prod_{i=1}^{n_{c}-1}(x+q_{i})
      &=& \prod_{a=1}^{n_{c}-2} (x-\phi_{a})
          - i \frac{\alpha_{n_{c}-1}^{2}}{\Lambda}
          \prod_{a=1}^{n_{c}-3} (x-\alpha_{a}).
\end{eqnarray}
Expanding both sides in terms of symmetric polynomials,
\begin{eqnarray}
      \sum_{k=0}^{n_{c}-2} s_{k}(p) x^{n_{c}-2-k}
      &=& \sum_{k=0}^{n_{c}-2}(-1)^{k}
          s_{k}(\phi) x^{n_{c}-k-2}
          + i \frac{\alpha_{n_{c}-1}^{2}}{\Lambda} \sum_{k=0}^{n_{c}-3}
          (-1)^{k} s_{k}(\alpha) x^{n_{c}-k-2}, \nonumber \\
      \sum_{k=0}^{n_{c}-1} s_{k}(q) x^{n_{c}-1-k}
      &=& \sum_{k=0}^{n_{c}-2}(-1)^{k}
          s_{k}(\phi) x^{n_{c}-k-2}
          - i \frac{\alpha_{n_{c}-1}^{2}}{\Lambda} \sum_{k=0}^{n_{c}-3}
          (-1)^{k} s_{k}(\alpha) x^{n_{c}-k-2}. \nonumber \\
\end{eqnarray}
Finally, we find the solutions
\beq
      (-1)^k s_{k}(p)
     =    s_{k}(\phi) - i \frac{\alpha_{n_{c}-1}^{2}}{\Lambda}
          s_{k-1}(\alpha), \qquad
      (-1)^k s_{k}(q)
    =   s_{k}(\phi) + i \frac{\alpha_{n_{c}-1}^{2}}{\Lambda}
          s_{k-1}(\alpha).
\eeq
Here and below, we use the notation that $s_{-1}=0$ identically.  By
setting $k=1$ and subtracting both sides, we find $\alpha_{n_{c}-1}^{2} =
-i\Lambda (s_{1}(p) - s_{1}(q))/2 = -i\Lambda \sum_{k=1}^{n_{f}/2}
(p_{i}-q_{i})/2$.  All $s_{k}(\phi)$ and $s_{k-1}(\alpha)$ are then given
in terms of $s_{k}(p)$ and $s_{k}(q)$, and hence we can write order
$n_{c}-3$ ($n_{c}-2$) polynomial equation for $\alpha$ ($\phi$) which
can always be solved to find all $\alpha$'s ($\phi$'s).

For smaller even $n_f$ with $n_c-n_f/2$ even, the curve reduces  to   the one
with $2n_c-n_f=4$.  To see this, we write the curve around the
Chebyshev point as
\begin{equation}
    y^2 = \prod_{a=1}^{n_f/2} (x-\phi_a)^2
    \prod_{k=1}^{n_c-n_f/2} (x-\phi_k)^2
    - 4 \Lambda^{2n_c-n_f} \prod_{i=1} (x+m_i),
\end{equation}
where $\phi_k = 2 \Lambda \cos \pi (k-1/2)/(n_c-n_f/2)$.  Therefore,
neglecting the fluctuations of    $\phi_k$ ($k\neq(n_c-n_f/2+1)/2$), the
curve reduces to
\begin{equation}
     y^2 = \prod_{a=1}^{n_f/2} (x-\phi_a)^2
    \prod_{k=1}^{n_c-n_f/2} \phi_k^2
    - 4 \Lambda^{2n_c-n_f} \prod_{i=1} (x+m_i).
\end{equation}
The product of $\phi_k$'s can be obtained as follows.  Recalling
$T_N(x) = 2^{N-1} \prod_{k=1}^{N} (x-w_k)$ with $w_k = \cos \pi
(k-1/2)/N$, we obtain for even $N$, $T_N(0) = 2^{N-1} \prod_{k}
(-w_k)$, while $T_N(x) = \cos(N \arccos x)$ and hence $T_N(0) = \cos N
\pi/2 = (-1)^{N/2}$.  Therefore, $\prod_{k=1}^{n_c-n_f/2}  \phi_k^2 =
(2T_N(0))^2    \cdot   \Lambda^{2N} = 4 \Lambda^{2N}$.  Therefore the
low-energy
curve is
\begin{equation}
     \frac{y^2}{4 \Lambda^{2n_c-n_f}}
     = \prod_{a=1}^{n_f/2} (x-\phi_a)^2
     - \prod_{i=1}^{n_f / 2} (x+m_i).
\end{equation}
   Note that  this  is     precisely the curve  of the  $ SU({n_f  \o 2}) $
theory.       It has the same form as the left hand side of
Eq.(\ref{eq:tosolve0}).

\item{\bf   iv)}    $r= {\tilde n}_c=n_f-n_c:$    Root of the baryonic branch

The case with $r= {\tilde n}_c=n_f-n_c$   also requires  a separate
consideration
since the unperturbed curve has a special    form,
Eq.(\ref{bbranch}),  Eq.(\ref{bbranch2}).
The curve is
\begin{equation}
        y^{2} = x^{2\tilde{n}_{c}}
        \left[ \prod_{k=1}^{n_{c}-\tilde{n}_{c}} (x-\Phi_{k})^{2}
                + 4 \Lambda^{2n_{c}-n_{f}} x^{n_{c}-\tilde{n}_{c}}
\right] =   x^{2{\tilde n}_c}
    ( x^{n_c - {\tilde n}_c } -   \Lambda^2   )^2.
\label{baryoniccc}  \end{equation}
with adjoint VEVS  taken as
$\phi_{1}, \cdots, \phi_{n_{c}-\tilde{n}_{c}}\neq 0$,
$\phi_{n_{c}-\tilde{n}_{c}+1}= \cdots =\phi_{n_{c}}=0$,
\begin{equation}
        (\Phi_{1}, \cdots, \Phi_{k}, \cdots, \Phi_{n_{c}-\tilde{n}_{c}})
        = \Lambda (\omega^2, \cdots, \omega^{2k}, \cdots,
        \omega^{2(n_{c}-\tilde{n}_{c})}),
\label{baryoniccc2}  \end{equation}
where
$\omega = e^{\pi i/(n_{c}-\tilde{n}_{c})}$.
The double zeros of the factor in the parenthesis are at
\begin{equation}
        x = \Lambda \omega, \Lambda \omega^{3}, \cdots,
                \Lambda \omega^{2k-1}, \cdots,
                \Lambda \omega^{2(n_{c}-\tilde{n}_{c})-1} .
                \label{eq:zeros}
\end{equation}

There are  two ways    for maintaining   the curve maximally
singular,  when generic  bare quark masses are added.
One is to keep all  ``large'' $n_{c}-\tilde{n}_{c}$ zeros doubled,
while allowing
$2\tilde{n}_{c}$ zeros at $x=0$ to decompose into $\tilde{n}_{c}-1$
double zeros and two single zeros.  The other is to take    all ``small''
zeros  doubled,    while keeping only $n_{c}-\tilde{n}_{c}-1$ ``large'' double
zeros.

\item {iv-a)}   {Keeping all of ``large''  zeros doubled}

Upon mass perturbation  the adjoint scalar  VEVS take  the form,  $(\phi_{a},
\Phi_{k} =
\Lambda
\omega^{2k} + \gamma_{k}),    $ with $\gamma_{k}, \phi_{a} \sim O(m)$.  The
constraint is therefore
\begin{equation}
        \sum_{a=1}^{\tilde{n}_{c}} \phi_{a}
        + \sum_{k=1}^{n_{c}-\tilde{n}_{c}}\gamma_{k} = 0.
        \label{eq:constraint}
\end{equation}
The perturbed curve is
\begin{equation}
        y^{2} = \prod_{a=1}^{\tilde{n}_{c}} (x-\phi_{a})^{2}
                \prod_{k=1}^{n_{c}-\tilde{n}_{c}}
                        (x-\Lambda \omega^{2k} - \gamma_{k})^{2}
                + 4 \Lambda^{2n_{c}-n_{f}} \prod_{i=1}^{n_{f}} (x+m_{i}).
                \label{eq:newcurve}
\end{equation}
The zeros (\ref{eq:zeros}) are also shifted to
\begin{equation}
        x = \Lambda \omega^{2\ell-1} + \delta_{\ell},
                \qquad \ell = 1, \cdots, n_{c}-\tilde{n}_{c}.
        \label{eq:newzeros}
\end{equation}
We substitute these zeros for each $\ell$ into the curve and require
that the r.h.s. of the curve (\ref{eq:newcurve}) vanishes at $O(m)$.

The first factor in the curve (\ref{eq:newcurve}) is expanded as
\begin{equation}
        \prod_{a=1}^{\tilde{n}_{c}}
        (\Lambda \omega^{2\ell-1} + \delta_{\ell}-\phi_{a})^{2}
        = (\Lambda \omega^{2\ell-1})^{\tilde{n}_{c}}
        \left[ 1 + 2 \sum_{a=1}^{\tilde{n}_{c}}
                \frac{\delta_{\ell}-\phi_{a}}{\Lambda
\omega^{2\ell-1}} \right].
                \label{eq:first}
\end{equation}

The second factor in the curve (\ref{eq:newcurve}) is expanded as
\begin{equation}
        \prod_{k=1}^{n_{c}-\tilde{n}_{c}}
                (\Lambda \omega^{2\ell-1} + \delta_{\ell}
                -\Lambda \omega^{2k} - \gamma_{k})^{2}
        = \prod_{k=1}^{n_{c}-\tilde{n}_{c}}
                (\Lambda \omega^{2\ell-1} -\Lambda \omega^{2k})^{2}
                \left[ 1 + 2 \sum_{k=1}^{n_{c}-\tilde{n}_{c}}
                        \frac{\delta_{\ell} - \gamma_{k}}
                                {\Lambda \omega^{2\ell-1} -\Lambda
\omega^{2k}} \right].
        \label{eq:second}
\end{equation}
This factor needs to be simplified.  We show that
\begin{equation}
        \prod_{k=1}^{n_{c}-\tilde{n}_{c}}
                (\Lambda \omega^{2\ell-1} -\Lambda \omega^{2k})
                = - 2 \Lambda^{n_{c}-\tilde{n}_{c}}.
\end{equation}
This can be proven by studying the polynomial
\begin{equation}
        \prod_{k=1}^{n_{c}-\tilde{n}_{c}}
                (t -\Lambda \omega^{2k})
                = t^{n_{c}-\tilde{n}_{c}} - \Lambda^{n_{c}-\tilde{n}_{c}}.
\end{equation}
By substituting $t = \Lambda \omega^{2\ell-1}$, and noting that
$(\Lambda \omega^{2\ell-1})^{n_{c}-\tilde{n}_{c}} =
\Lambda^{n_{c}-\tilde{n}_{c}} e^{\pi i (2\ell-1)} = -
\Lambda^{n_{c}-\tilde{n}_{c}}$.  The second factor (\ref{eq:second})
of the curve (\ref{eq:newcurve}) is now simplified to
\begin{equation}
        4 \Lambda^{2(n_{c}-\tilde{n}_{c})}
        \left[ 1 + 2 \sum_{k=1}^{n_{c}-\tilde{n}_{c}}
                        \frac{\delta_{\ell} - \gamma_{k}}
                                {\Lambda \omega^{2\ell-1} -\Lambda
\omega^{2k}} \right].
        \label{eq:second2}
\end{equation}

The last term in the curve (\ref{eq:newcurve}) is expanded as
\begin{equation}
        4 \Lambda^{2n_{c}-n_{f}} \prod_{i=1}^{n_{f}}
                (\Lambda \omega^{2\ell-1} + \delta_{\ell} + m_{i})
        = 4 \Lambda^{2n_{c}} (\omega^{2\ell-1})^{n_{f}}
        \left[ 1 + \sum_{i=1}^{n_{f}}
                \frac{\delta_{\ell} + m_{i}}{\Lambda \omega^{2\ell-1}} \right].
                \label{eq:last}
\end{equation}
Note further that
\begin{equation}
        (\omega^{2\ell-1})^{n_{f}} =
        (\omega^{2\ell-1})^{2\tilde{n}_{c}+(n_{c}-\tilde{n}_{c})} =
        (\omega^{2\ell-1})^{2\tilde{n}_{c}} e^{\pi i (2\ell -1)} =
        -(\omega^{2\ell-1})^{2\tilde{n}_{c}}.
\end{equation}
The last term therefore is
\begin{equation}
        - 4 \Lambda^{2n_{c}} (\omega^{2\ell-1})^{2\tilde{n}_{c}}
        \left[ 1 + \sum_{i=1}^{n_{f}}
                \frac{\delta_{\ell} + m_{i}}{\Lambda \omega^{2\ell-1}} \right].
                \label{eq:last2}
\end{equation}

By putting together the expansion
Eqs.(\ref{eq:first},\ref{eq:second2},\ref{eq:last2}) up to $O(m)$
into the curve (\ref{eq:newcurve}) and requiring it to vanish, we find
\begin{equation}
        0 = 2 \sum_{a=1}^{\tilde{n}_{c}}
                \frac{\delta_{\ell}-\phi_{a}}{\Lambda \omega^{2\ell-1}}
                + 2 \sum_{k=1}^{n_{c}-\tilde{n}_{c}}
                        \frac{\delta_{\ell} - \gamma_{k}}
                                {\Lambda \omega^{2\ell-1} -\Lambda \omega^{2k}}
                - \sum_{i=1}^{n_{f}}
                \frac{\delta_{\ell} + m_{i}}{\Lambda \omega^{2\ell-1}}.
\end{equation}
Multiply the equation by $\Lambda \omega^{2\ell-1}$ to simplify it to
\begin{equation}
        0 = 2 \sum_{a=1}^{\tilde{n}_{c}} (\delta_{\ell}-\phi_{a})
                + 2 \sum_{k=1}^{n_{c}-\tilde{n}_{c}}
                        \frac{\delta_{\ell} - \gamma_{k}}
                                {1 - \omega^{2k-2\ell+1}}
                - \sum_{i=1}^{n_{f}} (\delta_{\ell} + m_{i}).
        \label{eq:zero}
\end{equation}

We try to further simplify this equation.  The sum over $a$ in the
first term gives
\begin{equation}
        2 \sum_{a=1}^{\tilde{n}_{c}} (\delta_{\ell}-\phi_{a})
        = 2 \tilde{n}_{c} \delta_{\ell} - 2 \sum_{a=1}^{\tilde{n}_{c}} \phi_{a}
        = 2 \tilde{n}_{c} \delta_{\ell} +
        2 \sum_{k=1}^{n_{c}-\tilde{n}_{c}}\gamma_{k},
        \label{eq:zerofirst}
\end{equation}
where the constraint Eq.~(\ref{eq:constraint}) was used in the last
equality.  To simplify the second term, we would like to prove that
\begin{equation}
        \sum_{k=1}^{n_{c}-\tilde{n}_{c}}
                \frac{1}{1 - \omega^{2k-2\ell+1}}
                = \frac{1}{2} (n_{c}-\tilde{n}_{c}).
        \label{eq:identity}
\end{equation}
Note that the sum over $k$ exhausts all possible odd powers in
$\omega$ in the denominator.  Therefore we can shift $k$ to $k+\ell$
and find
\begin{equation}
        \sum_{k=1}^{n_{c}-\tilde{n}_{c}}
                \frac{1}{1 - \omega^{2k-2\ell+1}}
        = \sum_{k=1}^{n_{c}-\tilde{n}_{c}} \frac{1}{1 - \omega^{2k+1}}.
\end{equation}
Now we distinguish two cases, when $n_{c}-\tilde{n}_{c} = 2m$ (even),
and $n_{c}-\tilde{n}_{c} = 2m+1$ (odd).  When $n_{c}-\tilde{n}_{c} =
2m$, the sum is separated in two pieces,
\begin{equation}
        \sum_{k=1}^{2m} \frac{1}{1 - \omega^{2k+1}}
        = \sum_{k=1}^{m} \frac{1}{1 - \omega^{2k+1}}
        + \sum_{k=m+1}^{2m} \frac{1}{1 - \omega^{2k+1}}.
\end{equation}
Change the variable $k$ in the second sum to $2m-k+1$, and using the
definition $\omega = e^{\pi i/2m}$,
\begin{equation}
        = \sum_{k=1}^{m} \frac{1}{1 - \omega^{2k+1}}
        + \sum_{k=1}^{m} \frac{1}{1 - \omega^{-(2k+1)}}.
\end{equation}
Adding terms for each $m$,
\begin{equation}
        = \sum_{k=1}^{m} \frac{1 - \omega^{2k+1}+1 - \omega^{-(2k+1)}}
                {1 - \omega^{2k+1} - \omega^{-(2k+1)} + 1}
        = m = \frac{1}{2} (n_{c}-\tilde{n}_{c}).
\end{equation}
This completes the proof of Eq.~(\ref{eq:identity}) for even
$n_{c}-\tilde{n}_{c}$.  Similarly, for $n_{c}-\tilde{n}_{c} = 2m+1$,
\begin{equation}
        \sum_{k=1}^{2m} \frac{1}{1 - \omega^{2k+1}}
        = \sum_{k=1}^{m} \frac{1}{1 - \omega^{2k+1}}
        + \frac{1}{1 - \omega^{2m+1}}
        + \sum_{k=m+2}^{2m+1} \frac{1}{1 - \omega^{2k+1}}.
\end{equation}
The middle term in the r.h.s. is $1/(1-(-1)) = 1/2$.  Change the
variable $k$ in the last sum to $2m+2-k$,
\begin{equation}
        = \sum_{k=1}^{m} \frac{1}{1 - \omega^{2k+1}}
        + \frac{1}{2}
        + \sum_{k=1}^{m} \frac{1}{1 - \omega^{-(2k+1)}}.
\end{equation}
Again by adding terms for each $m$,
\begin{equation}
        = \frac{1}{2} +
                \sum_{k=1}^{m} \frac{1 - \omega^{2k+1}+1 - \omega^{-(2k+1)}}
                {1 - \omega^{2k+1} - \omega^{-(2k+1)} + 1}
        = \frac{1}{2} + m = \frac{1}{2} (n_{c}-\tilde{n}_{c}).
\end{equation}
This completes the proof of Eq.~(\ref{eq:identity}) for odd
$n_{c}-\tilde{n}_{c}$.  Therefore the identity (\ref{eq:identity}) is
proven.  The second term in Eq.~(\ref{eq:zero}) is then given by
\begin{equation}
        2 \sum_{k=1}^{n_{c}-\tilde{n}_{c}}
                \frac{\delta_{\ell} - \gamma_{k}}{1 - \omega^{2k-2\ell+1}}
        = (n_{c}-\tilde{n}_{c}) \delta_{\ell}
                - 2 \sum_{k=1}^{n_{c}-\tilde{n}_{c}}
                \frac{\gamma_{k}}{1 - \omega^{2k-2\ell+1}}.
        \label{eq:zerosecond}
\end{equation}
Finally the last term in Eq.~(\ref{eq:zero}) is simply
\begin{equation}
        - \sum_{i=1}^{n_{f}} (\delta_{\ell} + m_{i})
        = - n_{f} \delta_{\ell} - \sum_{i=1}^{n_{f}} m_{i}.
        \label{eq:zerolast}
\end{equation}
Putting together
Eqs.~(\ref{eq:zerofirst},\ref{eq:zerosecond},\ref{eq:zerolast}) in
(\ref{eq:zero}), we find
\begin{eqnarray}
        0 &=& 2 \tilde{n}_{c} \delta_{\ell} +
                2 \sum_{k=1}^{n_{c}-\tilde{n}_{c}}\gamma_{k}
                + (n_{c}-\tilde{n}_{c}) \delta_{\ell}
                - 2 \sum_{k=1}^{n_{c}-\tilde{n}_{c}}
                \frac{\gamma_{k}}{1 - \omega^{2k-2\ell+1}}
                - n_{f} \delta_{\ell} - \sum_{i=1}^{n_{f}} m_{i}
                \nonumber \\
        &=& 2 \sum_{k=1}^{n_{c}-\tilde{n}_{c}}\gamma_{k}
                - 2 \sum_{k=1}^{n_{c}-\tilde{n}_{c}}
                \frac{\gamma_{k}}{1 - \omega^{2k-2\ell+1}}
                - \sum_{i=1}^{n_{f}} m_{i},
        \label{eq:zero2}
\end{eqnarray}
and $\delta_{\ell}$ disappeared from the equation.  Finally, we add
over $\ell$.  Only the second term depends on $\ell$, and the sum
over $\ell$ is simplified again by using the identity
Eq.~(\ref{eq:identity}).  We find
\begin{equation}
        0 = 2 (n_{c}-\tilde{n}_{c}) \sum_{k=1}^{n_{c}-\tilde{n}_{c}} \gamma_{k}
        - 2 \frac{1}{2} (n_{c}-\tilde{n}_{c})
                \sum_{k=1}^{n_{c}-\tilde{n}_{c}}\gamma_{k}
        - (n_{c}-\tilde{n}_{c}) \sum_{i=1}^{n_{f}} m_{i},
\end{equation}
and therefore
\begin{equation}
        \sum_{a=1}^{\tilde{n}_{c}} \phi_{a}
        = - \sum_{k=1}^{n_{c}-\tilde{n}_{c}} \gamma_{k}
        = - \sum_{i=1}^{n_{f}} m_{i}.
        \label{eq:constraint2}
\end{equation}

The fluctuation around $x \sim 0$ is described by a low-energy curve
by neglecting $x, \gamma_{k} \ll \Lambda$ in Eq.~(\ref{eq:newcurve}),
\begin{equation}
        y^{2} = \prod_{a=1}^{\tilde{n}_{c}} (x-\phi_{a})^{2}
                \prod_{k=1}^{n_{c}-\tilde{n}_{c}}
                        (-\Lambda \omega^{2k})^{2}
                + 4 \Lambda^{2n_{c}-n_{f}} \prod_{i=1}^{n_{f}} (x+m_{i}) ,
\end{equation}
or
\begin{equation}
        \frac{y^{2}}{\Lambda^{2(n_{c}-\tilde{n}_{c})}}
        = \prod_{a=1}^{\tilde{n}_{c}} (x-\phi_{a})^{2}
                + \frac{4}{\Lambda^{n_{c}-\tilde{n}_{c}}}
                \prod_{i=1}^{n_{f}} (x+m_{i}),
        \label{eq:LEcurve}
\end{equation}
where $\phi_{a}$ are subject to an unusual constraint
Eq.~(\ref{eq:constraint2}).  Apart from the constraint, the problem
is similar to earlier    study of the $r$-root, where we take
$\phi_{a} = - m_{a}$ to have double zeros at $x = -m_{a}$.  The
maximum number of $\phi_{a}$'s that can be chosen this way is,
however, $\tilde{n}_{c}-1$ because of the constraint.  Let us choose
$r$ out of $\tilde{n}_{c}$ to coincide with $r$ of $-m_{i}$
(${}_{n_{f}}\!C_{r}$ choices), {\it e.g.}\/,
\begin{equation}
        \phi_{1} = -m_{1}, \, \cdots, \, \phi_{r} = -m_{r},
\end{equation}
While the remaining $\phi_{a}$ are
\begin{equation}
        \phi_{k} = - \frac{1}{\tilde{n}_{c}-r} \sum_{i=r+1}^{n_{f}} m_{i}
                + \delta \phi_{k}, \qquad
                k = 1, \cdots, \tilde{n}_{c}-r.
\end{equation}
The fluctuations are subject to the constraint
\begin{equation}
        \sum_{k=1}^{\tilde{n}_{c}-r} \delta \phi_{k} = 0.
        \label{eq:LEconstraint}
\end{equation}
The low-energy curve Eq.~(\ref{eq:LEcurve}) then becomes
\begin{equation}
        \frac{y^{2}}{\Lambda^{2(n_{c}-\tilde{n}_{c})}}
        = \prod_{a=1}^{r} (x-m_{a})^{2}
                \prod_{k=1}^{\tilde{n}_{c}-r}
                \left(x + \frac{1}{\tilde{n}_{c}-r} \sum_{i=r+1}^{n_{f}} m_{i}
                        - \delta \phi_{k}\right)^{2}
                + \frac{4}{\Lambda^{n_{c}-\tilde{n}_{c}}}
                \prod_{i=1}^{n_{f}} (x+m_{i}).
\end{equation}
Now shift $x$ to $x - \frac{1}{\tilde{n}_{c}-r} \sum_{i=r+1}^{n_{f}}
m_{i}$, and we assume that the remaining fluctuations $x, \delta \phi_{k}
\ll m$ which will be justified {\it a posteriori}\/.  The curve is
further approximated as
\begin{eqnarray}
        \frac{y^{2}}{\Lambda^{2(n_{c}-\tilde{n}_{c})}}
        &=& \prod_{a=1}^{r}
                \left(\frac{1}{\tilde{n}_{c}-r} \sum_{i=r+1}^{n_{f}} m_{i}
                        + m_{a}\right)^{2}
                \prod_{k=1}^{\tilde{n}_{c}-r}
                (x - \delta \phi_{k})^{2}
        \nonumber \\
        & &     + \frac{4}{\Lambda^{n_{c}-\tilde{n}_{c}}}
                \prod_{i=1}^{n_{f}}
                \left(m_{i} - \frac{1}{\tilde{n}_{c}-r} \sum_{i=r+1}^{n_{f}}
                        m_{i}\right).
\end{eqnarray}
Up to an overall constant, this is the curve of pure
$SU(\tilde{n}_{c}-r)$ Yang--Mills theories.  The dynamical scale of
this theory is
\begin{eqnarray}
        \Lambda_{\it pure}^{2(\tilde{n}_{c}-r)}
                &=& \frac{1}{\Lambda^{n_{c}-\tilde{n}_{c}}}
                \frac{\prod_{i=1}^{n_{f}}
                \left(m_{i} - \frac{1}{\tilde{n}_{c}-r} \sum_{i=r+1}^{n_{f}}
                        m_{i}\right)}
                        {\prod_{a=1}^{r}
                        \left(\frac{1}{\tilde{n}_{c}-r}
\sum_{i=r+1}^{n_{f}} m_{i}
                        + m_{a}\right)^{2}}
                \nonumber \\
        &\sim &  m^{2(\tilde{n}_{c}-r)}
                        \left(\frac{m}{\Lambda}\right)^{n_{c}-\tilde{n}_{c}}
                        \ll m^{2(\tilde{n}_{c}-r)}.
\end{eqnarray}
Therefore the singularities of this curve are located at $x, \delta
\phi_{k} \sim \Lambda_{\it pure} \ll m$ and hence the approximation
is justified.  There are $\tilde{n}_{c}-r$ such singularities
(basically the Witten index) for each $r$, and hence the total number
of vacua obtained in this subsection are
\begin{equation}
        {\cal N}_{2}
        = \sum_{r=0}^{\tilde{n}_{c}-1} (\tilde{n}_{c}-r) {}_{n_{f}}C_{r}.
        \label{eq:N2}
\end{equation}

\item {iv-b)}   {Keeping all of ``small'' double zeros}

Another possibility is that   all    $\tilde{n}_{c}$  ``small'' zeros
near $x=0$   are doubled,  while  one of  the
$n_{c}-\tilde{n}_{c}$ ``large'' double zeros   is split into two single  zeros.

The crucial difference from the  previous case   is that
we keep one less ``large'' double zeros and hence fluctuations
$\phi_{1}, \cdots, \phi_{\tilde{n}_{c}}$ are not subject to a
constraint.  Even though the low-energy curve
\begin{equation}
        \frac{y^{2}}{\Lambda^{2(n_{c}-\tilde{n}_{c})}}
        = \prod_{a=1}^{\tilde{n}_{c}} (x-\phi_{a})^{2}
                + \frac{4}{\Lambda^{n_{c}-\tilde{n}_{c}}}
                \prod_{i=1}^{n_{f}} (x+m_{i}),
\end{equation}
is the same as Eq.~(\ref{eq:LEcurve}) in the previous subsection,
$\phi_{a}$ can all freely vary and hence can all be matched to the
quark masses  as in the cases  i)   above,     as
\begin{equation}
        \phi_{1} = -m_{1}, \, \cdots, \,
        \phi_{\tilde{n}_{c}} = -m_{\tilde{n}_{c}},
\end{equation}
and there are ${}_{n_{f}}C_{\tilde{n}_{c}}$ choices.  Recall that
there are $n_{c}-\tilde{n}_{c}$ ``large'' double zeros at the baryonic
root and we can give up one of them; therefore there are actually
$(n_{c}-\tilde{n}_{c}) \,  {}_{n_{f}}  \!  C_{\tilde{n}_{c}}$ choices.

\item{\bf  v)}   Summary of the   vacuum counting in $SU(n_c)$ theories

The ``nonbaryonic" branch roots   Eq.(\ref{rbranch111}),
Eq.(\ref{chebi222}),    yield,  upon quark mass perturbation,
\beq     ( 2 n_c -n_f ) \cdot 2^{n_f-1}  -  (n_{c}-\tilde{n}_{c}) \,
{}_{n_{f}}  \!  C_{\tilde{n}_{c}} = {\cal N}_1 -
(n_{c}-\tilde{n}_{c}) \,
{}_{n_{f}}  \!  C_{\tilde{n}_{c}}  \eeq vacua.
The baryonic root,  Eq.(\ref{baryoniccc}),  Eq.(\ref{baryoniccc2}),   leads to
\beq
\sum_{r=0}^{\tilde{n}_{c}-1} (\tilde{n}_{c}-r) {}_{n_{f}}C_{r}  +
(n_{c}-\tilde{n}_{c}) \,  {}_{n_{f}}  \!  C_{\tilde{n}_{c}}
= {\cal N}_2   +  (n_{c}-\tilde{n}_{c}) \,  {}_{n_{f}}  \!
C_{\tilde{n}_{c}} \eeq
   vacua.   Their
sum coincides  with the total number of $N=1$ vacua  found from the
semiclassical analysis   as well as  from the analyses at   large
$\mu$.

\end{description}

\subsection  {Perturbation  around  CFT  points  of the   $USp(2n_c)$
Curve   \label{sec:masspert2}}

We start from the    CFT  points  described by the   Chebyshev
polynomial, Eq.(\ref{cheby0})-Eq.(\ref{cheby3}),
   and  add generic quark masses $m_i$.

\begin{description}

   \item{\bf   i)}   Chebyshev point:    Odd $n_{f}$

Let us  take   $n_{f}$    odd  first.
We first study the specific case of $2n_{c} = n_{f}-1$.
The more general cases will  be discussed later on.

The Chebyshev solution is obtained in the massless limit by setting
all of $\phi_{a}=0$:
\begin{equation}
      x y^{2} = \left[ x^{n_{c}+1} \right]^{2}
          - 4\Lambda^{2} x^{2n_{c}+1}
          = x^{2n_{c}+1}(x-4\Lambda^{2}).
\end{equation}
The zero at $x=0$ is of degree $2n_{c}$, and there is another
isolated zero at $x=4\Lambda^{2}$.  There is also a branch point at
$x=\infty$.

Under the perturbation by generic quark masses, we go back to the
original curve.  The only way that the curve can be arranged to have
$n_{c}$ double zeros as
\begin{equation}
      x y^{2} = x (x-4\Lambda^{2} -\beta) \prod_{a=1}^{n_{c}}
      (x-\alpha_{a})^{2}
\end{equation}
is by assuming
\begin{equation}
      \alpha_{a} \sim m^{2}, \qquad \phi_{1}^{2} \sim m \Lambda, \qquad
      \phi_{2}^{2} \sim \cdots \sim \phi_{n_{c}}^{2} \sim m^{2}.
\end{equation}
By neglecting $\beta, x \ll \Lambda^{2}$, we need to solve the
equation
\begin{equation}
      \left[ x \prod_{a=1}^{n_{c}} (x-\phi_{a}^{2})
          + 2\Lambda m_{1} \cdots m_{n_{f}} \right]^{2}
          - 4\Lambda^{2} \prod_{i=1}^{n_{f}}(x+m_{i}^{2})
          = -4\Lambda^{2} x \prod_{a=1}^{n_{c}} (x-\alpha_{a})^{2}.
          \label{eq:tosolve}
\end{equation}

To address this question, we introduce a few preliminary facts.  First
of all, consider a generic polynomial
\begin{equation}
      \prod_{i=1}^{N} (z-\rho_{i})
      = \sum_{k=0}^{N} (-1)^{k} s_{k}(\rho) z^{N-k}.
\end{equation}
The symmetric polynomials $s_{j}(\rho)$ are given by $s_{0}(\rho) =
1$, $s_{1}(\rho) = \sum_{i=1}^{N} \rho_{i}$, $s_{2}(\rho) =
\sum_{i<j} \rho_{i} \rho_{j}$, etc.  This defines the notation $s_{k}$.
Then Eq.~(\ref{eq:tosolve}) is written as
\begin{equation}
      \left[ x \sum_{k=0}^{n_{c}} (-1)^{k} s_{k}(\phi^{2}) x^{n_{c}-k}
          + 2\Lambda m_{1} \cdots m_{n_{f}} \right]^{2}
          - 4\Lambda^{2} \prod_{i=1}^{n_{f}}(x+m_{i}^{2})
          = -4\Lambda^{2} x \prod_{a=1}^{n_{c}} (x-\alpha_{a})^{2}.
\end{equation}
Note that $s_{k}(\phi^{2}) = O(m^{2k-1}\Lambda)$ for $k\neq 0$
because one of the
$\phi^{2}$'s is $O(m\Lambda)$.  This allows us to neglect
$s_{0}(\phi^{2})x^{n_{c}} = x^{n_{c}}$ term in the sum, and the
equation becomes
\begin{equation}
      \left[\sum_{k=1}^{n_{c}} (-1)^{k} s_{k}(\phi^{2}) x^{n_{c}-k+1}
          + 2\Lambda m_{1} \cdots m_{n_{f}} \right]^{2}
          - 4\Lambda^{2} \prod_{i=1}^{n_{f}}(x+m_{i}^{2})
          = -4\Lambda^{2} x \prod_{a=1}^{n_{c}} (x-\alpha_{a})^{2}.
\end{equation}
Now rewrite it as
\begin{eqnarray}
      \lefteqn{
      \prod_{i=1}^{n_{f}}(x+m_{i}^{2}) }\nonumber \\
      &=& x \left[ \sum_{k=0}^{n_{c}} (-1)^{k} s_{k}(\alpha) x^{n_{c}-k}
          \right]^{2}
          + \left[\sum_{k=1}^{n_{c}} (-1)^{k}
          \frac{s_{k}(\phi^{2})}{2\Lambda} x^{n_{c}-k+1}
          + m_{1} \cdots m_{n_{f}} \right]^{2} \nonumber \\
      &=& x \left[ \sum_{k=0}^{n_{c}} (-1)^{k} s_{k}(\alpha) x^{n_{c}-k}
          \right]^{2}
          + \left[-\sum_{k=0}^{n_{c}-1} (-1)^{k}
          \frac{s_{k+1}(\phi^{2})}{2\Lambda} x^{n_{c}-k}
          + m_{1} \cdots m_{n_{f}} \right]^{2} . \nonumber \\
          \label{eq:tosolve2}
\end{eqnarray}

Now consider the following polynomial
\begin{equation}
      F(x) = \prod_{i=1}^{n_{f}} (\sqrt{x} + i m_{i})
      = \sum_{k=0}^{n_{f}} i^{k} s_{k}(m) \sqrt{x}^{n_{f}-k}.
\end{equation}
This polynomial can be divided into the ``real'' and ``imaginary''
parts (this is not strictly true because $m$'s are complex, but what is meant
   here is the division between terms of odd powers in $i$
and of even powers in $i$),
\begin{eqnarray}
      F(x) &=& \sum_{k=0}^{(n_{f}-1)/2} (-1)^{k} s_{2k}(m)
      \sqrt{x}^{n_{f}-2k}
      + \sum_{k=0}^{(n_{f}-1)/2} i (-1)^{k} s_{2k+1}(m)
      \sqrt{x}^{n_{f}-2k-1} \nonumber \\
      &=& \sqrt{x} \sum_{k=0}^{n_{c}} (-1)^{k} s_{2k}(m) x^{n_{c}-k}
      + i \sum_{k=0}^{n_{c}} (-1)^{k} s_{2k+1}(m)
      x^{n_{c}-k},
\end{eqnarray}
where we   used $n_{f} = 2n_{c}+1$.  Similarly consider the polynomial
\begin{equation}
      G(x) = \prod_{i=1}^{n_{f}} (\sqrt{x} - i m_{i})
      = \sqrt{x} \sum_{k=0}^{n_{c}} (-1)^{k} s_{2k}(m) x^{n_{c}-k}
      - i \sum_{k=0}^{n_{c}} (-1)^{k} s_{2k+1}(m) x^{n_{c}-k}.
\end{equation}
  From the definition,
\begin{equation}
      F(x)    G(x) = \prod_{i=1}^{n_{f}} (x + m_{i}^{2}).
\end{equation}
On the other hand, this product is also given by
\begin{equation}
      F(x) G(x) = \left[ \sqrt{x} \sum_{k=0}^{n_{c}} (-1)^{k}
      s_{2k}(m) x^{n_{c}-k} \right]^{2}
      + \left[ \sum_{k=0}^{n_{c}} (-1)^{n_{f}-k} s_{2k+1}(m) x^{k}
      \right]^{2}.
\end{equation}
This is precisely the same as Eq.~(\ref{eq:tosolve2}) upon
identifications
\begin{equation}
      s_{k}(\alpha) = s_{2k}(m), \qquad
      s_{k+1}(\phi^{2}) = - 2\Lambda (-1)^{n_{c}} s_{2k+1}(m).
      \label{eq:solutions}
\end{equation}
In the last identification, we  used the fact that $s_{2n_{c}+1}(m) =
s_{n_{f}}(m) = m_{1} \cdots m_{n_{f}}$.  This gives explicit solutions
to the vacuum.

Once we have this solution, however, we can obtain other
$2^{n_{f}-1}-1$ solutions as follows.  First note that the curve
Eq.~(\ref{eq:curve}) is invariant under changing signs of even number
of masses.  Therefore, we can change signs of even number of masses
from the solution (\ref{eq:solutions}).  This gives $2^{n_{f}-1}$
solutions in total agreeing with ${\cal N}_{1}$ in the large $\mu$
analysis.  This solution therefore decomposes under $U(n_{f})$ for the
equal mass case as $2^{n_{f}-1}={}_{n_{f}} \! C_{0} + {}_{n_{f}} \!
C_{2}  + \cdots
{}_{n_{f}} \! C_{n_{f}-1}$, reminiscent of the spinor representation.

The case of {\it    equal mass}   deserves further    comments.  As
noted above, we can
flip the signs of even number of quark masses, and therefore a general
situation has $2r$ negative masses $-m$ and $n_{f}-2r$ positive
masses $m$.  Let us study the location of branch points in this
situation.  The curve studied above has the form
\begin{equation}
          y^{2} = -4 \Lambda^{2} \prod_{a=1}^{n_{c}} (x-\alpha_{a})^{2}
\end{equation}
near $x \sim 0$.  Given the solutions Eq.~(\ref{eq:solutions}), we can
write
\begin{eqnarray}
       \lefteqn{ 2\Lambda \prod_{a=1}^{n_{c}} (x-\alpha_{a})
          = 2\Lambda \sum_{k=0}^{n_{c}} (-1)^{k} s_{k}(\alpha) x^{n_{c}- k}
      }    \non \\
     &&   =   2\Lambda \sum_{k=0}^{n_{c}} (-1)^{k} s_{2k}(m) x^{n_{c}- k}
          \nonumber \\
          & & = 2\Lambda \frac{1}{2 \sqrt{x}}
          \left[ \prod_{i=1}^{n_{f}} (\sqrt{x} + im_{i})
                  + \prod_{i=1}^{n_{f}} (\sqrt{x} - im_{i}) \right]
          \nonumber \\
          & & = \Lambda \frac{1}{\sqrt{x}}
          \left[ (\sqrt{x} + im_{i})^{n_{f}-2r}(\sqrt{x} - im_{i})^{2r}
          + (\sqrt{x} - im_{i})^{n_{f}-2r}(\sqrt{x} + im_{i})^{2r } \right].
          \nonumber \\
\end{eqnarray}
Note that $1/\sqrt{x}$ does not introduce a singularity at $x=0$.  If
$4r < n_{f}$, we can factor $(x + m^{2})^{2r}$ from above and hence
$(x+m^{2})^{4r}$ from the curve.  We interpret this factor as the
emergence of $SU(2r)$ gauge theory.  If $4r > n_{f}$, we can factor
$(x+m^{2})^{n_{f}-2r}$ from above and hence $(x+m^{2})^{2n_{f}-4r}$
from the curve.  We interpret this factor as the emergence of
$SU(n_{f}-2r) = SU(2(n_{c}-r)+1)$ gauge theory.  Combining both, we
see that gauge groups up to $SU((n_{f}-1)/2)$ are possible.  This fact
can also be understood from the Higgs branch picture.  When quark
masses are large and equal, one can cancel quark masses by the adjoint
VEVS classically (squark singularity) and obtain $U(k)$ gauge theories
$k = 0, 1, \cdots, (n_{f}-1)/2$ depending on how many components of
the adjoint field is used to cancel the quark masses.  These are all
infrared-free theories and the gauge fields survive the quantum
effects.  While smoothly decreasing the quark masses, the Higgs branch
emanating from the squark singularity does not change due to the
non-renormalization theorem and hence the $U(k)$ theories survive to
the small mass studied using the Coulomb branch.  The effective
low-energy Lagrangian which describes physics around the squark
singularity is therefore nothing but that of  Argyres--Plesser--Seiberg
for $SU(n_{c})$ theories at the   non-baryonic roots \cite{ArPlSei}.
We have shown  (Section
\ref{shown})  that these theories indeed produce ${}_{n_{f}} \!
C_{k}$ vacua upon
$\mu \neq 0$ and generic quark mass perturbation.  Therefore the whole
picture nicely fits together.  On the other hand, the strictly massless
case has the high singularity $x^{2n_{c}}$ and appears to give a new
superconformal theory with a global symmetry $SO(2n_{f})$.  We do not
know of a weakly coupled description of this singularity in terms of a
   local field theory.  We have checked that this
singularity indeed produces mutually non-local dyons for $n_{c}=2$ and
$n_{f}=5$.

Now we come back to the case of smaller $n_{f}$.  Around the Chebyshev
solution in the presence of quark masses,  the curve  is
\begin{eqnarray}
      x y^{2} &=& \left[ x \prod_{a=1}^{(n_{f}-1)/2} (x-\phi_{a}^{2})
                  \prod_{k=1}^{n_{c}-(n_{f}-1)/2}
(x-\phi_{k}^{2}-\kappa_{k}^{2})
          + 2\Lambda^{2n_{c}+2-n_{f}} m_{1} \cdots m_{n_{f}} \right]^{2}
                  \nonumber \\
          & & - 4\Lambda^{2(2n_{c}+2-n_{f})} \prod_{i=1}^{n_{f}}(x+m_{i}^{2}),
\end{eqnarray}
where $\phi_{k}^{2}$ are given by those for the Chebyshev polynomial.
However, for the purpose of studying the behavior around $x=0$, both
$x$ and the shifts $\kappa_{k}^{2}$ can be ignored relative to
$\phi_{k}^{2}$.  We need to know $\prod_{k=1}^{n_{c}-(n_{f}-1)/2}
(-\phi_{k}^{2})$.  This can be calculated as
\begin{eqnarray}
          \prod_{k=1}^{n_{c}-(n_{f}-1)/2} (-\phi_{k}^{2})
          & = & \left. \frac{2\Lambda^{2n_{c}+1-n_{f}}}{\sqrt{x}}
                  T_{2n_{c}+2-n_{f}} \left(\frac{\sqrt{x}}{2\Lambda}\right)
                  \right|_{x\rightarrow 0}
          \nonumber  \\
           & = & \Lambda^{2n_{c}+1-n_{f}}\lim_{t\rightarrow 0} \frac{1}{t}
           \cos\left( (2n_{c}+2-n_{f}) \arccos t\right)
          \nonumber  \\
           & = & (-1)^{n_{c}-(n_{f}-1)/2} (2n_{c}+2-n_{f})
\Lambda^{2n_{c}+1-n_{f}}.
\end{eqnarray}
Then the curve is approximated as
\begin{eqnarray}
      \frac{x y^{2}}{\Lambda^{2(2n_{c}+1-n_{f})}}
          &=& \left[ (-1)^{n_{c}-(n_{f}-1)/2} (2n_{c}+2-n_{f})
                  x \prod_{a=1}^{(n_{f}-1)/2} (x-\phi_{a}^{2})
          + 2\Lambda m_{1} \cdots m_{n_{f}} \right]^{2}
                  \nonumber \\
          & & - 4\Lambda^{2} \prod_{i=1}^{n_{f}}(x+m_{i}^{2}).
\end{eqnarray}
This is nothing but the curve of $USp(2n'_{c})$  theory with
$n'_{c}=(n_{f}-1)/2$    upon
changing normalizations of $x$, $y$, $\phi_{a}^{2}$.  The rest of the
analysis therefore follows exactly the same as in $2n_{c}=n_{f}-1$
case.  Even when other Chebyshev solutions obtained by
$Z_{2n_{c}+2-n_{f}}$ are used, they simply amount to the change of
phase of $\Lambda$ in the above approximate curve and the analysis
remains the same.

\item{ \bf  ii)}  Chebyshev point:  even $n_{f}$

Consider now  even   $n_{f}$ cases.
Again, let us  study the specific case of $n_{f}= 2n_{c}$ first.
    We  shall come back to the more general cases later on.

The Chebyshev solution is obtained in the massless limit by setting
all but one of $\phi_{a}=0$:
\begin{equation}
      x y^{2} = \left[ x^{n_{c}}(x-\phi_{n_{c}}^{2}) \right]^{2}
          - 4\Lambda^{4} x^{2n_{c}}
          = x^{2n_{c}}(x-\phi_{n_{c}}^{2}-2\Lambda^{2})
                  (x-\phi_{n_{c}}^{2}+2\Lambda^{2}).
\end{equation}
We take $\phi_{n_{c}}^{2} = \pm 2\Lambda^{2}$ so that the zero at $x=0$
has degree $2n_{c}$. There is another
isolated zero at $x=\pm 4\Lambda^{2}$.  There is also a branch point at
$x=\infty$.  We first consider the case $\phi_{n_{c}}^{2} = +
2\Lambda^{2}$ and will come back to the case $\phi_{n_{c}}^{2} = -
2\Lambda^{2}$ later on.

Under the perturbation by generic quark masses, we go back to the
original curve.  The only way that the curve
\begin{equation}
      x y^{2} = \left[ x \prod_{a=1}^{n_{c}-1} (x-\phi_{a}^{2})
                  (x-2\Lambda^{2}-\beta)
          + 2\Lambda^{2} m_{1} \cdots m_{n_{f}} \right]^{2}
          - 4\Lambda^{4} \prod_{i=1}^{n_{f}}(x+m_{i}^{2})
\label{arranged}  \end{equation}
can be arranged to have $n_{c}$ double zeros as
\begin{equation}
      x y^{2} = x (x-4\Lambda^{2} -\gamma) \prod_{a=1}^{n_{c}}
      (x-\alpha_{a})^{2},
\end{equation}
is by assuming
\begin{equation}
      \phi_{a} \sim m^{2},
          \qquad \alpha_{n_{c}} \sim m \Lambda, \qquad
      \alpha_{1}^{2} \sim \cdots \sim \alpha_{n_{c}-1}^{2} \sim m^{2}.
\end{equation}
By neglecting $\beta, \gamma, x \ll \Lambda^{2}$, we must    solve the
equation
\begin{eqnarray}
          \lefteqn{
      \left[ -2\Lambda^{2} x \prod_{a=1}^{n_{c}-1} (x-\phi_{a}^{2})
          + 2\Lambda^{2} m_{1} \cdots m_{n_{f}} \right]^{2}
          - 4\Lambda^{4} \prod_{i=1}^{n_{f}}(x+m_{i}^{2})
          } \nonumber \\
          & & = -4\Lambda^{2} x \prod_{a=1}^{n_{c}} (x-\alpha_{a})^{2}
          = 4\Lambda^{2} \alpha_{n_{c}}^{2} x
                  \prod_{a=1}^{n_{c}-1} (x-\alpha_{a})^{2}.
          \label{eq:tosolve3}
\end{eqnarray}
It is interesting to note that this is the curve for the
superconformal $USp(n_{f}-2)$ theory with $n_{f}$ flavors with a
special choice of $g=-1$.  This can be rewritten as
\begin{equation}
      \left[ -x \sum_{k=0}^{n_{c}-1}
                  (-1)^{k} s_{k}(\phi^{2}) x^{n_{c}-1-k}
          + m_{1} \cdots m_{n_{f}} \right]^{2}
          - \prod_{i=1}^{n_{f}}(x+m_{i}^{2})
          = -\frac{\alpha_{n_{c}}^{2}}{\Lambda^{2}} x
                  \prod_{a=1}^{n_{c}-1} (x-\alpha_{a})^{2}.
\end{equation}
By moving terms, we find
\begin{eqnarray}
      \lefteqn{
      \prod_{i=1}^{n_{f}}(x+m_{i}^{2}) }\nonumber \\
      &=& \left[ -\sum_{k=0}^{n_{c}-1} (-1)^{k} s_{k}(\phi^{2}) x^{n_{c}-k}
          + m_{1} \cdots m_{n_{f}} \right]^{2}
          + \frac{\alpha_{n_{c}}^{2}}{\Lambda^{2}} x
                  \left[ \sum_{k=0}^{n_{c}-1} (-1)^{k} s_{k}(\alpha)
x^{n_{c}-1-k
}
                  \right]^{2}. \nonumber \\
          \label{eq:tosolve4}
\end{eqnarray}

Now consider the following polynomial
\begin{equation}
      F(x) = \prod_{i=1}^{n_{f}} (\sqrt{x} + i m_{i})
      = \sum_{k=0}^{n_{f}} i^{k} s_{k}(m) \sqrt{x}^{n_{f}-k}.
\end{equation}
This polynomial can be divided into the ``real'' and ``imaginary''
parts (this is not strictly true because $m$'s are complex, but what is
meant  here is the division between terms of odd powers in $i$ and of
even powers in $i$),
\begin{eqnarray}
      F(x) &=& \sum_{k=0}^{n_{f}/2} (-1)^{k} s_{2k}(m)
      \sqrt{x}^{n_{f}-2k}
      + \sum_{k=0}^{n_{f}/2-1} i (-1)^{k} s_{2k+1}(m)
      \sqrt{x}^{n_{f}-2k-1} \nonumber \\
      &=& \sum_{k=0}^{n_{c}} (-1)^{k} s_{2k}(m) x^{n_{c}-k}
      + i \sqrt{x} \sum_{k=0}^{n_{c}-1} (-1)^{k} s_{2k+1}(m)
      x^{n_{c}-k-1},
\end{eqnarray}
where we   used $n_{f} = 2n_{c}$.  Similarly consider the polynomial
\begin{equation}
      G(x) = \prod_{i=1}^{n_{f}} (\sqrt{x} - i m_{i})
      = \sum_{k=0}^{n_{c}} (-1)^{k} s_{2k}(m) x^{n_{c}-k}
      - i \sqrt{x} \sum_{k=0}^{n_{c}-1} (-1)^{k} s_{2k+1}(m)
      x^{n_{c}-k-1}.
\end{equation}
  From the definition,
\begin{equation}
      F(x) G(x) = \prod_{i=1}^{n_{f}} (x + m_{i}^{2}).
\end{equation}
On the other hand, this product is also given by
\begin{equation}
      F(x) G(x) = \left[ \sum_{k=0}^{n_{c}} (-1)^{k} s_{2k}(m) x^{n_{c}-k}
                  \right]^{2}
      + \left[ \sqrt{x} \sum_{k=0}^{n_{c}-1} (-1)^{k} s_{2k+1}(m)
      x^{n_{c}-k-1} \right]^{2}.
\end{equation}
This is precisely the same as Eq.~(\ref{eq:tosolve4}) upon
identifications
\begin{equation}
      \frac{1}{\Lambda} \alpha_{n_{c}} s_{k}(\alpha) = s_{2k+1}(m), \qquad
      s_{k}(\phi^{2}) = - (-1)^{n_{c}} s_{2k}(m).
      \label{eq:solutions2}
\end{equation}
The first equation gives $\alpha_{n_{c}} = \Lambda s_{1}(m)$
by setting $k=0$ and using $s_{0} = 1$.  In the last identification, we
used the fact that $s_{2n_{c}}(m) = s_{n_{f}}(m) = m_{1} \cdots
m_{n_{f}}$.  Note that $s_{k}(\phi^{2})$ excludes $\phi_{n_{c}}^{2}$
and hence are different from the conventional gauge-invariant
polynomials.  This gives explicit solutions to the vacuum.

Once we have this solution, however, we can obtain other
$2^{n_{f}-1}-1$ solutions as follows.  First note that the curve
Eq.~(\ref{eq:curve}) is invariant under changing signs of even number
of masses.  Therefore, we can change signs of even number of masses
from the solution (\ref{eq:solutions2}).  This gives $2^{n_{f}-1}$
solutions in total agreeing with ${\cal N}_{1}$ in the large $\mu$
analysis.  This solution therefore decomposes under $U(n_{f})$ for the
equal mass case as $2^{n_{f}-1}={}_{n_{f}} \! C_{0} + {}_{n_{f}} \!
C_{2}  + \cdots
{}_{n_{f}} \! C_{n_{f}}$, {\it i.e.}\/, to even-rank anti-symmetric
tensors, reminiscent of the spinor representation.

Now we come back to the other solution $\phi_{n_{c}}^{2} =
-2\Lambda^{2}$.  This choice changes Eq.~(\ref{eq:tosolve4}) to
\begin{eqnarray}
      \lefteqn{
      \prod_{i=1}^{n_{f}}(x+m_{i}^{2}) }\nonumber \\
      &=& \left[ +\sum_{k=0}^{n_{c}-1} (-1)^{k} s_{k}(\phi^{2}) x^{n_{c}-k}
          + m_{1} \cdots m_{n_{f}} \right]^{2}
          - \frac{\alpha_{n_{c}}^{2}}{\Lambda^{2}} x
                  \left[ \sum_{k=0}^{n_{c}-1} (-1)^{k} s_{k}(\alpha)
x^{n_{c}-1-k
}
                  \right]^{2}. \nonumber \\
\end{eqnarray}
This is again the curve for the superconformal $USp(2n_{f}-2)$ theory
with $n_{f}$ flavors with a different choice of $g=+1$.  The sign
changes can be absorved if we flip the sign of $m_{1}$ and change
$\alpha_{n_{c}}$ to $i \alpha_{n_{c}}$.  The sign flip of one of the
quark masses implies that we have an odd number of minus signs in
quark masses.  This solution therefore decomposes under $U(n_{f})$ for
the equal mass case as $2^{n_{f}-1}={}_{n_{f}} \! C_{1} + {}_{n_{f}}
\! C_{3}  + \cdots
{}_{n_{f}} \! C_{n_{f}-1}$,, {\it i.e.}\/, to odd-rank anti-symmetric
tensors, reminiscent of the anti-spinor
representation.

Comments   on the case of {\it  equal mass}  are  in order.  As noted
above, we can
flip the signs of even number of quark masses, and therefore a general
situation has $2r$ negative masses $-m$ and $n_{f}-2r$ positive
masses $m$.  Let us study the location of branch points in this
situation.  The curve studied above has the form
\begin{equation}
          y^{2} = -4 \Lambda^{2} \alpha_{n_{c}}^{2}
                  \prod_{a=1}^{n_{c}-1} (x-\alpha_{a})^{2}
\end{equation}
near $x \sim 0$.  Given the solutions Eq.~(\ref{eq:solutions2}), we can
write
\begin{eqnarray}
          \lefteqn{2\Lambda \alpha_{n_{c}} \prod_{a=1}^{n_{c}-1}
          (x-\alpha_{a})
          = 2\Lambda \alpha_{n_{c}} \sum_{k=0}^{n_{c}-1}
          (-1)^{k} s_{k}(\alpha) x^{n_{c}-1- k}
          = 2\Lambda^{2}
          \sum_{k=0}^{n_{c}-1} (-1)^{k} s_{2k+1}(m) x^{n_{c}-1-k} }
          \nonumber \\
          & & = 2\Lambda^{2}\frac{1}{2 i} \sqrt{x}
          \left[ \prod_{i=1}^{n_{f}} (\sqrt{x} + im_{i})
                  - \prod_{i=1}^{n_{f}} (\sqrt{x} - im_{i}) \right]
          \nonumber \\
          & & = -i \Lambda^{2} \sqrt{x}
          \left[ (\sqrt{x} + im_{i})^{n_{f}-2r}(\sqrt{x} - im_{i})^{2r}
          - (\sqrt{x} - im_{i})^{n_{f}-2r}(\sqrt{x} + im_{i})^{2r } \right].
          \nonumber \\
\end{eqnarray}
If $4r \leq n_{f}$, we can factor $(x + m^{2})^{2r}$ from above and
hence $(x+m^{2})^{4r}$ from the curve.  We interpret this factor as
the emergence of $SU(2r)$ gauge theory.  If $4r > n_{f}$, we can
factor $(x+m^{2})^{n_{f}-2r}$ from above and hence
$(x+m^{2})^{2n_{f}-4r}$ from the curve.  We interpret this factor as
the emergence of $SU(n_{f}-2r) = SU(2(n_{c}-r))$ gauge theory.  This
fact can also be understood from the Higgs branch picture.  When quark
masses are large and equal, one can cancel quark masses by the adjoint
VEV classically (squark singularity) and obtain $U(k)$ gauge theories
$k = 0, 1, \cdots, n_{f}/2=n_{c}$ depending on how many components of
the adjoint field is used to cancel the quark masses.  These are all
infrared-free or scale-invariant theories and the gauge fields survive
the quantum effects.  While smoothly decreasing the quark masses, the
Higgs branch emanating from the squark singularity does not change due
to the non-renormalization theorem and hence the $U(k)$ theories
survive to the small mass studied using the Coulomb branch.  The
effective low-energy Lagrangian which describes physics around the
squark singularity is therefore the one given   by
Argyres--Plesser--Seiberg for $SU(n_{c})$ theories at the   non-baryonic
roots \cite{ArPlSei}. These theories indeed produce
${}_{n_{f}} \! C_{k}$ vacua upon $\mu \neq 0$ and generic quark mass
perturbation.  Therefore the whole picture nicely fit together.  On
the other hand, the strictly massless case has the high singlarity
$x^{2n_{c}}$ and appears to give a new superconformal theory with a
global symmetry $SO(2n_{f})$.  We do not know of a weakly coupled
description of this singularity in terms of a weakly coupled local
field theory.  We have checked that this singularity indeed produces
mutually non-local dyons for $n_{c}=2$ and $n_{f}=4. $

Now we come back to the case of smaller $n_{f}$.  Around the Chebyshev
solution in the presence of quark masses, we study the curve
\begin{eqnarray}
      x y^{2} &=& \left[ x \prod_{a=1}^{n_{f}/2-1} (x-\phi_{a}^{2})
                  \prod_{k=1}^{n_{c}+1-n_{f}/2} (x-\phi_{k}^{2}-\kappa_{k}^{2})
          + 2\Lambda^{2n_{c}+2-n_{f}} m_{1} \cdots m_{n_{f}} \right]^{2}
                  \nonumber \\
          & & - 4\Lambda^{2(2n_{c}+2-n_{f})} \prod_{i=1}^{n_{f}}(x+m_{i}^{2}),
\end{eqnarray}
where $\phi_{k}^{2}$ are given by those for the Chebyshev polynomial.
However, for the purpose of studying the behavior around $x=0$, both
$x$ and the shifts $\kappa_{k}^{2}$ can be ignored relative to
$\phi_{k}^{2}$.  We need to know $\prod_{k=1}^{n_{c}+1-n_{f}/2}
(-\phi_{k}^{2})$.  This can be calculated as
\begin{eqnarray}
          \prod_{k=1}^{n_{c}+1-n_{f}/2} (-\phi_{k}^{2})
          & = & \left. 2\Lambda^{2n_{c}+2-n_{f}}
                  T_{2n_{c}+2-n_{f}} \left(\frac{\sqrt{x}}{2\Lambda}\right)
                  \right|_{x\rightarrow 0}
          \nonumber  \\
           & = & \left. 2\Lambda^{2n_{c}+2-n_{f}}
                  \cos \left[(2n_{c}+2-n_{f}) \arccos
\frac{\sqrt{x}}{2\Lambda}\right]
                  \right|_{x\rightarrow 0}
          \nonumber  \\
           & = & 2 (-1)^{n_{c}+1-n_{f}/2} \Lambda^{2n_{c}+2-n_{f}}.
\end{eqnarray}
Then the curve is approximated as
\begin{eqnarray}
      \frac{x y^{2}}{4\Lambda^{2(2n_{c}+2-n_{f})}}
          &=& \left[(-1)^{n_{c}+1-n_{f}/2}
                  x \prod_{a=1}^{n_{f}/2-1} (x-\phi_{a}^{2})
          + m_{1} \cdots m_{n_{f}} \right]^{2}
                  \nonumber \\
          & & - \prod_{i=1}^{n_{f}}(x+m_{i}^{2}),
\end{eqnarray}
This is nothing but the left hand side of Eq.(\ref{eq:tosolve3}) for
the effective curve of  $USp({n_f } )$   theory
upon changing normalizations of $x$, $y$, $\phi_{a}^{2}$.  The rest of the
analysis therefore is   exactly the same as in the
$2n_{c}=n_{f}$ case.  With other Chebyshev solutions obtained by
$Z_{2n_{c}+2-n_{f}}$, the relative sign between the two terms in the
square bracket changes.  This sign change corresponds to two different
solutions $\phi_{n_{c}}^{2} = \pm 2 \Lambda^{2}$ in the $2n_{c}=n_{f}$
case and hence they give decompositions into even-rank (odd-rank)
anti-symmetric tensors under $U(n_{f})$ for the equal mass
perturbation, respectively.


\item{ \bf iii)}   Special  (baryonic-like)   point

$USp(2n_c)$ theories also have   special  Higgs branch root
(Eq.(\ref{baryonliketext})-Eq.(\ref{baryonliketext3})),   similar to the
baryonic roots of the $SU(n_c)$ theories.   This point is obtained by setting
$\phi_{1}, \cdots, \phi_{n_{c}-\tilde{n}_{c}}\neq 0$, and
$\phi_{n_{c}-\tilde{n}_{c}+1}= \cdots =\phi_{n_{c}}=0$
   in the
original
$USp(2n_c)$  curve
\begin{equation}
    x y^2 = \left[ x \prod_{a=1}^{n_c} (x-\phi_a^2)
      + \Lambda^{2n_c + 2 - n_f} \prod_{i=1}^{n_f} m_i \right]^2
    - \Lambda^{4n_c + 4 - 2 n_f} \prod_{i=1}^{n_f} (x+m_i^2),
    \label{eq:USpcurve}
\end{equation}
   (here and
below, $\tilde{n}_c = n_f - n_c - 2$), leading to
\begin{equation}
          y^{2} = x^{2\tilde{n}_{c}+1}
          \prod_{k=1}^{n_{c}-\tilde{n}_{c}} (x-\Phi_{k})^{2}
                  - 4 \Lambda^{4n_c + 4 - 2 n_f} x^{n_f-1} .
\label{baryonlike} \end{equation}
Nonvanishing $\Phi$'s are taken as
\begin{equation}
          (\Phi_{1}^2, \cdots, \Phi_{k}^2, \cdots,
\Phi_{n_{c}-\tilde{n}_{c}}^2)
          = \Lambda^2 (\omega, \cdots, \omega^{2k-1}, \cdots,
          \omega^{2(n_{c}-\tilde{n}_{c})-1}),
\label{baryonlike2} \end{equation}
where $\omega = e^{\pi i/(n_{c}-\tilde{n}_{c})}$.  Note that our
$\omega$ is the square root of $\omega$ in Argyres--Plesser--Seiberg
paper because of later convenience.  Then the product
$\prod_{k=1}^{n_{c}-\tilde{n}_{c}} (x-\Phi_{k})$ can be rewritten as
$x^{n_{c}-\tilde{n}_{c}} + \Lambda^{2(n_{c}-\tilde{n}_{c})}$, and the
curve becomes
\begin{eqnarray}
    y^{2} &=& x^{2\tilde{n}_{c}+1}
    \left[ (x^{n_{c}-\tilde{n}_{c}} + \Lambda^{2(n_{c}-\tilde{n}_{c})})^{2}
    - 4 \Lambda^{4n_{c}+4-2n_{f}} x^{n_{c}-\tilde{n}_{c}} \right]
\nonumber \\
    &=& x^{2\tilde{n}_{c}+1}
    (x^{n_{c}-\tilde{n}_{c}} - \Lambda^{2(n_{c}-\tilde{n}_{c})})^{2}.
\end{eqnarray}
The double zeros of the factor in the parenthesis are at
\begin{equation}
          x = \Lambda^2 \omega^2, \Lambda^2 \omega^{4}, \cdots,
                  \Lambda^2 \omega^{2k}, \cdots,
                  \Lambda^2 \omega^{2(n_{c}-\tilde{n}_{c})}=\Lambda^2.
\end{equation}

One crucial difference from the $SU(n_c)$ case is that there is no
choice but keep all of the ``large'' double zeros because the zeros at
$x=0$ has an odd power $2\tilde{n}_c+1$ and hence leaves one of the
zeros not doubled anyway.  This explains why the separation between
${\cal N}_1$ and ${\cal N}_2$ works out nicely with the $USp(2n_c)$
theories, but one of the $r$-roots gets mixed up with the baryonic
root for $SU(n_c)$ theories.  Another crucial difference is that there
is no constraint among $\phi_a$'s.  Therefore we can just fix
``large'' ones and study the low-energy curve.  Using
\begin{equation}
    \prod_{k=1}^{n_c-\tilde{n_c}} (x- \Lambda^2 \omega^{2k-1})
    = x^{n_{c}-\tilde{n}_{c}} + \Lambda^{2(n_{c}-\tilde{n}_{c})}
    \simeq \Lambda^{2(n_{c}-\tilde{n}_{c})},
\end{equation}
the curve is approximated as
\begin{eqnarray}
    x y^2 &=& \left[ x \prod_{a=1}^{\tilde{n}_c} (x-\phi_a^2)
      \Lambda^{2(n_{c}-\tilde{n}_{c})}
      + \Lambda^{2n_c + 2 - n_f} \prod_{i=1}^{n_f} m_i \right]^2
    - \Lambda^{4n_c + 4 - 2 n_f} \prod_{i=1}^{n_f} (x+m_i^2)
    \nonumber \\
    &=& \Lambda^{4(n_{c}-\tilde{n}_{c})} \left\{
    \left[ x \prod_{a=1}^{\tilde{n}_c} (x-\phi_a^2)
      + \frac{1}{\Lambda^{n_c -\tilde{n}_c}} \prod_{i=1}^{n_f} m_i \right]^2
    - \frac{1}{\Lambda^{2(n_c -\tilde{n}_c)}} \prod_{i=1}^{n_f}(x+m_i^2)
      \right\}. \nonumber \\
    \label{eq:USpLEcurve}
\end{eqnarray}
This curve clearly describes an infrared free $USp(2\tilde{n}_c)$
theory.

Now we choose $r$ out of $\tilde{n}_c$ $\phi_a$'s to match $r$ of the
mass-squared (${}_{n_f}C_r$ choices), {\it e.g.}\/,
\begin{equation}
          \phi_{1}^2 = -m^2_{1}, \, \cdots, \, \phi_{r}^2 = -m^2_{r},
\end{equation}
while the other $\phi_a^2$ are still allowed to fluctuate but with
magnitudes much less than $m^2$.  Note that the absence of constraints
among $\phi_a$ allows us to have $r \leq \tilde{n}_c$ while we had $r
< \tilde{n}_c$ in $SU(n_c)$ theories.  Then the low-energy curve
(\ref{eq:USpLEcurve}) can further be approximated as
\begin{equation}
    x y^2 = \Lambda^{4(n_{c}-\tilde{n}_{c})}
    \left[ x \prod_{a=1}^{\tilde{n}_c-r} (x-\phi_a^2) \prod_{i=1}^{r} m_i^2
      + \frac{1}{\Lambda^{n_c -\tilde{n}_c}} \prod_{i=1}^{n_f} m_i \right]^2
    - \frac{1}{\Lambda^{2(n_c -\tilde{n}_c)}} \prod_{i=1}^{n_f} m_i^2 ,
\end{equation}
which is nothing but the same as the curve of pure
$USp(2(\tilde{n}_c-r+1))$ Yang--Mills theories with the dynamical
scale
\begin{equation}
    \Lambda_{\it pure}^{2(\tilde{n}_c-r+1)}
    = \frac{1}{\Lambda^{n_c -\tilde{n}_c}}
    \frac{\prod_{i=1}^{n_f} m_i}{\prod_{i=1}^{r} m_i^2}
    \sim m^{2(\tilde{n}_c-r+1)}
    \left( \frac{m}{\Lambda} \right)^{n_c -\tilde{n}_c} \ll
    m^{2(\tilde{n}_c-r+1)} .
\end{equation}
This justifies the assumption of $\phi_a^2 \ll m^2$.  Since this is
the curve of pure $USp(2(\tilde{n}_c-r+1))$ Yang--Mills theories, it
gives $(\tilde{n}_c-r+1)$ vacua, and hence in total we find
\begin{equation}
    {\cal N}_2 = \sum_{r=0}^{\tilde{n}_c} (\tilde{n}_c-r+1) {}_{n_f}C_r.
\end{equation}

\item{\bf  iv)}   Summary of the   vacuum counting in $USp(2n_c)$ theories

There are thus two groups of $N=1$ vacua  predicted by the
Seiberg-Witten curve in  $USp(2n_c)$ theories.   The
Chebyshev point  Eq.(\ref{chevy1}), Eq.(\ref{chevy2}),  gives rise to
${\cal N}_1=(2n_{c}+2-n_{f}) \cdot 2^{n_f-1} $  vacua upon  mass
perturbation ,  while   the special point
Eq.(\ref{baryonliketext}),  Eq.(\ref{baryonliketext2}),  leads to   $
{\cal N}_2 =
\sum_{r=0}^{\tilde{n}_c} (\tilde{n}_c-r+1) {}_{n_f}C_r$  vacua.
Their sum coincides  with the total number of $N=1$ vacua
found from the semiclassical as well as from   large $\mu$ analyses.

\end{description}

\newpage

\noindent{\bf Acknowledgment}

One of the authors (K.K.) thanks Lawrence Berkeley National
Laboratory,  University   of California, Berkeley, and ITP, University of
California Santa Barbara, for their   warm hospitality.  Part of the
work was done during the workshop,
``Supersymmetric Gauge Dynamics and String Theory"  at ITP,  UCSB,
to which  two of us (G.C. and K.K.)
participated.     The authors
acknowledge useful discussions with Philip Argyres,    Alex Buchel,
S. Elitzur,   Amit  Giveon,    Prem Kumar,  Rita Pardini, Misha
Shifman and Arkady Vainshtein.  This  research was supported in part by
the National Science Foundation under Grant No. PHY-94-07194,
PHY-95-14797, and in part by the Director, Office of Science, Office
of High Energy and Nuclear Physics, Division of High Energy Physics of
the U.S. Department of Energy under Contract DE-AC03-76SF00098.

\newpage

\appendix

\section{$SO(2N)\cap  USp(2N) = U(N)$ \label{grouptheory}}

    The $SO(2N)$ generators are the most general pure
imaginary
anti-symmetric matrices.  Break it down to the $N\times N$ blocks,
and write them down as:
\begin{equation}
          \left( \begin{array}{cc}
                  E & F \\ -^{t}F & D
          \end{array}
          \right),
\end{equation}
where $D$, $E$, $F$ are all pure imaginary $N \times N$ matrices, with
the constraints $^{t}E = -E$, $^{t}D = -D$.  The generators of
$USp(2N)$
are given by
\begin{equation}
          \left( \begin{array}{cc}
                  B & A \\ C & -^{t}B
          \end{array}
          \right),
\end{equation}
with the constraints, $^{t}A=A$, $^{t}C=C$, $A^{*}=C$,
$B^{\dagger}=B$.

The way to compare them is to go to the bases of $SO(2N)$ where it
naturally breaks to $N+\bar{N}$ under $U(N)$.  This can be done by the
following rotation,
\begin{eqnarray}
          \lefteqn{
          \left( \begin{array}{cc}
                  1/\sqrt{2} & -i/\sqrt{2}\\
                  -i/\sqrt{2} & 1/\sqrt{2}
          \end{array}
          \right)
          \left( \begin{array}{cc}
                  E & F \\ -^{t}F & D
          \end{array}
          \right)
          \left( \begin{array}{cc}
                  1/\sqrt{2} & i/\sqrt{2}\\
                  i/\sqrt{2} & 1/\sqrt{2}
          \end{array}
          \right) } \nonumber \\
          & &
          = \frac{1}{2}
          \left( \begin{array}{cc}
                  (E+D) + i(F+^{t}F) & i(E-D)+(F-^{t}F)\\
                  -i(E-D)+(F-^{t}F) & (E+D) -i(F+^{t}F)
          \end{array}
          \right) .
\end{eqnarray}
Since both $E$, $D$ are anti-symmetric, $(E+D)$ in the 1st block is
the most general anti-symmetric imaginary matrix, while $i(F+^{t}F)$
is the most general symmetric real matrix.
Their sum gives the most general hermitian matrix.
Comparing to the $USp(N)$ generators, the off-diagonal blocks are
completely symmetric for $USp(N)$ and completely anti-symmetric for
$SO(2N)$, and hence there is no overlap.
While the diagonal blocks are the most general $N\times N$ hermitian
matrices, and overlap completely, hence $SO(2N)\cap USp(2N) = U(N)$.

\section{Semiclassical Monopole States
\label{sec:semicmon}}

The Jackiw--Rebbi zero mode   has the form
\beq   \psi_{L }^{(0)}= i   b   \,  \eta(x); \qquad  \psi_{R }^{(0)} =  b
\, \eta(x).     \label{chzero1}\eeq
in the chiral representation, where
the  commutation  relations are the standard one:
\beq    \{  b^i,    (b^j)^{\dagger}   \}  =
\delta^{ij}.   \label{anticom}\eeq
Given the monopole state  $|\Omega \ket$,    one can construct $2^{n_f}$
positive-norm  states by acting
various  number of   creation operators upon it,
\beq    (b^{i_1} ) ^{ \dagger}  (b^{i_2})^{ \dagger} \ldots  (b^{i_k} ) ^{
\dagger} |\Omega\ket,  \label{monopoles}\eeq
which are all spinless  bosons \cite{SpinIsospin}.
In general $SU(n_c)$ theories  ($n_c \ge 3$)      with $n_f$ flavors
the  Noether currents are:
\beq   J_{\mu}^A=    {\bar   \psi}_D^i \,{\gamma}_\mu  \, \lambda^A_{ij} \,
\psi_D^j    \eeq
so that the charge operators   are
\beq   Q^A =   (b^i)^{\dagger} \,  \lambda_{ij}^A  \,   b_j  +   {\hbox{\rm
non zero modes}}.
\eeq
The semiclassical monopole multiplets are formed by the state $
|\Omega \ket $ which is a singlet, the $n_f $ states $ (b^i)^{\dagger}
|\Omega \ket $ belonging to a ${\underline n_f}$, the $\, _{n_f}C_2 $
states $ (b^i)^{\dagger} (b^j)^{\dagger} |\Omega \ket $ which is the
second rank antisymmetric tensor, etc.  Although semi-classically $2^{
n_f}$ states (\ref{monopoles}) are all degenerates, higher quantum
effects lift such a degeneray in general, and only those belonging to
the same irreducible representation will have the same mass (monopole
multiplets).

In the $USp(2n_c)$ gauge theories the fermionic part of the lagrangian
is
\beq
\sum_{i=1}^{2 n_f} \, \left[ \, {\bar \psi}^i_a i {\bar {\sigma
}}^{\mu}({\cal D}_{\mu})_{ab} \psi_b^i + {1 \o \sqrt 2} \, {\psi
}^i_a\, \phi_{ab}\, \psi^i_c \, J^{bc}+ \mathrm{h.c.} \, \right] \, ,
\eeq
where all fermions are pure left--handed.  In this basis of fermions
the $SO(2 n_f)$ symmetry is manifest and the global symmetry current
is simply
\beq
J_{\mu}^{ij} = {\bar \psi}^i_a {\bar {\sigma }}^{\mu} \psi_a^j - (i
\to j)\, ,
\eeq
their charges ($SO(2 n_f)$ generators) are
\beq
Q_{\mu}^{ij} = {\bar \psi}^i_a \psi_a^j - (i \to j) \, .
\label{songen}
\eeq
The zero mode operators in $USp(2n_c)$ theories have the form
($\gamma^i$ are just the name of these operators, not gamma matrices)
\beq
\psi^i = \gamma^i \, \eta(x) + \ldots, \qquad (i=1,2,\ldots 2n_f) \, ,
\label{sonzer}
\eeq
where
\beq
\gamma^i = b^i + {b^{i}}^\dagger \, ; \qquad \gamma^{n_f +i} = {1 \o
i} (b^i - {b^i}^{\dagger}) \, ,  \qquad (i = 1, 2, \ldots n_f).
\eeq
This particular form of the zero mode contribution reflects the fact
that the fermion basis in which the standard Jackiw--Rebbi solution
Eq.~(\ref{chzero1}) applies and the one which transforms as an
$SO(2n_f)$ vector, are related by:
\beq
{\psi}_a^{i}=\frac{1}{\sqrt{2}}(\hat\psi_a^{2i-1}+\hat\psi_a^{2i}), \quad
{\psi}_a^{n_f+i}=\frac{1}{i\sqrt{2}}(\hat\psi_a^{2i-1}-\hat\psi_a^{2i}), \qquad
(i=1,2,\ldots n_f).
\label{hatbasis}\eeq
where
\beq   \hat\psi_a^{2i-1}   \equiv \psi_{L a}^i   \qquad
      \hat\psi_a^{2i}  \equiv {\bar \psi_{R
a}^i }, \qquad  (i=1,...,n_f).   \label{chzero3}\eeq
Note that $\gamma^i$'s are all real.
These relations show that $\gamma^i$'s  obey the Clifford algebra,
\beq
\{ \gamma^i, \gamma^j \}  = 2 \delta_{ij}.
\eeq
By substituting Eq.~(\ref{sonzer}) into Eq.~(\ref{songen}) we find
that the generators of $SO(2n_f)$ symmetry are (by renormalizing by a
constant):
\beq
\Sigma_{ij} ={ 1 \o 4i  } \, [ \gamma^i, \gamma^j ]\, ,
\eeq
which obviously satisfies the standard $SO(2n_f)$ algebra.

This shows that various monopole states:
\beq
(b^{i_1})^{\dag}  (b^{i_2})^{\dag} \ldots  (b^{i_k})^{\dag} |\Omega
\ket \, .
\label{mon}
\eeq
transform as spinor representations of $SO(2n_f)$.
Furthermore one notes  that  states with   odd or even numbers  of
creation operators have  definite ``chirality''   with respect to
\beq
\gamma^{2N_f+1} =(-i)^{N_f} \gamma^1 \gamma^2 \cdots \gamma^{2N_f}
=\prod_{i=1}^{N_f} (1 - 2b^{i \dagger}b^i)
\eeq
so that each of them  transform independently.
Each  monopole state thus   belongs to    a    spinor representation    of
definite chirality  of the global $SO(2n_f)$  group.

\section{Explicit Formulae for $a_{Di}$, $a_{i}$, ${\de a_{Di} \o \de
u_j},$ and ${\de a_{i} \o \de u_j}$\label{sec:formulas} }

\begin{equation}
    y^{2} = \prod_{k=1}^{n_{c}}(x-\phi_{k})^{2} + 4 \Lambda^{2n_{c}-n_{f}}
    \prod_{j=1}^{n_{f}}(x+m_{j}), \qquad   SU(n_c), \, \,\,  n_f \le 2n_c-2,
\label{curveA1} \end{equation}
and
\begin{equation}
    y^{2} = \prod_{k=1}^{n_{c}}(x-\phi_{k})^{2} + 4 \Lambda
    \prod_{j=1}^{n_{f}}\left(x+m_{j} + {\Lambda \o n_c} \right), \qquad
SU(n_c),
\, \,\,  n_f = 2n_c-1,
\label{curveA2} \end{equation}
with $\phi_{k}$ subject to the constraint $\sum_{k=1}^{n_{c}}\phi_{k} =
0$, and
\begin{equation}
    x y^{2} = \left[ x \prod_{a=1}^{n_{c}} (x-\phi_{a}^{2})^{2}
      + 2 \Lambda^{2n_{c}+2-n_{f}} m_{1} \cdots m_{n_{f}} \right]^{2}
    - 4 \Lambda^{2(2n_{c}+2-n_{f})} \prod_{i=1}^{n_{f}}(x+m_{i}^{2}),
    \quad USp(2 n_c). \label{curveA3}
\end{equation}
In each case, these represent a genus $g=n_c-1$ ($g=n_c$  for the
$USp(2n_c)$  case)   hypertorus, which are
characterized by $2 g$ homology cycles $\alpha_{i}$, $\beta_{i}$,
$i=1,2, \ldots,  g$.  These cycles are taken in the doubly sheeted
$x-$ plane to
surround  two    branch points of $y$, and such that they intersect
pairwise, in the canonical way, $(\alpha_{i} \cdot  \beta_{j}) =
\delta_{ij}.  \,$ ${\de a_{Di}/ \de u_j},$ and ${\de a_{i}/\de u_j}$
are given by   the   $ g  \times  2g$   period integrals of the
holomorphic differentials
(neglecting the normalization constant)  \cite{SW1}-\cite{SUN}
\beq
{\de  {a_{Di}} \o \de u_j} = \oint_{\alpha_{i}}  {dx   \,  x^{j-1} \o
    y};\qquad    {\de  a_{i} \o \de u_j} = \oint_{\beta_{i}}  {dx   \,
x^{j-1} \o
y}\, ;
\eeq
whereas    $a_{Di}$, $a_{i}$ are given by    the integrals of the meromorphic
differential $\lambda$ (defined such as $ d \lambda / du_j= {dx \,
x^{j-1}\o y}; $
\beq
a_{Di}  =  \oint_{\alpha_{i}} \, \lambda,  \qquad
   a_{i} =\oint_{\beta_{i}}  \,  \lambda, \,
   \eeq
where some additive terms proportional  to the bare quark masses are
neglected.

\newpage
\section{Absence of the   Non-Baryonic Root with $r={\tilde n}_c =
n_f-n_c$  \label{sec:absence}  }

In this Appendix we prove the absence of the   non-baryonic root with
$r={\tilde n}_c =   n_f-n_c$ for
$SU(n_c)$   gauge theory.    The nonbaryonic branch root
in question  is characterized by the adjoint   VEVS
\beq    {\hbox {\rm diag}} \, \phi =(\,  {\underbrace  {0,0, \ldots,
0}_{ n_f-n_c}  },   \phi_1, \ldots \phi_{2 n_c- n_f} ),
\qquad
\sum_{a=1}^{2 n_c-n_f}  \phi_a=0:  \eeq
the curve has the  form
\bea y^2  &=& x^{ 2 {\tilde n}_c  }      \prod_{a=1}^{2 n_c- n_f }
(x- \phi_a)^2  - 4 \Lambda^{2 n_c- n_f } x^{n_f}   \non \\
    &=&   x^{ 2 {\tilde n}_c  }   \left[ \prod_{a=1}^{N} (x- \phi_a)^2  - 4
\Lambda^{N} x^{N}\right],       \qquad  N= 2n_c-n_f.  \label{square} \eea
We prove below that  this curve  cannot be put  (for whatever  $\{
\phi_a\}$)     in the form,
\beq    y^2 =   x^{ 2 {\tilde n}_c  }    \prod_{i=1}^{N-1} (x-
\alpha_i)^2  (x- \gamma)(x- \delta), \qquad \gamma\ne \delta.
\label{impossible} \eeq

\begin{description}

\item{1. {\it  Theorem:}  }

{\it   The function
\beq F(x) = \prod_{a=1}^{N} (x- \phi_a)^2  - 4   x^{N},     \qquad
\prod_{a =1}^{N}  \phi_a \ne 0,
\label{fx}  \eeq
$x$, $\{\phi\}$  complex,    cannot have exactly   $N-1$ double  factors.     }

\item{2. $N=2,3,4$}

For   $N=2,3,4$,  we   have checked explicitly that   there are
indeed   no $\phi$   configurations such that
$F(x)$        has    exactly   $N-1$  pairs
of    double factors.  There   are either   $N$ pairs,  as   can be
realized by taking
\beq  \phi_a=   (\omega_0)^a,\qquad    \omega_0= e^{2 \pi i  / N},  \eeq
or   less than   $N-1$  pairs of
double factors.

\item{3.  $N=  2n_c-n_f$  even }

In this case,
\beq F(x)=  F_+(x) F_-(x), \qquad     F_{\pm}(x)  =
\prod_{a=1}^{N} (x- \phi_a)  \pm  2    x^{N/2}.\eeq
   First of all,   there cannot be any common factor   between $F_+(x)$
and $ F_-(x)$.    For if there is one, $(x-x^*)$,
$F_+(x^*)= F_-(x^*)=0,$
    hence  $x^*=0$. It means that   there is an extra power of $x^2  $
in front  (an extra $\phi_a=0$),   which is not possible
because       $\prod_{a=1}^{N} \phi_a \ne 0$.

Since   there are no common factor in $F_+(x)$ and $ F_-(x)$,  in
order to get  at least   $N-1$  double  factors,   one of
$F_+(x)$ and $ F_-(x)$  must be fully doubled up,  say:
\beq   F_+(x) =  \prod_{a=1}^{N} (x- \phi_a)  +  2     x^{N/2}  =
\prod_{a=1}^{N/2} (x- \alpha_a)^2. \label{suppose} \eeq
We wish to prove that in this case  $ F_-(x)$ is   a perfect square also.

In order to show it,  note that  $F(x)$    is   invariant  under the
transformation,
\beq      x \to \omega  x; \quad   \phi_a \to  \omega  \phi_a,
\label{transfo} \eeq
where   $\omega = \exp{2 \pi i /N}$.     Note that    under this
transformation,  $F_+(x)$ and $ F_-(x)$  get exchanged:
\beq     F_+(x)  \to  F_-(x);  \quad      F_-(x)  \to  F_+(x).
\label{accord} \eeq

Assume  now    that  a  configuration $\{\phi\}$  such that
(\ref{suppose} )      holds  was found.   $\alpha_i$'s are
functions of   $\{\phi\}$ :
\beq         \alpha_i=   \alpha_i(\{\phi\}). \eeq
According to   (\ref{accord}),   $ F_-(x)$   can be found by  the
$\omega$  transformation from  $ F_+(x)$:
\beq     F_-(x)   =  \prod_{i=1}^{N/2} [x- \omega^{-1}  \alpha_i
(\{\omega\phi\})   ]^2. \eeq
Thus we have proved   that   if  $F(x) $    has   at least $N-1$
double factors, then
it has $N$  of them.

\item{ 4.   General     $N=  2n_c-n_f$ }

Assume  that  $\{ \phi_a\}$ 's are found such that
\beq
   F(x) =  \prod_{a=1}^{N} (x- \phi_a)^2  - 4  x^{N}\equiv
    \prod_{A=1}^{N-1} (x- \alpha_A)^2  (x- \gamma)(x- \delta), \qquad
\gamma\ne \delta,
\label{impos} \eeq
where   $\alpha_A$'s are all different  among each other and  none of
them coincides  either with
$\gamma$    or with  $\delta$.
The left hand side of Eq.(\ref{impos})  is invariant under the
transformation   Eq.(\ref{transfo}),  so must be also the
right hand side.  That is
\bea  && \prod_{A=1}^{N-1} (x- \omega^{-1}
\alpha_A(\{\omega\phi\}))^2  (x- \omega^{-1}
\gamma(\{\omega\phi\}))(x-
\omega^{-1}  \delta  (\{\omega\phi\})) \non \\
&=&
\prod_{A=1}^{N-1} (x- \alpha_A)^2  (x- \gamma)(x- \delta); \eea
it implies  however
\beq \prod_{A=1}^{N-1} (x- \omega^{-1}  \alpha_A(\{\omega\phi\}))
=\prod_{A=1}^{N-1} (x- \alpha_A),  \label{however}\eeq
\beq
    (x- \omega^{-1}  \gamma(\{\omega\phi\}))(x-
\omega^{-1}  \delta  (\{\omega\phi\})) =
   (x- \gamma)(x- \delta).   \label{however2}\eeq

Now  Eq.(\ref{however}) and    Eq.(\ref{however2}),  which   are
equivalent to
\beq \prod_{A=1}^{N-1} ( \omega  x-   \alpha_A(\{\omega\phi\})) =
\omega^{N-1}  \prod_{A=1}^{N-1} (x-
\alpha_A(\{\phi\})),
\label{however3}\eeq
\beq
    (\omega x- \gamma(\{\omega\phi\}))(   \omega   x-
       \delta  (\{\omega\phi\})) =
    \omega^2   (x- \gamma(\{\phi\}))(x- \delta(\{\phi\})),
\label{however4}\eeq
   imply that   the polynomials
\beq    H_1(x, \phi )= \prod_{A=1}^{N-1} (x- \alpha_A), \eeq
\beq   H_2(x, \phi )= (x- \gamma)(x- \delta), \eeq
of order  $N-1$ and $2$,     are   both   homogeneous   in
$(\{\phi\},   x)$:  namely,
\beq     H_1(x,\phi)=   x^{N-1}+  \sum_{i=1}^{N-1}  s_i(\phi)  \,
x^{N-1-i}; \eeq
\beq     H_2(x,\phi)=   x^{2}+  \sum_{i=1}^{2}  t_i(\phi)  \,  x^{2-i}; \eeq
where   $ s_i(\phi) $ and $ t_i(\phi) $ satisfy
\beq   s_i( \omega  \phi) = \omega^i  s_i(\phi); \quad    t_i( \omega
\phi) = \omega^i  t_i(\phi).\eeq

This is so because  $H_1, H_2$,   being polynomials  in $x$   of
order less than $N$,   have each term  in them  transformed non
trivially under the   $\omega$ transformation.

     It follows now that
\beq  F(x)  =   H_1(x)^2 \cdot    H_2(x) \eeq
is    a   homogeneous   expression in   $(\{\phi\},   x)$  with
nontrivial coefficients in the expansion in $x$,     which contradicts
the    form    of $F(x)$,    Eq.(\ref{fx}).   We have thus shown that
the assumption Eq.(\ref{impos}) is impossible.

\end{description}

\newpage

\section{Monodromies in $SU(3)$ Theories with $n_f=4$ \label{sec:monodromy}}

In this Appendix we  briefly describe the analysis of monodromy transformation
around various  singularities for     $SU(3)$ gauge theory    with $n_f=4$.
To study the monodromies one  sets  $(u,v)$  slightly off the
singularity interested, and lets $(u,v)$  make  a small circle
   around it  in the parameter space (QMS).   From the way  the branch
points  move around and the branch cuts get   entangled
one easily finds the manodromy matrix for $a_{D1}, a_{D2},a_{1},
a_{2}$.         One must also
   study     how the positions of the
branch   points and cuts  are  varied as  one goes  from one singularity to
another.   This allows one to determine  the homology cycles defining
various periods,
$a_{D1}, a_{D2},a_{1}, a_{2}$,  in a globally   consistent manner.
The quantum numbers of the massless states
at each singularity are found  from the nonvanishing eigenvectors  of
the monodromy matrix thus obtained.
   In many cases,  the movements of the branch points can be studied
analytically  as well: we
illustrate below  how such a check can be made, in some examples.

The $SU(3)$  gauge theory  with four flavor has  $17$ vacua
for generic quark  masses and for a nonvanishing adjoint mass.
They  collapse to six
vacua in the limit of equal  quark masses,  three singlets,  two
quartets and one sextet.      For
$\Lambda=2$ and $m=2^{-6}$  they are at:

\vspace{1cm}
\begin{tabular}{|c|l|c|}
      \hline
      1. & $(u,v)=(0,0)$ & sextet  \\
      \hline
      2. & $(u,v)=(-0.92,-0.14)$ & singlet  \\
      \hline
      3. & $(u,v)=(-0.85,0.09)$ & singlet  \\
      \hline
      4. & $(u,v)=(-1.05,-0.02)$ &  quartet  \\
      \hline
      5. & $(u,v)=(-0.95,-0.01)$ &  quartet \\
      \hline
      6. & $(u,v)=(-1.00,-0.06)$ &  singlet   \\
      \hline
\end{tabular}
\vspace{1cm}

\begin{figure}[tb]
      \begin{center}
\epsfig{file=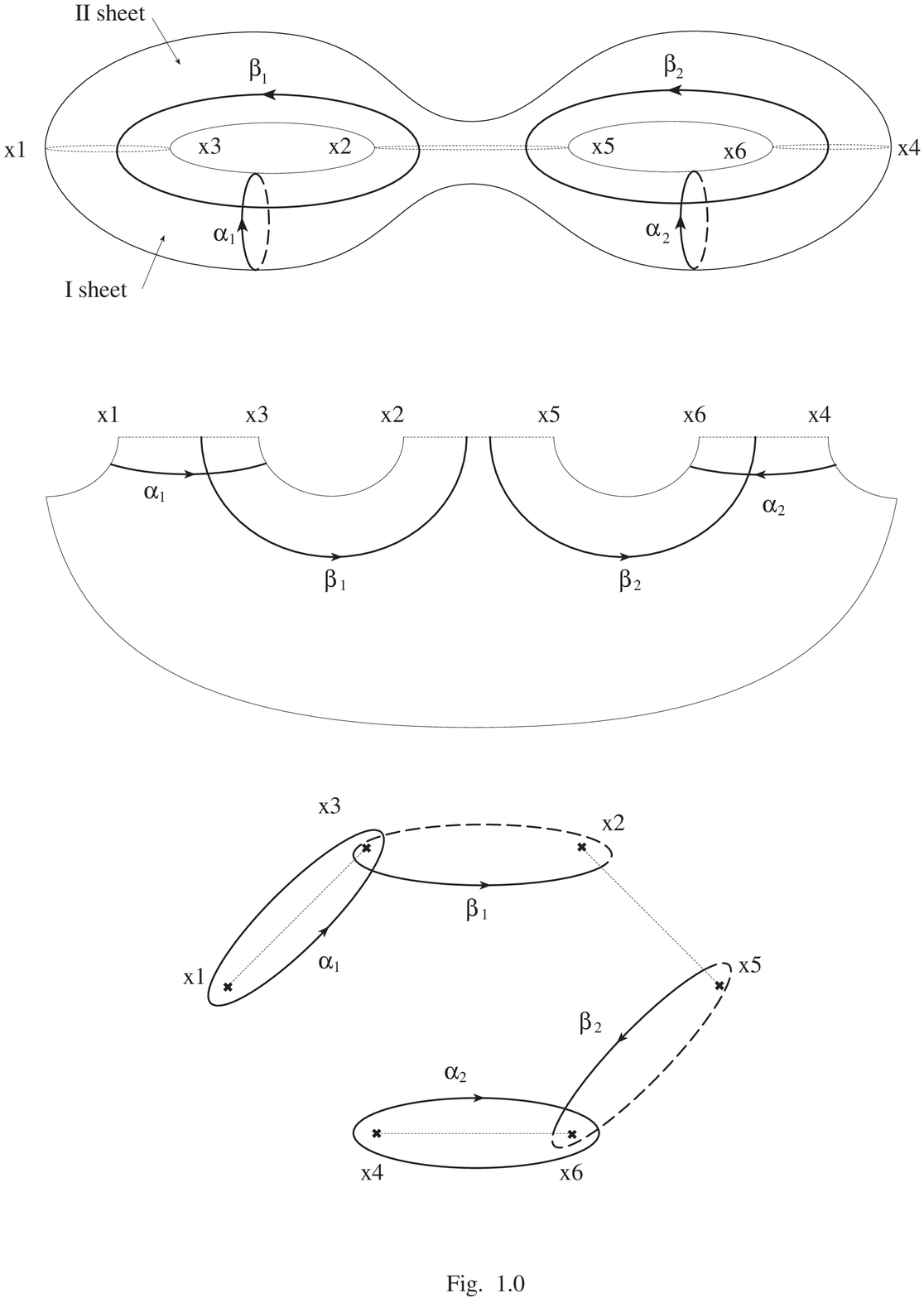,width=10cm}
      \end{center}
      \caption{}
      \label{figsu3nf4monod1}
\end{figure}

      The branch points and  cuts
(dotted lines) are chosen as shown in Fig.  \ref{figsu3nf4monod1}.
Let us  analyze each singularity, starting from  the
singularity 2.

\smallskip
\noindent{ \bf  Singularity 2.}

\begin{figure}[tb]
      \begin{center}
        \epsfig{file=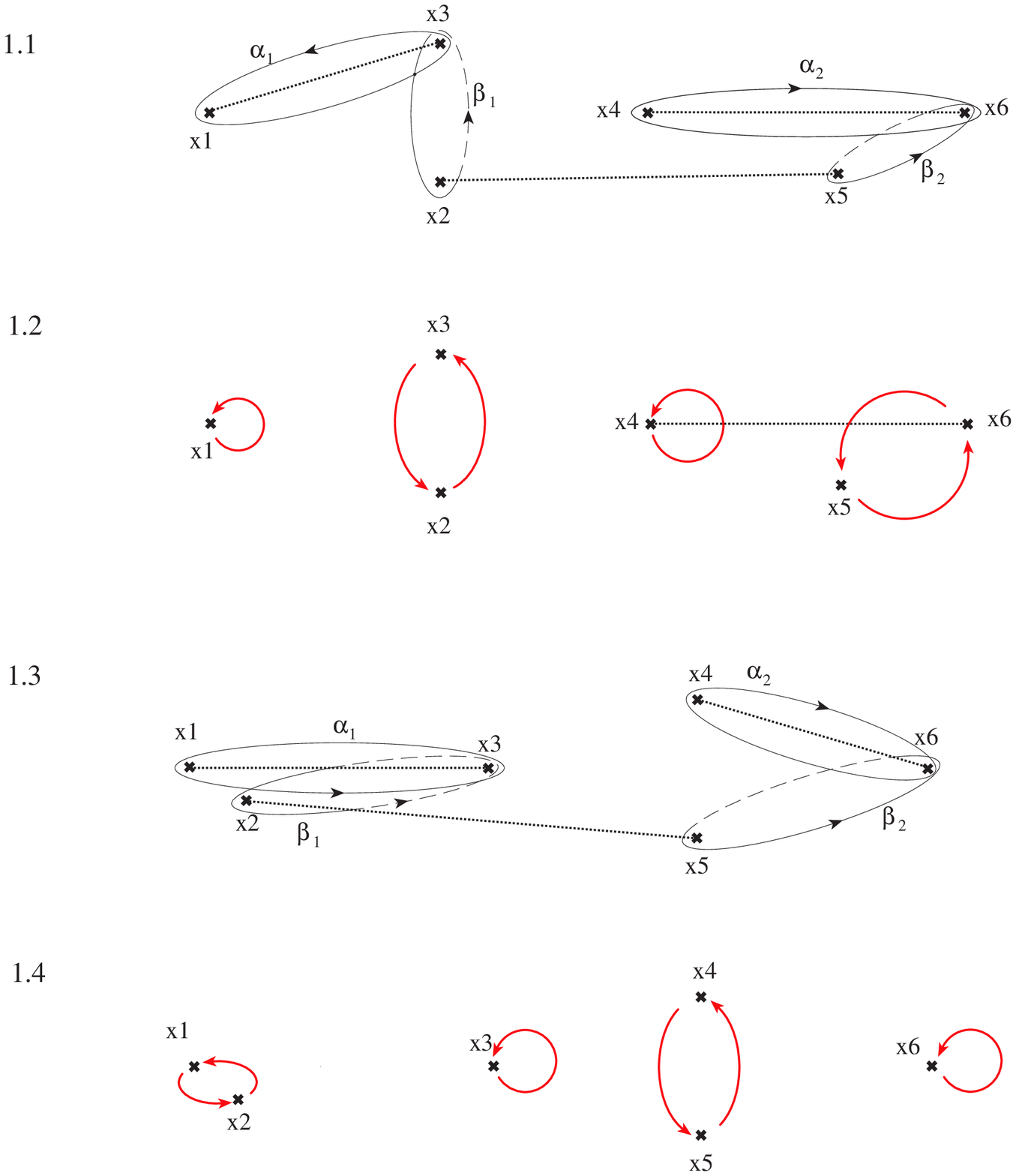,width=10cm}
      \end{center}
      \caption{}
      \label{figsu3nf4monod2}
\end{figure}

      The branch points near this singularity are located as in  (1.1)    in
      Fig. \ref{figsu3nf4monod2}     with $x_{2} \equiv x_{3}$ and
$x_{5}\equiv x_{6}$ on the
      singularity.

      To determine the massless BPS states condensing on the
      singularity, we perform a small circle around the singularity
      itself in the parameter space.  The branch points transform as
      in  1.2 (Fig. \ref{figsu3nf4monod2})       therefore the
monodromy matrix is:
      \begin{equation}
        M_{2}=
        \left( \begin{array}{cccc}
        1 & 0 & 0 & 0 \\
        0 & 1 & 0 & 0 \\
        1 & 0 & 1 & 0 \\
        0 & -1 & 0 & 1
        \end{array} \right) \, .
      \end{equation}
      The eigenvectors, with unimodular eigenvalues, of the (transpose
      of the) monodromy matrix give the charges of the massless particles
      condensing in the singularity.  In this case, we have:
      \begin{equation}
        (n_{m1}, n_{m2}, n_{e1}, n_{e2}) = (1,0,0,0) , \, (0,1,0,0)  \, ,
      \end{equation}
      i.e., the two monopoles of the two abelian factors.

   \smallskip
\noindent{ \bf  Singularity 3.}

   The $x$--plane is shown in   1.3   of  Fig. \ref{figsu3nf4monod2}.  On
      the singularity: $x_{1} \equiv x_{2}$ and $x_{4} \equiv x_{5}$.
      The monodromy matrix is (see 1.4    Fig. \ref{figsu3nf4monod2}):
      \begin{equation}
        M_{3}= \left(
        \begin{array}{cccc}
            0 & 0 & -1 & 0 \\
            0 & 0 & 0 & -1 \\
            1 & 0 & 2 & 0 \\
            0 & 1 & 0 & 2
        \end{array} \right) \, .
      \end{equation}
      The charges of the condensing particles are:
      \begin{equation}
        (n_{m1}, n_{m2}, n_{e1}, n_{e2}) = (0,1,0,1) , \, (1,0,1,0) \, ,
      \end{equation}
      i.e. two dyons of   the two abelian factors.
By an appropriate redefinition of
$       (n_{m1},   n_{m2},  n_{e1},   n_{e2})
$ they become
      \begin{equation}
        (n_{m1}^{'}, n_{m2}^{'}, n_{e1}, n_{e2}) = (0,1,0,0) , \,
(1,0,0,0) \, :
      \end{equation}
two monopoles of $U(1)^2$.

\smallskip
\noindent{ \bf  Singularity 6.}

\begin{figure}[tb]
      \begin{center}
        \epsfig{file=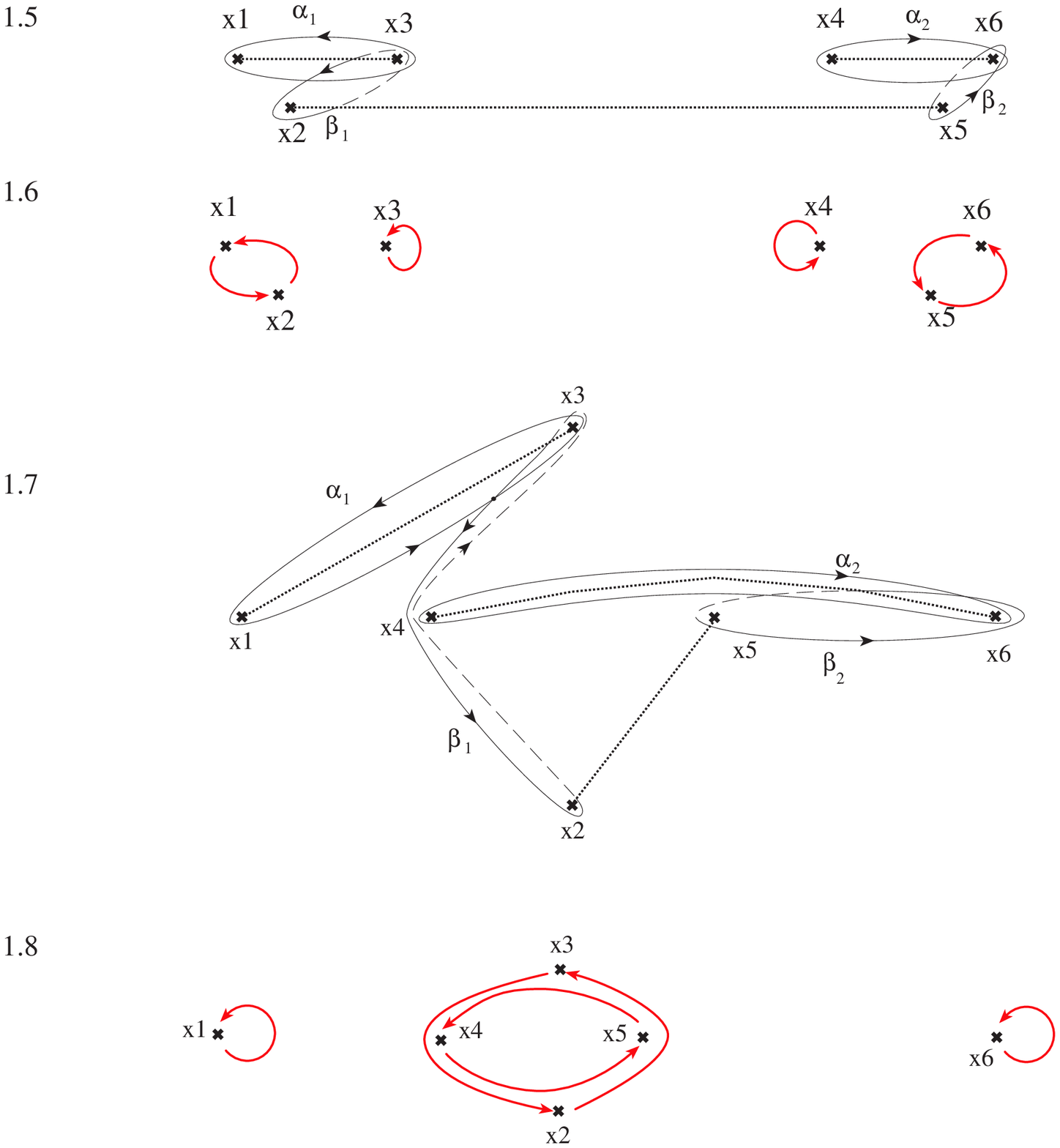,width=10cm}
      \end{center}
      \caption{}
      \label{figsu3nf4monod3}
\end{figure}

   At the singularity $6$  the coalescing branch points are:   $x_{1} \equiv
      x_{2}$ and $x_{5} \equiv x_{6}$.      See (1.5), (1.6)  of Figs.
\ref{figsu3nf4monod3}.       The
monodromy matrix is:
      \begin{equation}
        M_{6}= \left(
        \begin{array}{cccc}
            0 & 0 & -1 & 0 \\
            0 & 1 & 0 & 0 \\
            1 & 0 & 2 & 0 \\
            0 & -1 & 0 & 1
        \end{array} \right) \, .
      \end{equation}
      The charges of the condensing states:
      \begin{equation}
        (n_{m1}, n_{m2}, n_{e1}, n_{e2}) = (1,0,1,0) , \, (0,1,0,0) \, .
      \end{equation}
      Note that, these objects are mutually local.
   By an appropriate redefinition of $  (n_{m1}, n_{m2}, n_{e1}, n_{e2}) $
they become again
      \begin{equation}
        (n_{m1}^{''}, n_{m2}, n_{e1}, n_{e2}) = (0,1,0,0) , \, (1,0,0,0) \, :
      \end{equation}
two monopoles of $U(1)^2$.

\smallskip
\noindent{ \bf  Singularity 1.}

This is the sextet singularity.  The
      branch points ($x$-plane) are in the positions depicted in the
      1.7  (Fig. \ref{figsu3nf4monod3}), with $x_{2} \equiv x_{3}
\equiv x_{4} \equiv x_{5}$
      exactly on the singularity.  Performing a small circle around the
      singularity (in the QMS) the branch points move as indicated in
      Fig.  1.8.   We find the monodromy matrix
     \begin{equation}
        M_{1}=
        \left( \begin{array}{cccc}
        1 & 2 & 0 & 2 \\
        -2 & -2 & -2 & -1 \\
        1 & 0 & 1 & 0 \\
        2 & 1 & 2 & 0
        \end{array} \right) \, .
      \end{equation}
so the massless particles have the quantum numbers:
      \begin{equation}
        (n_{m1}, n_{m2}, n_{e1}, n_{e2}) = (0,1,0,1) , \, (1,2,2,0),
\, (1,1,0,1).
      \end{equation}
The first and the  second   of these states
are relatively nonlocal, hence this is a conformally invariant vacuum.
This was to be expected since this singularity   corresponds to a
class 3 conformal theory
of Eguch et. al. (see the main text).

\smallskip
\noindent{ \bf  Singularity 4, 5.}

\begin{figure}[tb]
      \begin{center}
        \epsfig{file=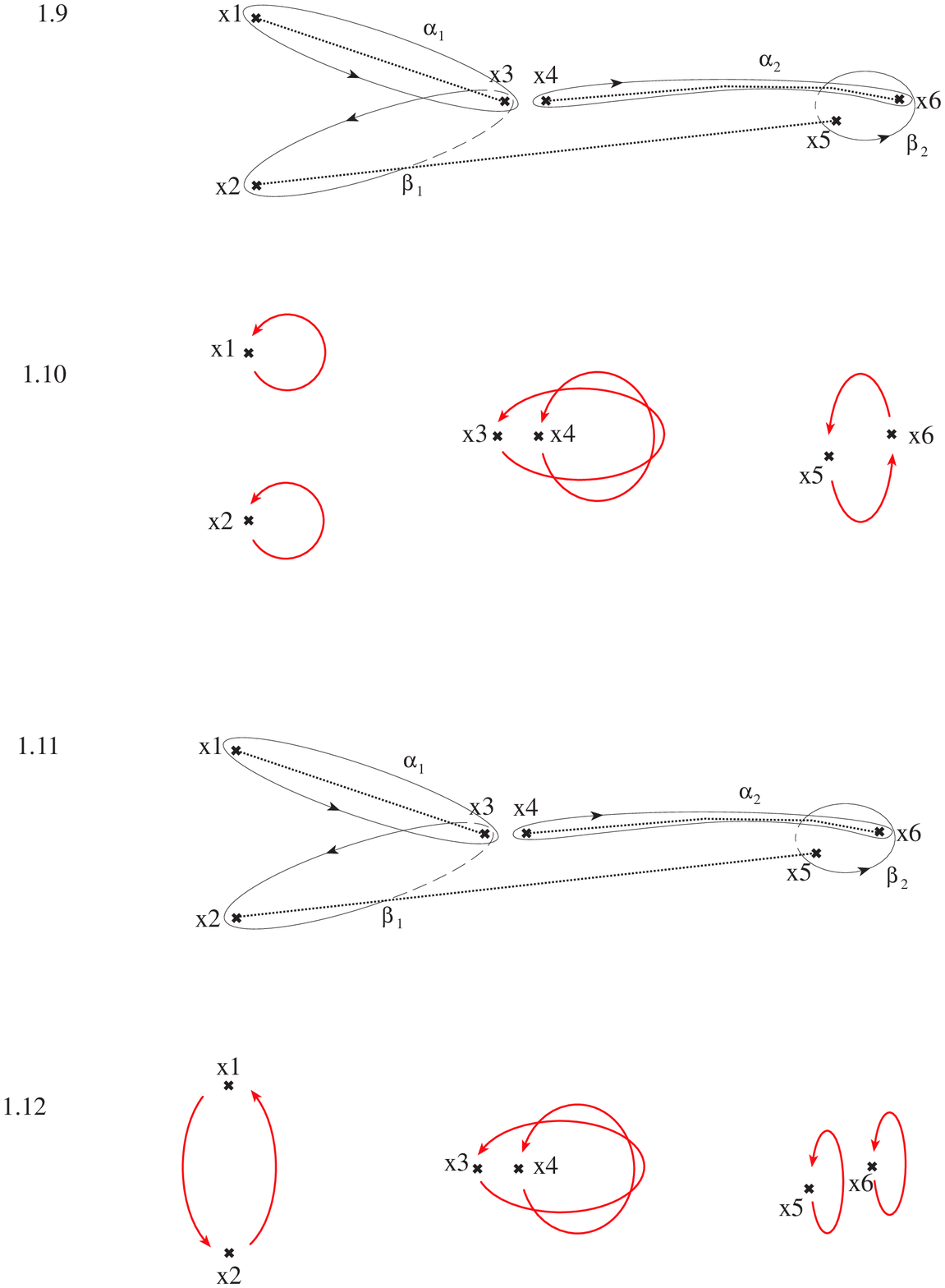,width=10cm}
      \end{center}
      \caption{}
      \label{figsu3nf4monod4}
\end{figure}

    For singularity 4, the $x$--plane
      looks like in  1.9 (Fig. \ref{figsu3nf4monod4})     and the
branch points rotates as in Fig.
      1.10.  The coincident branch points, on the singularity are:
      $x_{3} \equiv x_{4}$ and $x_{5} \equiv x_{6}$.

      As for the  singularity 5, the $x$--plane looks like in  1.11
      and the branch points rotates as in   1.12.  The coincident
      branch points, on the singularity are: $x_{1} \equiv x_{2}$ and
      $x_{3} \equiv x_{4}$.

The monodromy matrix at the singularity $4$ turns out to be
     \begin{equation}
        M_{4}=
        \left( \begin{array}{cccc}
-3 & -4 & -4 & 0 \\
        0 & 1 & 0 & 0 \\
        4 & 4 & 5 & 0 \\
        4 & 5 & 4 & 1
        \end{array} \right) \, .
      \end{equation}
so the massless particles have the quantum numbers:
      \begin{equation}
        (n_{m1}, n_{m2}, n_{e1}, n_{e2}) = (1,0,1,0) , \, (0,1,0,0).
     \label{charges}    \end{equation}
They are relatively local.   Note that
\beq    M_{4} = T^4 A, \qquad       T=  \left( \begin{array}{cccc}  0
& -1  & -1 & 0 \\
     0 & 1 & 0 & 0 \\   1 & 1  &    2 & 0 \\   1 & 2   &  1 & 1
        \end{array} \right), \, \quad
   A=   \left( \begin{array}{cccc}  1 & 0  & 0 & 0 \\
     0 & 1 & 0 & 0 \\   0 & 0  &    1 &0 \\ 0 & -3   &  0 & 1
        \end{array} \right), \, \eeq
with the matrix  $T$ having the same charge  eigenvectors
(\ref{charges})  and $A$
representing    a possible change of homology cycles.  This shows
that the this  singularty correspond to a quartet of
singularity.

The monodromy matrix at the singularity $5$ is
     \begin{equation}
        M_{5}=
        \left( \begin{array}{cccc}
-4 & -4 & -5 & 0 \\
        0 & 1 & 0 & 0 \\
        5 & 4 & 6 & 0 \\
        4 & 5 & 4 & 1
        \end{array} \right) \, .
      \end{equation}
so the massless particles have the same  quantum numbers as  at $4$:
      \begin{equation}
        (n_{m1}, n_{m2}, n_{e1}, n_{e2}) = (1,0,1,0) , \, (0,1,0,0).
      \end{equation}
They are again  relatively local and the same as at the point 4.

\smallskip
\noindent{ \bf  Analytic determination of the monodromy}

One  can actually  study the movement of the branch points
analytically in many cases.
      For instance  take  the
      singularities $4$   or   $5$    and condsider the double branch
point at $-m=  -{1 \o 64}$.
(At both singularities,  one of the double branch point occurs at  $x=-m$.)
   The auxiliary curve
\beq    y^{2}= - 4 \left(x + {1 \o 64} \right)^{4} + (-v-u x + x^3)^2,
\eeq
   can be rewritten as:
      \begin{equation}
        y^{2}= - 4 \left(x + {1 \o 64} \right)^{4} + \left\{ \left(x+ {1 \o
64}\right) f(x) - (u - u_{0})
        x - (v-v_{0}) \right\}^{2}
     \label{curvenc3nf4}     \end{equation}
      with
\beq
f(x) = x^{2} - {1 \o 64} x - 64 v_{0}.
\eeq
The positions of  the singularities $4$ and $5$ are given by
\beq  u_0= -{17149 \o 16384};\quad  v_0= { u_0 \o 64 }= -  {17149 \o 1048576}
\eeq
(the vacuum $4$), and  by
\beq  u_0= -{15613 \o 16384};\quad  v_0= { u_0 \o 64 }= -  {15613 \o
1048576} \eeq
(the vacuum     $5$).

Now, shift $x$ by
\beq
x+ {1 \o 64} \equiv x^{\prime}
\eeq
and rewrite the curve as:
\beq
y^{2} = - 4 {x^{\prime}}^{4} + \left\{ x^{\prime} f\left(x^{\prime}-{1 \o
64}\right) - (u - u_{0}) \left(x^{\prime} - {1 \o 64} \right) - (v - v_{0})
\right\}^{2} \, .
\eeq
For small but nonzero values of
    $u - u_{0}$ and/or  $v - v_{0}$  of order $\epsilon$,
the second term of the right hand side has the form
\beq
\{ (c_{0} x^{\prime} + c_{1} {x^{\prime}}^{2} + \ldots)  + \epsilon
x^{\prime} + \epsilon \}^{2}. \,  \eeq
and   the curve looks like
\beq
y^{2} \simeq {x^{\prime}}^{4} + (x^{\prime} + \epsilon)^{2}.\eeq
Shifting further  $x^{\prime}$  as   $\tilde x = x^{\prime} + \epsilon$
one gets
\beq
y^{2} \simeq \tilde x^{2} + (\tilde x - \epsilon)^{4} \, .
\eeq
The approximately  doublet zeros of the right hand side are then found from
\beq
\tilde x^{2} + \epsilon^{4} - 4 \epsilon^{3} \tilde x + 6
\epsilon^{2} \tilde x^{2} - 4 \epsilon \tilde x^{3} + \tilde x^{4}
=0 \non
\eeq
to be of order  $\tilde x \sim \pm \epsilon^{2}$.     The splitting   of
the double branch point  $-{1 \o 64}$  is therefore     given by:
\beq
x^{\prime} \sim \epsilon \pm \epsilon^{2},  \qquad   .^.. \quad
x\sim -{1 \o 64} +\epsilon \pm \epsilon^{2}.
\eeq
A small circular motion   in QMS     around
$u_{0}$  and/or  $v_0$  yields a convoluted  circular  movements of
the  split  double zeros:
their relative position makes a $4\pi$ rotation (this is relevant to
the monodromy analysis)  while their center of mass
performs a single
$2 \pi$ rotation.

This movement of the branch points  has been confirmed by the
numerical analysis (Fig.
      1.10.).

Now, consider the other double zero at $x_{0} \simeq 1$  (in the case
of the singularity No. $4$).
The curve Eq.(\ref{curvenc3nf4})  becomes
      \begin{equation}
        y^{2}= \left(x + {1 \o 64} \right)^{2}  \, \left[   - 4 \left(x + {1 \o
64} \right)^{2}
+  f(x)^{2} \right]
      \end{equation}
   at the singularity.
The double zero  at $x_{0} \simeq 1$  come from (as can be seen
checked explicitly):
\beq
   f(x) + 2 \left(x+ {1 \o 64} \right)   = (x-x_{0})^{2} \, .
\eeq
Near the singularity the curve looks like
\beq
y^{2} = -4 \left(x+ {1 \o 64}\right)^{4} + \left[ \left(x+ {1 \o 64}\right)
f(x) - \epsilon \right]^2
\eeq
($u-u_0 \sim \epsilon,$ $v-v_0 \sim \epsilon.$)
Near $x=x_0 \sim1$,  the double zero  is   split by the presence of
terms  linear in $\epsilon$:
\beq
(x - x_{0})^{2} -  2 \epsilon \left(x + {1 \o 64}\right) f(x) \simeq 0
\eeq
that is:
\beq
x \simeq x_{0} \pm 2 \epsilon^{1/2} \, .
\eeq
In conclusion, a small $2 \pi$ circle around the singularity implies
that the two branch points (double zero splitted) simply   exchange between
themselves.
Again these movements of the branch points have been confirmed by a
numerical  analysis (1.10 in   (Fig.
\ref{figsu3nf4monod4})).

%
%

\newpage

\begin{thebibliography}{100}

\bibitem{SW1}
N. Seiberg and E. Witten, Nucl.Phys. {\bf B426} (1994) 19; Erratum
\textit{ibid.} \textbf{B430} (1994) 485, hep-th/9407087.

\bibitem{SW2}
N. Seiberg and E. Witten, Nucl. Phys. {\bf B431} (1994) 484,
   hep-th/9408099.

\bibitem{SUN}
P.~C.~Argyres and A.~F.~Faraggi, Phys. Rev. Lett {\bf 74} (1995)
3931, hep-th/9411047;
A. Klemm, W. Lerche, S. Theisen and S. Yankielowicz, Phys. Lett.
{\bf B344} (1995) 169, hep-th/9411048;
Int. J. Mod. Phys. A11 (1996) 1929-1974, hep-th/9505150;   A. Hanany
and Y. Oz, Nucl. Phys. {\bf B452} (1995) 283,
hep-th/9505075;
P.  C.  Argyres, M.  R.  Plesser and A.  D.  Shapere, Phys.  Rev.
Lett.  {\bf 75} (1995) 1699, hep-th/9505100;
P. C. Argyres and A. D. Shapere, Nucl. Phys. {\bf B461} (1996) 437,
hep-th/9509175;   A. Hanany,
Nucl.Phys. {\bf B466}   (1996) 85,  hep-th/9509176.


\bibitem{Sei}
N.~Seiberg,   Nucl. Phys. {\bf B435} (1995) 129, hep-th/9411149.

\bibitem{IntSei}
K. Intriligator and N. Seiberg, Nucl. Phys. {\bf B444}  (1995) 125,
hep-th/9503179;  K. Intriligator and P. Pouliot, Phys. Lett. {\bf
B353} (1995) 471,
hep-th/9505006.


\bibitem{Altri}
T.~Kutasov, Phys. Lett. {\bf B351} (1995) 230, hep-th/9503086;
T.~Kutasov and  A.~Schwimmer, \textit{ibid.} {\bf B354} (1995) 315,
hep-th/9505004;
T.~Kutasov, A.~Schwimmer and N.~Seiberg Nucl.Phys.  {\bf B459} (1996)
455, hep-th/9510222.

\bibitem{AD}
P. C. Argyres and M.R. Douglas, Nucl. Phys. {\bf B452} (1995) 283,
hep-th/9505062.

\bibitem{APSW}
P. C. Argyres,  M. R. Plesser,  N. Seiberg and E. Witten, Nucl. Phys.
{\bf 461} (1996) 71, hep-th/9511154.

\bibitem{EHIY}
T. Eguchi,  K. Hori, K. Ito and S.-K. Yang, Nucl. Phys. {\bf B471}
(1996)
430, hep-th/9603002.

\bibitem{EHsu}
N. Evans, S. D. H. Hsu, M. Schwetz and S. B. Selipsky, Nucl. Phys. {\bf B456}
(1995) 205,
    hep-th/9508002.

\bibitem {KK}  M. Di Pierro and K. Konishi,  Phys. Lett. {\bf B388} (1996) 90,
hep-th/9605178;
K. Konishi, Phys. Lett. {\bf B392} (1997) 101, hep-th/9609021.

\bibitem{LAZ}
   L.  {\'A}lvarez-Gaum\'e, M.  Marino and F.  Zamora, Int.  J.  Mod.
   Phys.  {\bf A13} (1998) 403, hep-th/9703072.


\bibitem{KT}
K. Konishi and H. Terao, Nucl. Phys. {\bf B511} (1998) 264,
hep-th/9707005;    G. Carlino, K. Konishi and H. Terao,  JHEP {\bf  04}
(1998) 003,
hep-th/9801027.


\bibitem {TH}  G. 't Hooft, {Nucl. Phys.} {\bf  B190}   (1981) 455.

\bibitem{NM} Y. Nambu, {Phys. Rev.} {\bf  D10}  (1974) 4262,
   S. Mandelstam, {Phys. Lett.} {\bf 53B}  (1975) 476;
                               {Phys. Rep.} {\bf 23C} (1976) 245.

\bibitem{TaKo}  K. Konishi and  K. Takenaga,   hep-th/9911097, Phys.
Lett B to appear.

\bibitem{ArPlSei}
P. C. Argyres, M. R. Plesser and N. Seiberg, Nucl. Phys. {\bf B471}
(1996)
159, hep-th/9603042.

\bibitem{APS2} P.C. Argyres, M.R. Plesser, and A.D. Shapere,
Nucl. Phys.  {\bf B483}   (1997) 172,   hep-th/9608129.


\bibitem{JR}
R. Jackiw and C. Rebbi, Phys. Rev. {\bf D13} (1976) 3398.


\bibitem{CKM}
G. Carlino, K. Konishi and H. Murayama,
   JHEP  {\bf  0002}  (2000) 004,     hep-th/0001036.
\bibitem{ColWit}
S. Coleman and E. Witten, Phys. Rev. Lett. {\bf 45} (1980) 100.



\bibitem {Hira}
T. Hirayama, N. Maekawa and S. Sugimoto, Progr. Theor. Phys. {\bf 99}
(1998)
843, hep-th/9705069.



\bibitem{Sei94}  N. Seiberg,  Phys. Rev. {\bf D49}  (1994) 6857,
hep-th/9402044.   K. Intriligator, R.G. Leigh  and N. Seiberg,
Phys. Rev. {\bf D50}  (1994) 1092,
hep-th/9403198.



\bibitem{AMKRV}
D. Amati, K. Konishi, Y. Meurice, G. C. Rossi and G. Veneziano, Phys.
Rep.{\bf C162} (1988) 169.


\bibitem{DS}  M.R. Douglas and  S.H. Shenker,  Nucl. Phys. {\bf B447}
(1995) 271,   hep-th/9503163;
   A. Brandhuber and K. Landsteiner,  Phys. Lett. {\bf  B358}
(1995) 73,   hep-th/9507008


\bibitem{Neme}
J. A. Minahan and D. Nemeschansky, Nucl. Phys. {\bf B464} (1996) 3,
hep-th/9507032.1


\bibitem{SpinIsospin}
R. Jackiw and C. Rebbi,  Phys. Rev. Lett {\bf 36} (1976) 1116;
P. Hasenfratz and G. `t Hooft,  {\it ibid.}  1119;
A. S. Goldharber,  {\it ibid.}  1122.




\end{thebibliography}
\end{document}